\documentclass[aps,nofootinbib, showpacs]{revtex4-1}
\usepackage[CJKbookmarks, pdftex, bookmarksnumbered, bookmarksopen, colorlinks, citecolor=blue, linkcolor=blue]{hyperref}
\usepackage[english]{babel}
\usepackage{amstext,amsmath,amssymb,amsfonts,bbm}
\usepackage[latin1]{inputenc}
\usepackage{graphicx}
\usepackage{epstopdf}
\usepackage{amsthm}
\usepackage{tocvsec2}
\usepackage{enumerate}
\usepackage{subfigure}
\usepackage{bm}
\usepackage{color}
\usepackage{tikz}
\usepackage[squaren,Gray]{SIunits}

\newcommand{\va}{\scriptscriptstyle}
\newcommand{\van}{\scriptstyle}

\usepackage{color}

\newcommand{\be}{\nopagebreak[3]\begin{equation}}
\newcommand{\ee}{\end{equation}}

\newcommand{\bee}{\nopagebreak[3]\begin{equation*}}
\newcommand{\eee}{\end{equation*}}
\newcommand{\ba}{\nopagebreak[3]\begin{eqnarray}}
\newcommand{\ea}{\end{eqnarray}}
\DeclareFontFamily{U}{rsfs}{}         
\DeclareFontShape{U}{rsfs}{m}{n}{<5> rsfs5 <6><7> rsfs7          %
  <8><9><10><10.95><12><14.4><17.28><20.74><24.88> rsfs10}{}     %
\DeclareMathAlphabet{\mathfs}{U}{rsfs}{m}{n}                     %
\newcommand{\mfs}[1]{\mathfs {#1}}                               %
\newcommand{\n}{{\nonumber}}

\newcommand{\sH}{{\mfs H}}

\newcommand{\sM}{{\mfs M}}

\newcommand{\sI}{{\mfs I}}
\newcommand{\sO}{{\mfs O}}

 \usepackage{braket}
\newcommand{\N}{\mathbb{N}}

\begin{document}
\count\footins = 1000
\title{A dialog on the fate of information in black hole evaporation}

\author{Alejandro Perez}
\affiliation{{Aix Marseille Univ, Universit\'e de Toulon, CNRS, CPT, Marseille, France}}

\author{Daniel Sudarsky}
\email{sudarsky@nucleares.unam.mx}
\affiliation{Instituto de Ciencias Nucleares,
Universidad Nacional Aut\'onoma de M\'exico, M\'exico D.F. 04510, M\'exico}

\date{\today}


\begin{abstract} 
We present two alternative perspectives for the resolution of Hawking's information puzzle in black hole evaporation. 
The  two views are  deeply contrasting, yet they share  several  common  aspects. One of them  is  the central role played by the existence of 
the interior singularity (whose physical relevance is implied by the singularity theorems of Penrose) that we expect to  be replaced by  a  region described by a more fundamental quantum gravity formulation. Both views  rely on the  notion  that  
the standard    effective quantum field theoretic perspective would require  some  deep modifications.
In this respect both of our scenarios are deeply influenced by  ideas that  Roger Penrose  has  advocated  at various times  and  thus  serves to  illustrate the lasting influence  that  his  deep  thinking on these   and  related matters  continues to have  on the   modern thinking  about  fundamental aspects of  both quantum theory and gravitation.   Despite that,   there is of course  no claim that    R.  Penrose  would   agree   with  any of the concrete proposals that will be  discussed here.  
\end{abstract}

\maketitle

\definecolor{mycolor}{rgb}{0.122, 0.435, 0.698}

\section{Introduction}
One of the most longstanding source of debates   among theoretical  physicists working  on the interface  of gravitation and  quantum theory, as  well as  one of the few existing  sources  of relatively  solid ground for speculations   about quantum gravity,   is  the, so-called, \emph{black hole information loss ``paradox"}. In  fact,  this  issue  was   at the core of  a   now  seminal   work \cite{Hawking:1996ny} in  which R. Penrose and   S.  Hawking  discuss their  differing views about the nature of quantum gravity, and  which   serves as   inspiration  for the format of the present work.
 
There  is actually   a certain  disagreement regarding whether or not the expected   black hole  evaporation,  resulting from  the  Hawking's   radiation effect , leads or not to  a real  problem.  After  briefly   clarifying  for the  benefit of the reader (and using as  a basis the  work \cite{Okon:2017pvc})   the   set of explicit  assumptions  under which one   faces a  true puzzle, we  present two different approaches   to deal  with  the  issue,   correspondingly  favored  by  each  one of  us.  After presenting them,   we  engage in a  debate   pointing out the  difficulties   faced  by  each one  of these views,  and  the price one must pay   in strictly adhering to them. In this last part,  each of  us  will try to convey why,   despite those  difficulties, we both  continue to  see some advantages of  one over the other.   In  so  doing,  we will be   taking  the opportunity to   air in some   detail the  source of our disagreement on this  issue, which has persisted   over  decades of fruitful collaborations among us, and   at the same time we will,  by following their lead,  be honoring  two of the greatest  figures  in our  field, two sources of   enduring  inspiration,  one of which  has recently and unfortunately passed away and one  which  luckily is,  and hopefully will be   with us   for a long time, and  who has brought recognition to the field  well before   his   highly  deserved  honoring with the Nobel Prize.

We not  turn to the problem, staring with a brief  presentation of the  subject, emphasizing  some aspects  that  are often   absent in the discussions  on the matter,  and  are in our view,  responsible in part for some of the  seemingly  unending discussions it generates. 

Sometimes, there is a debate about whether black hole evaporation, presents a puzzle requiring further analysis for its clarification.   Here is a list of assumptions whose acceptance would exhibit an incompleteness in our understanding of the problem of black hole formation and evaporation and hence the existence of what we call a `paradox'.

\begin{enumerate}
\item 
 The energy carried by the Hawking radiation generates a back reaction that decreases the mass of the black hole until  it evaporates  completely or leaves behind a small (in Planck units) mass remnant.
 
\item In the case where there is a small mass remnant,  its number of degrees of freedom cannot possibly encode the information contained in an  initial state with  an arbitrarily  large  mass.

\item  Information is not transferred to a causally disconnected universe.

\item  As a result of quantum gravity effects, the internal singularities within black holes  is   {\it cured}  ( i.e.,   replaced  at the fundamental  level by something that is non singular,  but rather   a {\it normal}  part of the new physics).   Moreover   we'll assume that   some  level of  causal notion   remains  meaningful,   so that  this   region can be said  to   be causally connected with future null infinity, even when the notion of geometry might be replaced by something dramatically different there.

\item Information is not encoded in low-energy degrees of freedom that go through the quantum gravity region. 

\item The complete characterization of the Hawking radiation is not unitarily related to the initial state (in particular, Hawking radiation is not pure,  even  if the initial state is pure).

\item  Quantum evolution is always unitary.
\end{enumerate}

Needless is to  say, that   in addressing the issue at hand,   it would not  be  nearly enough to simply indicate which of the premises  above  one  would chose to reject. For example, if one would deny, say, assumption (3), only then, resolving the information puzzle would require that convincing evidence is provided for how and why a parallel bubble universe is created in the process of gravitational collapse,  and why and how this parallel universe carries the degrees of freedom purifying the Hawking radiation.  Although the new account might  involve important    modifications of present theories, it will have to fit consistently within the established understanding without  doing away with the successes of well tested physical theories, and include potentially testable new predictions.     Of course, for reasons that we will discuss in this article,  a fully satisfactory  account can only  be expected   to arise in the context  of a fully workable theory  unifying  gravitation  and  the  quantum  aspect of nature,  together with a   suitable  version of  quantum theory free of  conceptual  ambiguities, and  a definite ontology.     

In this  work,  we  will not  even attempt   to  provide   such a  complete  final  answer,  but   simply to   describe  and to contrast  two  approaches,  which each   one of   us deems   as  most promising. These two alternatives are based on denying assumption (5), in a way that is different from traditional radiating remnant scenarios in the case of Alejandro's proposal, while denying assumption (7), i.e., the very need to preserve information, in the case of Daniel's proposal. In the rest of the article, we will develop each other's perspective, while  attempting to point  out  their weaknesses and open issues, and discussing the manner  in  which,  we  envision, the ultimate theory might resolve them.  
    
\section{General considerations and the framing of the problem}

It is  worth noting  that there  are, in fact,  three aspects of the black hole information puzzle   which  we find  convenient to consider separately,  before presenting our individual views, while  also pointing out their  linkage. These are:
1) The  question of  whether  the  standard   unitary  evolution  is  somehow  broken  during the process. 
2)  The issue  of the  ultimate  fate of the  information encoded in the initial state  characterizing the situation previous to the  formation of the black hole\footnote{It  is worthwhile pointing out that this is  a different issue  from  that in point  1):  That is,  in principle,  it  is conceivable that   somehow  the  information  about the precise initial  state  be  encoded in  some manner, while  the  actual state of the system is  not unitarily related  to the   original    one.  I.e., bijective  mappings  (not  necessarily  linear)   between   two subsets of a   Hilbert space  need  not  be unitary.}  
3) The nature of the entropy associated to the  black hole,  its interior,   its exterior, as well as that   connected  with the   entanglement  between   both sets of  degrees of freedom.
We  will now  offer  a brief   discussion of various  aspects   related to  these questions in order to    set the ground for  our individual perspectives on the  subject.

\subsection{ The  nature of  the  Entropy of a Black hole}

The discovery  of Hawking radiation, together with the laws of black hole mechanics (including the generalized second law)  implies that thermodynamics is applicable to suitable situations involving black holes. Such state of affairs calls for a more fundamental description, and thus  promotes black holes as a testing ground for quantum gravity where, in particular,  black holes are endowed an entropy given by a quarter if their  horizon  area.   
 Such identification opens up new technical, as well as conceptual, issues for candidate quantum fundamental descriptions. In fact,  some of the most prominent approaches to the subject  have obtained,   within   certain regimes and  making use  of  further   assumptions,   the  correct  result 
 \cite{Strominger:1996sh, G.:2015sda}, while  some independent lines  of analysis have  offered possible general perspectives for accounting for the  proportionality of  black hole entropy and  area
 \cite{Bombelli:1986rw}. However, at present there seem not to exist  a widespread consensus on  the   precise nature of the  {\it entropy},  when talking about situations involving black holes
(views on this questions according to one of us are discussed in \cite{Sudarsky:2002yi}). 

As  we  will see, this is a point  on   which  we  have  rather   different perspectives,  thus    we  will offer the  partial   answers preferred  by  each one of us  in our respective sections, and  then confront the two perspectives  in   the   final discussion. 

\subsection{ The language used in the discussion.}\label{language}

It is   a fact  that,    despite   the substantial progress in several of the most   promising  approaches, we  do not have  at this moment any  fully  workable  and   satisfactory    theory of  quantum gravity.  We are,  for instance, unable to   describe    in precise mathematical terms what  would be the superposition of two space-times (corresponding, say, to  two distinct   quantum  states of  matter \footnote{In some approaches, like in loop quantum gravity, quantum superpositions of (quantum) geometries make sense at the Planck scale as allowed by a well defined kinematical Hilbert space structure. However, the issue on how to bridge to four dimensional  physical states (solutions of the quantum constraints) remains open. Not to mention, the issue of how to recover a continuum smooth spacetime (four dimensional) interpretation which requires dealing with difficult technical, as well as conceptual, problems mixing up questions related to the definition of Dirac observables and the interpretation of quantum mechanics.}).  Thus,  the    language  in which our discussion    will be  framed,   must be  essentially a  semi-classical one, where  we  describe most of the gravitational degrees of   freedom   in    terms of a classical   space-time   with   a  classically valued  metric  tensor  $ (M, g_{ab})$,  while   we  should think  we  are  describing matter in  quantum  terms\footnote{ We  might use  a classical  characterization sometimes,   but  we  must think of that  as  shorthand for  some  suitable  and  well localized  excitation   matter quantum field  in states that  correspond    to something close to coherent   state and  suitable     wave packets.  It should be noted that even  here things   can become highly nontrivial,  if we think  of the effective  spacetime  geometry as that   which   characterizes   the   behavior of matter,   which  is itself  described  in terms   of some  sort of  coarse graining. This point  and related issues are  discussed  in    some  detail  in \cite{Bonder:2017jcu}.}.  Added to this 
    semi-classical   description  of the space-time  geometry,  we will often consider  additional  features representing those  aspects of the fundamental gravitational   degrees of  freedom which are not really describable  in  the  semi-classical  language.   That would be  the case,  for  instance,  when  we talk about the region that replaces the {\it  would be  singularity}. That will be thought  as having a  suitable   quantum  gravity  description that,  in general,  can  not be reduced to  that given  by  a   metric  tensor $ g_{ab}$.
    

\subsection{Mixed states and black holes}
\label{Mix}

Another important issue  that we must clarify   concerns   the notion of a mixed state. In fact, the statement of whether or  
not, during the evaporation process, pure states evolve into mixed states is commonplace in  discussions of this topic. However, what is often  not  sufficiently clarified   is the question of what exactly is  meant  by  such  a mixed state.   { The issue here  is that}, in general, mixed states have two different uses, and one must be rather  careful  concerning  which one is being contemplated  in each situation. On the one hand, mixed states are used to describe either ensembles of identical systems, in which not all members of the ensemble are  necessarily prepared in the same state, or situations in which one does not have full information regarding the actual pure state of the closed system one wants to study (and then, relies on a probabilistic characterization of that state).
These are the, so-called, proper mixtures { (see \cite{d2018conceptual} for   the terminology)}. On the other hand, mixed states are also used to 
effectively describe a \emph{subsystem} of a closed quantum system, which is itself, in a 
pure state. These are the, so-called, improper mixtures. In the first case, the described 
systems  do possess at all times a well-defined quantum state, even though it might be 
unknown. In the second case, if the subsystem in question is entangled  with the rest,  and generically,  by itself,  it simply does not 
possess a well-defined state\footnote{Some researchers  take the view,  often  grounded  
on  the algebraic   approach to QFT,   that  the states of a system,  even if  isolated  and  not  
entangled  with  anything  else,  might not be pure,   with  the   lack of purity being  of   
ontological  rather  than   epistemic  nature.  We regard that posture as  rather 
problematic  for various reasons:  On  the one  hand,   such  enlargement of the possible 
states of a system would signify  an   enormous   explosion in the number of  distinct  physical states  available  to a system without,  at the same time,   providing   a  clear  interpretation   of the nature  of   this enlargement, and,  of course,  calling for  a  general principle indicating  under    exactly which  physical   conditions  would the  state  of a  system   be  pure  and  when mixed.  Could  one,   for instance,  assume  without further   analysis   that    a state of a system   that   leads to the formation of a black hole   might be     arranged to  be initially  pure? In fact,  within the  algebraic  approach to QFT, it is  
known that  the   so  called  "GNS  construction"  (see for instance \cite{Wald:1995yp})-by which one recovers the  standard type of formulation of the QFT in  terms of  a Hilbert space-, based on a  state that is  not pure,  leads to a reducible  representation of the  algebra of field operators, a feature that is usually understood  as indicating  that   states  belonging to  the Hilbert space  of different   
irreducible  components can not  be  connected  to each other,  simply  because   the 
operators  capable of doing so correspond to  alterations of the  state of the system  that  
are associated  to inaccessible   degrees of freedom. However, that interpretation  would 
be impossible if we  hold  to the view  that a complete description of  a system that is 
not entangled  with  other system  might correspond  to a mixed state. In that  case,  it 
seems   to us  it would  be reasonable to expect  some  alternative physical explanation for 
the above mentioned impediment. Lacking that,  we would be forced to conclude  that  our 
theory is  deficient.
 It is  worthwhile noting  that here  we face a situation  where unfortunate  linguistic  ambiguities  have  contributed to    obscure  the  issues.
 We  refer,  of course, to the fact that  when referring to the characterization of  a physical system   the word  ``state" is  used  with various    slightly different   meanings.
 We  might use it to  refer to the  actual  ontological   physical situation,  or   to its  mathematical  characterization in   terms  of a    particular theory.  Moreover,  we might refer to   a  complete  such  characterization or to one that  is  incomplete due to     a lack of  knowledge on our  part (which  might result from  having obtained  only partial information about  that  specific  system),  or as  a result of our   choosing  to   describe in a collective   manner  a certain  ensemble  of similar  systems using  a
 compacted notation,   instead of  describing each  and  every element of the systems in the ensemble  separately).
  Alternatively the  reason  might be that  the theory  does not allow such complete  characterization in the situation of interest ( for instance  in the  case of Quantum Theory  we might be talking about a subsystem of a  larger  system  which is  entangled  wit other subsystems. The reader is thus   warned to  read  this  footnote   with due care.    
   In  short,  should one contemplate other type of situations where a non-pure state is taken to characterize a system,  one would need to explain  precisely  under what conditions something like that might occur, and one would need to understand under what circumstances  could a  generic  system be  characterized by  a \emph{pure} state, as the lack of entanglement with other systems would not be a sufficient condition. To our knowledge, no such account  has been proposed or carefully considered, so far. Therefore, will not dwell further into the subject, taking the conservative posture  we have detailed above,   bypassing the   issues that would have to be clarified if  an alternative  posture   were to be adopted.
   Some  readers  might dismiss  this whole  discussion as  trivial and  unnecessary  however in our  experience  a great deal of  confusion arises  precisely  in  this regard,  and  it  creeps up in the  discussions about the   black hole  information puzzle. }.

In our discussion, the contexts in which mixed states can occur are:  i)   the {\it proper mixtures}, where the non-purity of the quantum state is understood in purely epistemic grounds and, ii)   the { \it improper mixtures} where the lack of purity of the state is taken to reflect the entanglement of such system with  external degrees of freedom.  

As for the usual presentation of the black hole evaporation problem, one often considers a system initially prepared in a well-known pure quantum state, which evolves, according to the unitary dynamics provided by quantum theory, into an equally pure state that, however, now contains a black hole. It is only when we decide to limit our description to those degrees of freedom lying in the black hole \emph{exterior} that we end up with a characterization of the subsystem in terms of a mixed state, which happens to have thermal characteristics. That state is, at this point, considered as an  improper mixture, with  the lack of purity being  just a result of  our  tracing over the degrees of freedom  associated with  the  black hole's  interior. Can one continue to regard this mixture as an improper mixture at very late times, when the situation involves a  black hole  that has evaporated completely?
   For that to make sense, we must identify  some suitable  degrees  of freedom appropriately entangled  with those of our  quantum fields, so to that the  whole system is,  in fact,  in a  pure state.   Alternatively,  we must understand how  that improper mixture  transmutes into a proper mixture.


 One might  
adopt a position where entanglement is  always possible  among  degrees of freedom  as  long as they are not  also related  with causally connected  regions ( except when the  DOF  pertain to the same   spacetime  point). That is  one   would  take as possible the entanglement  between DOF  associated  with regions   which they  can not be  considered  as part of an ordinary  Cauchy surface,    a   situation  which  seems  closely related to the idea  that  somehow ``the interior of the black hole continues to exist''   after the  black hole  evaporation, in the sense of the picture presented  in \cite{Maudlin:2017lye}, (despite   some of the unusual   features that  such posture  entails (see \cite{Okon:2017pvc}).  It seems that,   in  order to do  this in  general, would  require   a mayor revision of  the  basis of standard quantum theory,  because one normally thinks  of entanglement as a feature of a quantum state,  and  a  quantum state as associated with an {\it "instant of time"} in non-relativistic  situations, and with  a  Cauchy  hypersurface in  its  special  or  general relativistic  generalizations\footnote{ The idea of  consider extending the notion of   entanglement to include that between, say,  a  system   {\it today} and  another  system (or even the  same  system) in the remote past or future seems  quite  difficult to make sense of. }.     


 Another possibility   is that the  purity of the state is the result of some novel type of correlations between the  standard   degrees of  freedom  in the field  quantum state and something   else.    Such  possibility will be considered  in some detail in Section \ref{Section-Ale}.  



 The  alternative  is to  view the mixed states one  seems to  obtain  after  the  complete evaporation of  a  black hole  as   proper mixtures.  I.e., one  would have  to take the   view  that  the  final state of the system is pure, but unknown.  A theory involving departures from unitary evolution seems then necessary in order to complete the picture,  by providing,  in  some detail, a characterization  of a  concrete  stochastic evolution for the system in question. That is, in following this  approach,  it seems that  one  would   need to take  the  view that  there is   some  deeper theory  providing  for  a specific (even if in-deterministic) unraveling
 \cite{Pearle:1988uh,Ghirardi:1989cn,Diosi:1998px,Gisin:1989sx} of, say,  the associated Lindblad equation or whatever provides the   detailed  evolution of the  system  described in terms of a density matrix . Natural candidates for  such theories  are spontaneous reduction theories  which  were  initially conceived  as  a path to resolve the  so called measurement problem in quantum theory.  We  will   discuss this  option in some detail  in   Section \ref{Section-Daniel}.

\subsection{What is a BH in quantum gravity?}\label{BH-def}

In classical general relativity, black holes are defined as the region that is causally disconnected from far away observers idealized, in asymptotically flat spacetimes, by future null infinity  $\sI^+$. One writes
 ${\rm BH}=\sM-J^-(\sI^+)$, 
where $J^{-} (\sI^+)$ denotes the chronological past of $\sI^+$ ( in the notation of 
\cite{Wald:1984rg}).
However, on the one hand, semiclassical arguments imply that black holes evaporate, while on the other hand,  quantum gravity considerations suggest that the singularity should be replaced by a quantum gravity region in a more fundamental description. Under such circumstances one can still  make sense of the notion of BH if one suitably adapts the classical definition to the quantum gravity context. Thus, one can, for instance, introduce the notion of {\em semiclassical past $J_{\va \rm C}^{-}(\sI^+)$} of $\sI^+$ as the collection of events in the spacetime that can be connected to $\sI^+$ by {\em semiclassical causal curves} defined as those that never go through the quantum gravity region. 
One might  still need to more  precisely   characterize such curves. A  possibility  that comes to mind  would  be    to require  curvature scalars   such  that the Kretschmann scalar  remains bounded   along  such curves, say $K\equiv R_{abcd}R^{abcd}\le {\rm C} \ell_p^{-4}$ for some constant $\rm C$ of order unity,  but  it   seems  that might   not  cover  all the  shaded  region  densely filled  with defects that,  as  we will  see, occurs  in Alejandro's  approach.
   Given such  definition of semiclassical past, then the black hole region can be defined as \be B \equiv \sM- J_{\va \rm C}^{-}(\sI^+). \ee
Its dependence on the constant $\rm C$ is not an important limitation in the discussion about information. Different $\rm C$ would lead to BH regions that coincide up to Planckian corrections.The previous criterion could be replaced to one based on fluctuations of geometric quantities instead of the quantity $K$ as the quantum region might not necessarily involve in all cases Planckian curvatures (that would be the case in the present scenarios for the future part of the QG region).

In some investigations, emphasis is  placed on local definitions of what the black hole would be arguing that the usual definition cannot be applied in the quantum realm. We do not agree with such perspective because, as our previous definition shows, there is a natural generalization of the standard definition (valid for classical black holes) valid when evaporation is taken into account. One of the concepts evoked to replace the such global definition by a local notion of event horizon is that of  {\it dynamical horizon}  \cite{Ashtekar:2004cn} or (the analogous notion of) holographic screens (later introduced in \cite{Bousso:2015qqa})  which are basically surfaces foliated by apparent horizons. Here we would like to point out that, even if  as mathematical notions these surfaces are certainly interesting, at the physical level such definitions do not present, in our view, clear advantages (for a contrasting opinion the reader is invited to refer to \cite{Ashtekar:2020ifw}). In the first place, these hypersurfaces can be timelike of spacelike depending on the ambient matter flow across them and hence they generally do not define a meaningful notion of the sort of causal separation at play in our definition that is central relevant for discussions about the fate of unitarity and information loss. 
Indeed,  while the black hole is evaporating, the dynamical horizons are timelike and lie outside of the null event horizon as defined here. Thus they do not correctly separate between the degrees of freedom that are bound to fall into the strong gravity regime from those escaping out to $\sI^+$.
Moreover, in the relevant physical situations where one would need to use the notion of dynamically horizons (for instance, when the system is sufficiently close to stationarity and the black hole is macroscopic) dynamical horizons approach (in the past direction) our causal horizon exponentially fast so that, in such contexts,   there is no relevant  distinction  of one from the other.

\begin{figure}[h] \centerline{\hspace{0.5cm} \(
\begin{array}{c}
\includegraphics[width=7cm]{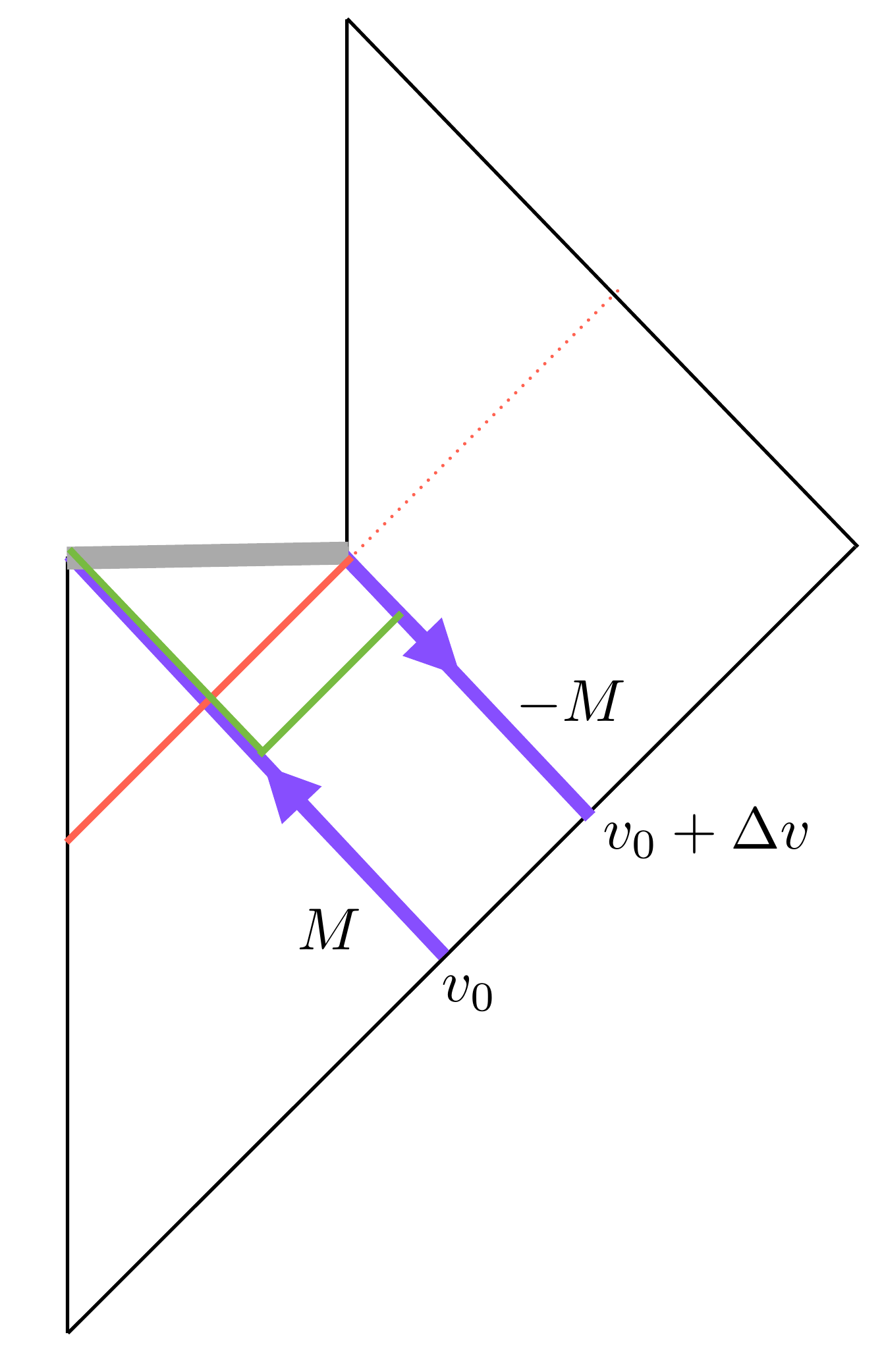} \end{array}
\)}
\caption{A Vaidya black hole formed and annihilated via null shock waves of mass $M$ and $-M$ sent at advanced time $v_0$ and $v_0+\Delta v$ respectively.
The dynamical horizon is represented in green. Our definition of event horizon in red. For large $\Delta v$ dynamical and event horizon have areas approaching exponentially at $v=v_0$. However, the two differ by about $\Delta r\approx 2M$ at $v=v_0+\Delta v$.}
\label{vai}
\end{figure}

{Some intuition about the distinction between our global definition and the local ones can be gained by means of simple spacetime models where the null geodesics can be explicitly integrated.  The simplest situation corresponds to a Vaidya spacetime produced by the collapse of a  shell of mass $M$ in an initially flat spacetime that produces a Schwarzschild black hole at advanced time $v_0$, and it is later  annihilated  by a new shell of mass $-M$ falling into the black hole at advanced time $v_0+\Delta v$.  The Penrose diagram corresponding to this situation is given in Figure \ref{vai}.
There is a dynamical horizon (in the sense of \cite{Ashtekar:2004cn}) that is piece-wise null whose area radius corresponds to the Schwarzschild radius $r_{\rm ah}=2M$ in the portion of spacetime between the two shells.  If we assume that the singularity is a boundary of spacetime, then our definition of black hole horizon coincides with the usual event horizon (the boundary of the past of $\sI^+$). A simple calculation shows that the radius of the event horizon $r_{\rm bh}(v_0)$ when the first shell collapses (at $v=v_0$) is given by the relation
\be -{\Delta v}=2r_{\rm bh}(v_0)+ 4M \log\left|\frac{r_{\rm bh}(v_0)-2M}{2M}\right|.\ee
Which implies that $r_{\rm bh}\le r_{\rm ah}$ and that $r_{\rm bh}\to r_{\rm ah}$ exponentially as $\Delta v\to \infty$.  Notice that the dynamical horizon deviates from the event horizon in the future and the difference is at its maximum of order $2M$!. 
This is because `evaportation' happens instantaneously at $v=v_0+\Delta v$. Nevertheless, it  shows that near the end of black hole evaporation when the process becomes fast one should expect important deviations between the two notions.  On the contrary, when the evaporation rate is slow (when the black hole is macroscopic) the previous model suggests that no significant difference between the dynamical horizon and the event horizon as defined here should exist. 
This can be illustrated in a perhaps more realistic (yet solvable) example where the negative mass shell is replaced by the a linear mass function $M(v)=M[1-(v-v_0)/\Delta v]$ for $v\in [v_0,v_0+\Delta v]$. The 
outgoing geodesics of this Vaidya metric are integrable 
\cite{Blau} and one finds that $|r_{\rm bh}-r_{\rm ah}|\approx 4 r_{\rm ah} M/\Delta v\lesssim  (m_p/M) \ell_p$ (where in the last estimate we used $\Delta v\approx M^3/m_p^2$ matching Hawking evaporation time).    A moment of reflection shows that the previous example captures well the qualitative behavior for  realistic black holes.}

These considerations imply that the simple generalization of the standard black hole notion introduced in this section captures the basic geometric notion that is necessary for our present discussion.

\subsection{Different alternatives for  the  spacetime global geometries  used  in  addressing the   problem}

\begin{figure}[h] \centerline{\hspace{0.5cm} \(
\begin{array}{c}
\includegraphics[width=5cm]{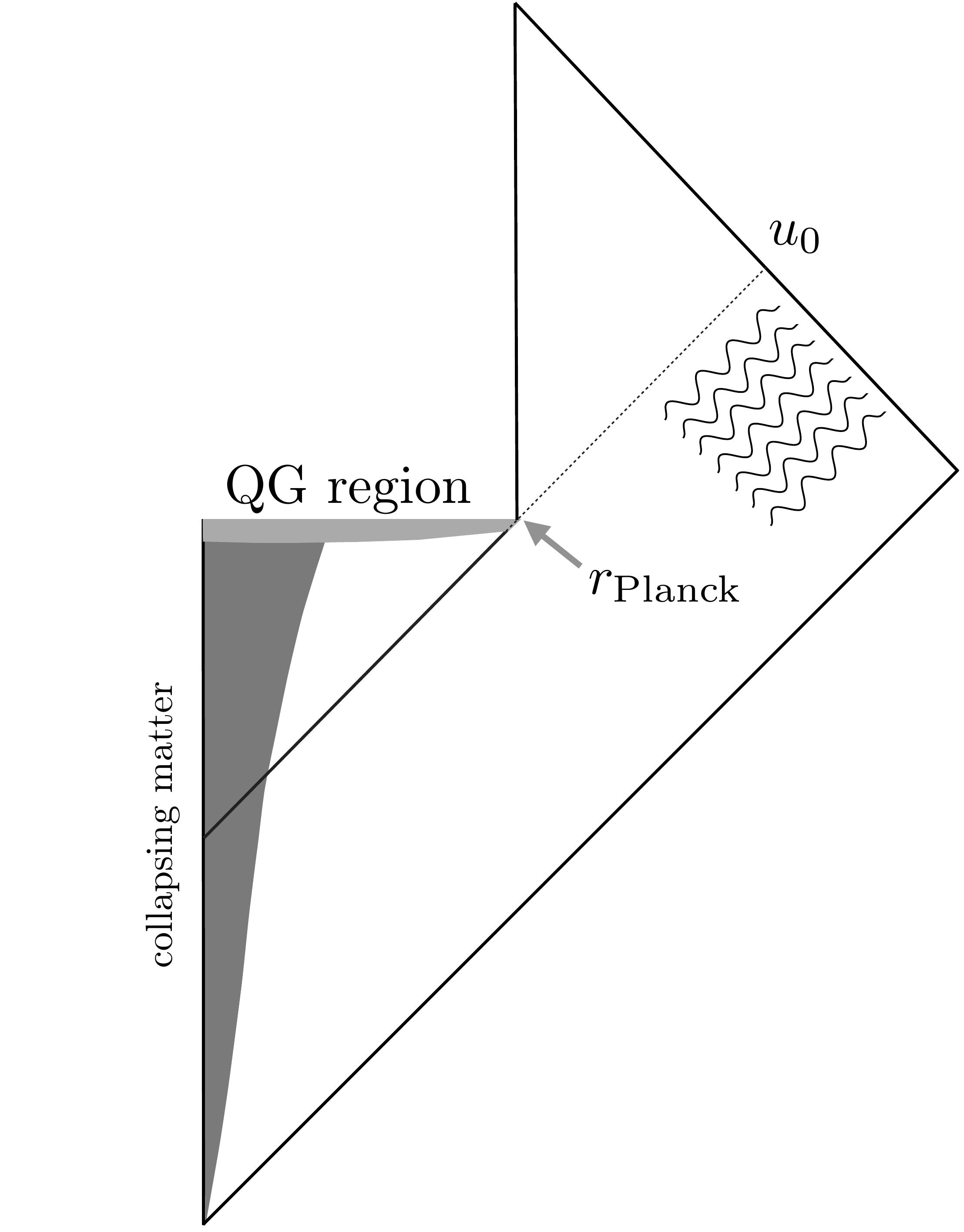} \end{array}\ \  \ \ \ \ \ \ \begin{array}{c}
 \includegraphics[width=5cm]{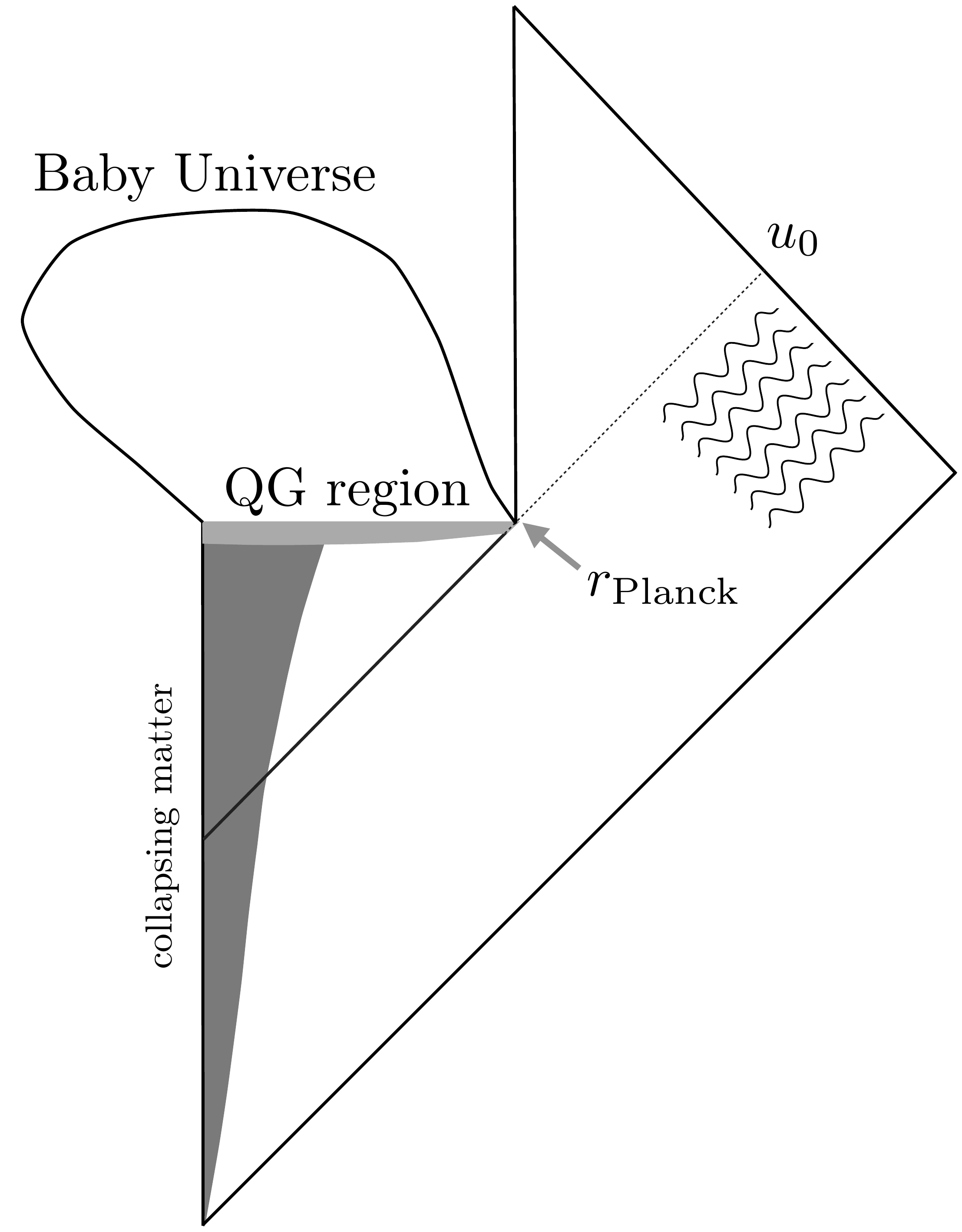} 
\end{array}
\)}
\caption{Two possibilities of global structures where the quantum gravity region remains inside the black hole region after evaporation. The quantum gravity region is represented by the light gray region in the figure.}
\label{PD}

\end{figure}

\begin{figure}[h] \centerline{\hspace{0.5cm} \(
\begin{array}{c}
\includegraphics[width=6cm]{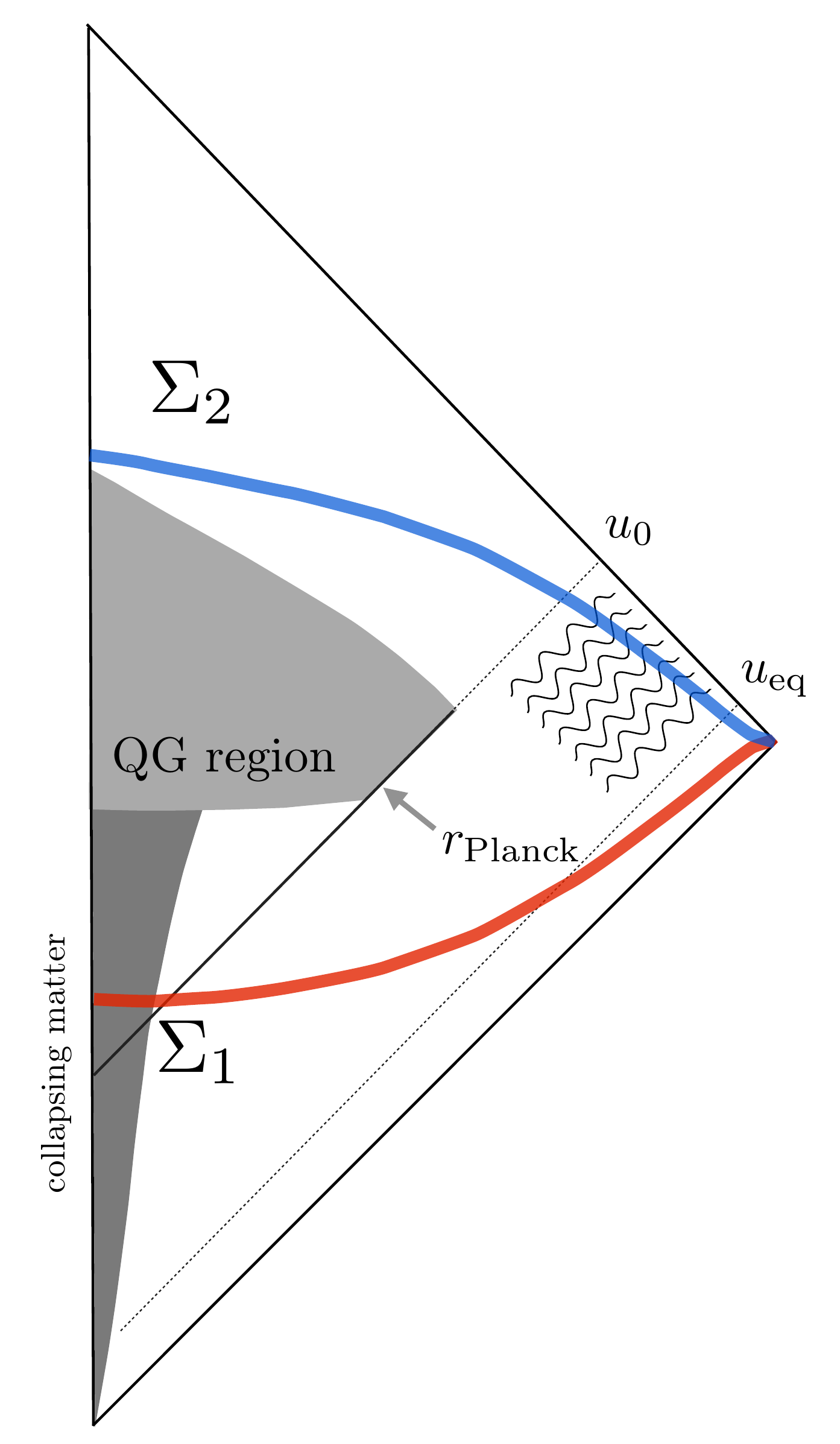} \end{array} \)}
\caption{Global structure where the quantum gravity region `exits' the black hole region at the end of the evaporation.  It  schematically  includes  the quantum gravity region (light gray region in the figure) as  it might be  described from the classical perspective. This is, generically speaking, the global structure of the standard remnant scenarios of unitarity resolution but also admit the alternative accounts. For simplicity of presentation, and given the actual proposals we will be considering in detail, this is the framework where most of the discussion in this article will be placed. Past and future generalized Cauchy surfaces $\Sigma_1$ and $\Sigma_2$ are represented.}
\label{PD-nous1}
\end{figure}

With the previous definition in mind---once the idealization of asymptotic flatness is used---three different global (effective) spacetime structures seem possible when representing the physics of black hole formation and evaporation. The later can be represented in terms of the Carter-Penrose diagrams in Figure \ref{PD} and \ref{PD-nous1}. The first case (left panel in Figure \ref{PD}) corresponds to the case where there is no (classical) future to the quantum gravity region that replaces the classical singularity, and where late observers inherit a spacetime geometry that is filled with matter fields in the vacuum state. The second qualitatively different situation is represented on the right panel of Figure \ref{PD} where there is a semiclassical region (for spacetime geometry and matter fields) to the future of the quantum gravity region. However, this region remains causally disconnected from the  asymptotically flat region representing the outside of the black hole. There a baby-universe emerging from the quantum gravity region that remains inside the black hole according to the previous definition.   


\subsection{Description of the evaporation process.}\label{bulb}

 The bulbous bow was a revolutionary engineering development allowing cargo ships to importantly reduce drag. The idea is that a bulb attached to the front of the ship produces a wave designed (through its geometry) to be in counter phase to the wave produced by the rest of the body of the ship. As seen from far away ( say,  `from infinity' ) the two waves interfere in a destructive manner with the consequence that no energy is `radiated' by the ship in surface waves. The net result is that (conservation of energy implies) the drag is reduced. 
But drag is about friction, thus there is a more fundamental description in terms of the microscopic degrees of freedom between the ship surface and the molecules of water.  Of course, such microscopic description is highly complicated and of no interest for the engineer. The account from infinity is enough for  the purpose  of  the designing  a well  performing  bulbous bow. 
 
Similarly, a key limitation for the full understanding of the black hole formation and  evaporation process is the fact that near the classical would-be-singularity strong quantum gravity effects require a dynamical description that is not available in present approaches to quantum gravity. However, as in the case of the bulbous bow,  some useful  information about the process can be obtained by considering the process from in terms of what  observers perceive at $\sI^+$. 
Such information certainly incomplete, in terms  of the  full understanding the issue at hand, but it provides a few solid landmarks to guide our attempts to produce a satisfactory theoretical model. In this section, we present the picture that emerges from this perspective and a few assumptions, such as the validity of the semiclassical analysis of the Hawking effect, and  the approximate validity of energy conservation.

Thus, the BH-formation and subsequent evaporation can be discussed in the context of an asymptotically flat idealization for which (as we argue now)  one should be able to represent some relevant features in terms of a Penrose diagram. The reason such representation is bound to be only partially correct is the  fact that in some `regions'  there is   no viable   description, even   at an approximated level,   due to the presence of strong quantum gravity effects. This is certainly the case for the regions where classical gravity predicts singularities,  but (as we shall argue), it might also apply to regions to the future of the latter where (in classical terms) we could talk about the presence of a naked-singularity: a quantum gravity region in causal contact with far away observers. 

We are assuming that the process of black hole formation and evaporation is well represented by the Figure \ref{PD-nous1}. In the far past initial conditions are given by an isolated and diluted cloud of matter distribution outside of which matter degrees of freedom are in a state well represented by the vacuum. This cloud has initial condition such that in the future,  a  gravitational collapse will take place and a BH (in the sense of the definition of Section \ref{BH-def} will form). It is important to point out that such cloud of matter in the initial state can very well be replaced by an initial state of suitably focused gravitational waves, i.e., pure geometric degrees of freedom could be the ones responsible for BH formation. This will not play {\it a} central role in either of our accounts of the process.   {In particular, one  should  recall that,   as   far  as our current understanding  of  matter  is concerned, at the fundamental level,  all matter consists of  quantum fields in various   kind of   states,  and  in the general  setting of curved spacetimes,  no  unique  canonical version of vacuum exists.}

The initially diluted matter cloud becomes denser towards the future and eventually undergoes gravitational collapse.  Trapped surfaces form and, assuming that classical energy conditions are satisfied by the initial cloud, singularities of the classical theory appear as granted by  singularity theorems 
\cite{Hawking:1973uf}. 
The previously evoked energy conditions are expected to be violated very strongly by quantum effects near the singularity where one expects a full quantum gravity treatment to be necessary (note that violations of energy conditions are also relevant  in other places, particularly, near the black hole horizon, and are, in fact,  responsible for the   slow horizon   area decease  that accompanies  the Hawking radiation.).
 At this stage some suitably defined version of  the cosmic censorship conjecture 
\cite{CC} 
adapted to the definition in Section \ref{BH-def} imply the formation of a black hole which we assume to have an {\it initially } macroscopic mass $M\gg m_p$. 
We  should recall that the   ADM  remains constant so that when we talk  about  a  mass  changing we are  referring to  the Bondi  mass. 

The formation process can be quite complicated and dynamically involved. Gravitational radiation as well as matter radiation is emitted during this process. Sufficiently late, the system settles to a quasi-stationary state where all the radiation  associated  with the gravitational collapse process has escaped out to $\sI^{+}$. This happens at some  late retarded (equilibrium) time $u=u_{\rm eq}$, where we are using the Bondi (rest) frame of the black hole that is well defined during this quasi-stationary phase. In fact using that the spacetime is extremely close to a stationary asymptotically flat spacetime there is a unique Bondi rest frame---up to rotations and $u$-retarded-time-translations encoded in an arbitrary choice of Bondi cut $u=0$. Once this is done using the quasi-stationary portion of spacetime, the rest Bondi system can be extended to the whole of $\sI^+$.
Thus we assume that, for sufficiently late times $u\ge u_{\rm eq}$ and until close to the retarded time $u_0$ (defined by the end of the evaporation) the spacetime geometry is well approximated by a member of the Kerr-Newman family of stationary black hole solutions with slowly varying mass $M(u)$, angular momentum $J(u)$, and electromagnetic charge $Q(u)$ which evolves according to the corresponding Bondi balance equations at $\sI^{+}$ where the fluxes are given by the Hawking fluxes computed from the Hawking radiation spectrum 
\cite{Hawking:1975vcx}. 

This quasi stationary situation lasts until the mass $M(u)$ becomes of the order of the Planck mass $m_p$ close to retarded time $u_0$. Then Hawking's calculation can certainly not be trusted any longer,  as the curvature close to the black hole horizon becomes of the order of the Planck scale. Only a full quantum gravity description {becomes}  necessary at this stage. However, a central point here is that, independently of the details of such final phase of evaporation,  the total energy available---as measured by our far away Bondi observers---is bounded by something  of the order of the Planck mass. The spacetime geometry and the state of matter fields {ought to be } very well approximated by a flat vacuum state for $u \ge u_0$.

{ We note that despite the expectation that the standard notions of spacetime geometry should   cease to be a viable   characterization of the  {\it quantum  gravity region}\footnote{Which  must include  the  {\it would-be-singularity}  one   encounters  if inappropriately  extending general relativity to regions that would have involve  arbitrarily large curvatures.}, for our  proposals to make  sense,  there must remain at least a notion of causal connectivity allowing,   for instance, to  talk   about  regions to the future of  the quantum gravity region. This is an assumption for both of our proposals which is explicit in the way we represent the Penrose diagrams of interest for our discussion. We are explicitly negating assumption (4) as stated in the introduction.}

\section{Diffusion and granularity: a common feature of both scenarios}

In what follows, we will describe two alternative scenarios for the resolution of the information puzzle. According to Alejandro's perspective, unitarity holds and information is not destroyed during the evaporation process but hidden in correlations with Planckian degrees of freedom that are not accounted for in the smooth effective field theory description. According to Daniel's perspective unitarity is broken and information is really lost due to dynamical collapse (localized) events whose frequency is controlled by entities or degrees of freedom that also escape the standard effective field theory description. A common consequence of this is the expected deviation from energy conservation or the existence of diffusion. Such common feature leads to possibly interesting phenomenology as illustrated by previous work \cite{Josset:2016vrq, Perez:2017krv, Perez:2018wlo, Perez:2019gyd, Perez:2020cwa, Amadei:2021aqd}.    

Turning back to the general line of thought,  in which  such   postures  might be   most  naturally framed, and having no realistically viable   and fully workable theory  of quantum  gravity\footnote{ Note, for instance, that none  of the  ongoing programs  trying to develop  a   theory of quantum gravity is  at this  time  able  to provide   a detailed   mathematical description of the  quantum  space-time   in which  something like,  say,  the moon would  be in a   superposition of two different locations.  The closest to  that which   we can  do,  as far as we know,  is  something like considering     quantizing   the   metric  perturbations   as   if they   were  any ordinary quantum field on a  fixed  space-time  background, a  process that is  conceptually unsatisfactory,     because the  causal structure, which plays  such   a fundamental  role  in the construction of any QFT,   would be that provided   by the  background  metric  and not the complete  space-time metric.}, we  will  argue  by analogy.
{To this  end  let us   consider  fluid dynamics as a useful analogy to the emergence of  what   we  describe  in  classical terms  as  space-time,  which seems to   underscore   a  general  picture  shared    explicitly or intuitively by many  researchers in quantum gravity .}

{The description of a fluid, in terms of the Navier-Stokes equations, is generally understood to emerge from an underlying theory of atoms and molecules interacting via electromagnetic forces; in some regime, such forces might be described in terms of Van der Waals potentials but, in even more general situations    the  full  use  of   quantum electrodynamics can be expected to    be required. When looking at fluid dynamics in such manner, it seems  evident   that such  concepts as fluid density or pressure cannot be taken to have any \emph{fundamental} character,  and that phenomena such as vorticity or turbulence, which at the level of fluid mechanics are essential, cannot be expected to have any special significance or play any crucial role at the fundamental level. Furthermore  we are looking at the   subject  from the standpoint of what  we take  as  this point  to play the role of the fundamental theory,  what the fluid dynamics theoretician calls vorticity and  or  turbulence must  be present,  at least  to some extent  in any real process.  The fact that, under certain conditions, those  can be safely ignored, is just a matter of the degree of approximation we are interested in and the spatio-temporal scales over which we  are interested in  the  study  of  the system. 
For instance  by solving  the equations for a viscosity-free liquid under laminar flow conditions,  in a circular tube,  one might  conclude that the motion will continue forever. It is,  nonetheless,   clear that such result does not represent what    goes on in  nature  over the long run, simply  because  the tiny, but ever-present viscosity, effects  will  inexorably lead to  microscopic vorticity-like behavior,  resulting in  energy dissipation into heat, eventually stopping the circulatory motion. 
}
 
{Now, when adopting the analogous view about the nature of spacetime, classical spacetime notions, such as the metric, the  general nature of  the space-time   casual structure, or even the notion of  what is a  Cauchy hypersurface can be   expected to become secondary or emergent. Thus, it follows that  a black hole should not be considered as a fundamental, or even  a  well  defined   object, any more than a tornado could be taken as  a fundamental notion by, say,  a fluid dynamics theoretician.  Of  course,  a tornado  is  a clear  and fundamental concept to a   student of  weather patterns, and, in fact,  could  be his/her major field of expertise.}

{Now  we might { push} the analogy further, suppose someone is concerned with the issue of energy conservation in fluid dynamics and considers a proposal in which energy is generally conserved, except  when  tornados  are involved ( let say that under such conditions, the proposal holds, part of the energy becomes inaccessible). We believe  that  such  proposal  will not be  taken  as even viable by most l specialists in hydrodynamics,   because people would be likely to suspect that, if during a hurricane some part of the energy went into a strange, inaccessible form, then similar effective losses of energy are likely to occur in other circumstances as well. People might even argue that, to some extent, small hurricanes can be thought of as taking place in almost any situation involving nontrivial fluid motion. Others might argue that, in fact, at a certain scale, the concept of a hurricane becomes simply inappropriate and a description of things in another language is necessary. Regardless, one should expect that  the   fundamental theory ought to  provide  an explanation in terms of the fundamental entities  and  their  dynamics,  of,  for instance,  how  what we call  {\it energy}  when using   the  fluid dynamics language effectively disappears when a hurricane is involved. Moreover, one will also  expect that similar processes,  \emph{not  necessarily } involving hurricanes, but also leading to energy non-conservation, should  be present in many other situations, although perhaps in a less significant fashion.}
 
{What about black holes? If the unitary evolution is, at least in practice, fundamentally lost when a black hole is involved, then, to researchers that do not see space-time as fundamental, this seems unavoidably tied with the conclusion that unitarity evolution {cannot be } exactly valid in every situation. { That is because the   very notion of  a  black hole is  one  that is only well-defined within a setting where the emergent notion of classical space-time becomes available,  and thus} the presence or absence of a black hole can have no implications regarding the behavior of the fundamental theory in generic situations. }

\section{ALEJANDRO'S  PERSPECTIVE}
\label{Section-Ale}
The mathematical models that so far define our successful physical theories are all reversible in the sense that they can predict the future value of the variables they use from their initial values, while conversely the past can be uniquely reconstructed from the values of these variables in the future. The memory of the initial condition is not lost in the dynamics and their information content remains. This is true for classical mechanics and field theory, and it is also true for quantum mechanics and quantum field theory, as long as we do not invoke the postulate of the collapse of the wave function (i.e.,  as long as we do not intervene from the outside via a measurement  or some form of objective collapse happens dynamically as in the framework evoked here by Daniel). In the quantum mechanical  setting, this property boils down to the fact that evolution is given by a unitary operator which can always be undone via its adjoint transformation.

However, in systems with many degrees of freedom an effective form of  irreversibility can arise in the sense that information can be degraded and become unavailable for an observer with limited probing capabilities for whom different configurations might appear indistinguishable. This typically happens when microscopic degrees are involved: a prototypical example is the burning of a newspaper. Common sense claims that the information that was contained in it is simply lost. However, the physicist, trained to find accounts of the facts that accommodate to the statement of  the previous paragraph, explains that the information is  degraded only or hidden (to the point of becoming unrecoverable in practice) in the humongous number of microscopic variables describing the molecular structure of the paper. The words in the newspaper remain `written' (it would be claimed) in the multiple correlations between the degrees of freedom of the molecules in the gas of the combustion diffusing in the atmosphere while transferring the information to even larger and yet pristine portions of the very large phase space of an unbounded universe.  Of course, the physicist cannot prove this; however, it is a consistent story in view of the strongly cherished principle of unitarity.   

Such effective irreversibility is clearly captured in the second law of thermodynamics stating that---for suitable situations involving large number of degrees of freedom---entropy can only increase. At the classical level, this clashes at first sight with the Liouville theorem stating that the phase space volume of the support of  a distribution in phase space is preserved by dynamical evolution. However, nothing restricts the shape of this volume to evolve into highly intricate forms that a macroscopic observer might be unable to resolve. More precisely,  suitable initial conditions (that the observer agent regards as special) are given with a certain uncertainty in accordance to the observers limited measurement capabilities. These special initial conditions are, in our example, the macroscopic configurations of ink particles defining words in the newspaper before the fire reached them.  Such initial conditions and the associated uncertainties are idealized by a distribution in phase space occupying an initial phase space volume of a regular shape. As time passes, the apparent phase space volume (but not the actual volume which remains constant), as measured by a `short-sighted' observer, seems to grow  because of its intrinsic inability to separate the points in phase space that the system actually occupies from the close neighboring ones where the system is not. As emphasized by Penrose---here illustrated by one of his pictures reproduced in Figure \ref{second}---this is possible in the part of the universe we live in because of the very special initial conditions in its remote past.  

I will argue that the general lines of this story remain the same when black hole evaporation is considered. In my view, it is possible that---once a more clear picture of the nature of the quantum gravity dynamics near the places where general relativity predicts singularities is available---the fate of information in black hole formation and evaporation will be explained in terms very similar to those used to describe what happens with the previous familiar  newspaper example. 
In the black hole case the singularity is the place where initially the low energy degrees of freedom---that as macroscopic agents, we describe as quantum field theory modes---are forced to interact with the fundamental building blocks of gravity and matter at the Planck scale (a scale at which effective quantum field theory is expected not to be a reliable description, I will argue). What we call singularity, in classical gravity is the analog of the fire in the previous familiar example: as such it allows for the system to explore a new humongous portion of phase space that was forbidden (because weakly interacting) in the past weak field and weak curvature configuration. In the view that I put forward, information is not destroyed or lost but simply degraded into correlations with Planckian microscopic states which are hidden to macroscopic agents trying to describe dynamics in terms of the limited tools of {\em effective quantum field theory}.
 
 \begin{figure}[h] \centerline{\hspace{0.5cm} \(
\begin{array}{c}
\includegraphics[width=8cm]{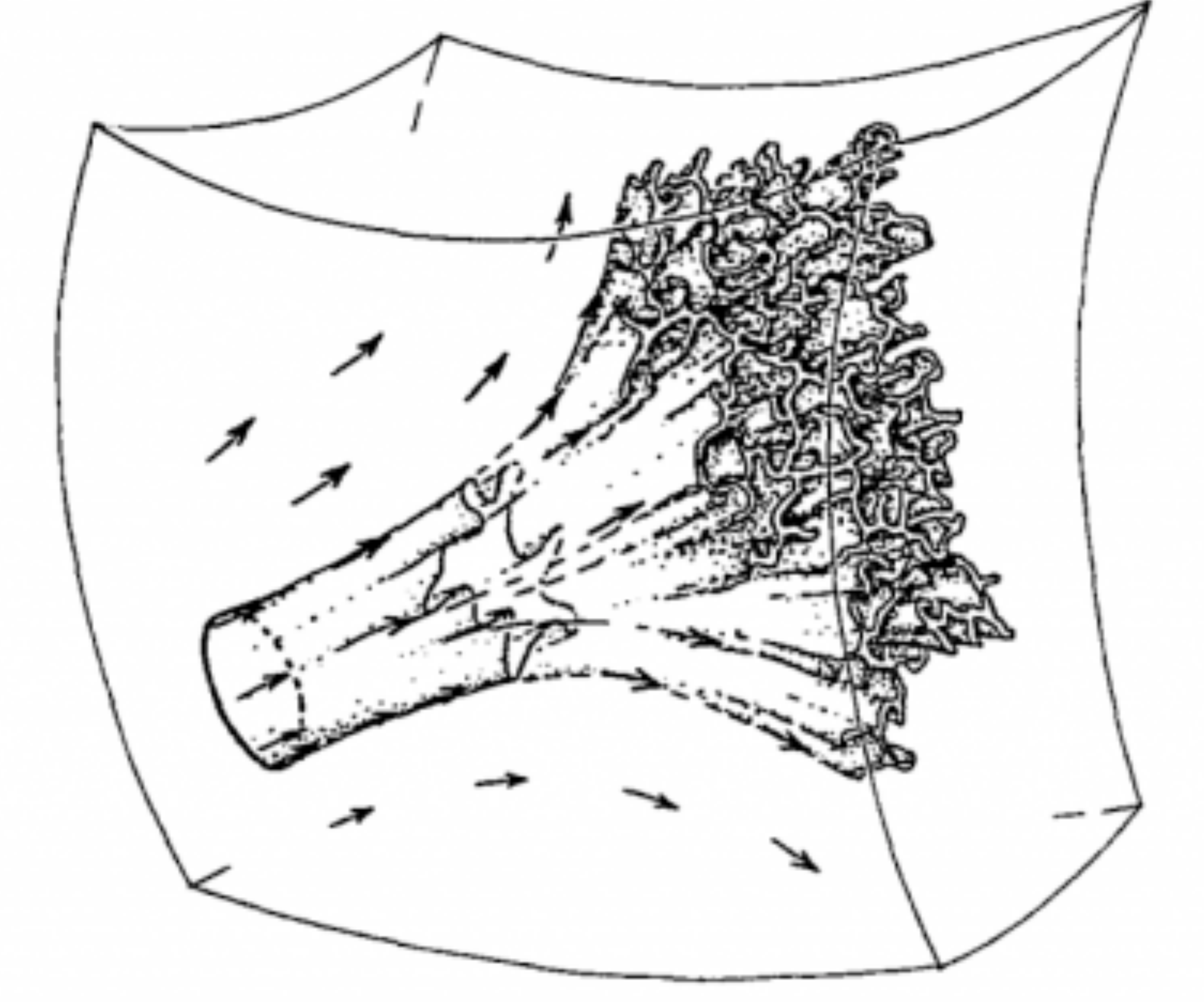}
\end{array}
\)}
\caption{Illustration of the second law in a picture by Penrose.}
\label{second}
\end{figure}

A first indirect indication of the pertinence of this view is that---according to 
general relativity combined with quantum field theory, in a regime where both are expected to be good approximations---large isolated black holes behave like thermodynamical systems in equilibrium. The first striking fact in this respect is the validity of the no-hair theorem 
implying that classical black holes (made of standard model type of matter) are all equivalent in the far future (and as seen from the outside) if their mass $M$, angular momentum $J$, and electric charge $Q$ coincide. The differences characterized in the infinitely many ways in which these black holes can be formed from the gravitational collapse of suitable initial conditions are to be found partly in the radiation flux of matter and gravity at $\sI^+$ before the black hole becomes stationary, but  also in the  `inside' degrees of freedom localized beyond their event horizons `falling' into the interior singularity and hence, apparently (or temporarily), lost to exterior observers. The later degrees of freedom evolve into the region where general relativity fails to be predictable and hence their dynamical fate can only be assessed in the framework of a full quantum gravity description.

Quantum mechanically, stationary black holes (the end result of gravitational collapse according to the classical description to which the no-hair theorem applies) are only close to equilibrium but never in actual equilibrium as---as soon as $\hbar$-effects are taken into account---they radiate particles at the Hawking temperature losing energy extremely slowly at least when they are macroscopic.  When perturbed, they come back to quasi-equilibrium to a new state and the process satisfies the first law of black hole  thermodynamics
\be\label{1st}
\delta M=T\delta S_{\rm bh}+\Omega \delta J+\Phi \delta Q,
\ee
 with an entropy equal  $S_{\rm bh} =A/(4\ell_p^2)$  where $A$ is the horizon area. Under such perturbations, it is expected that the total entropy of the outside universe can only increase. Namely, according to the so-called generalized second law  (GSL)
\be  \label{gsl}
\delta S=\delta S_{\rm matter}+\frac{\delta A}{4}\ge 0, 
\ee
where $\delta S_{\rm matter}$ represents the entropy of whatever is outside the black hole (including for instance the emitted radiation). The fact that black holes are thermal systems satisfying the previous thermodynamical properties compels to consider the existence of microscopic degrees of freedom responsible for the the origin of their entropy. These microscopic degrees of freedom are actually well identified in certain approaches to quantum gravity, such as loop quantum gravity (see \cite{BarberoG:2015xcq, Perez:2017cmj} and references therein).

Now, if such microscopic degrees of freedom at the Planck scale are the ones responsible for black hole entropy, then---in analogy to the molecules that constitute the newspaper in our introductory example---they must be essential for   understanding the problem of information in black hole evaporation. This is the central postulate of my view which, I will argue,  allows for the process of evaporation to remain compatible with unitarity of quantum evolution and avoids (as we will see) the draw backs of traditional remnant scenarios or poorly motivated (yet very popular) holographic descriptions. 

\subsubsection{Black hole entropy is the UV contribution to entanglement entropy} \label{uv-ir}
Let me first review how the previous semiclassical properties of black holes, combined with generic features of certain approaches to quantum gravity, suggest that such underlying  granularity should actually be part of a more fundamental description of black holes.  To do this, let us first consider the semiclassical evaporation process in more detail. Namely, we can first ask the question of how much entropy is carried by the Hawking radiation during evaporation. There is perhaps no consensus about the meaning of black hole entropy, and we will expand on this presenting my view in our latter discussions; however, we certainly know what the radiation entropy is. Thus, from the Hawking radiation  spectrum and energy conservation at $\sI^+$ one can calculate the entropy of the radiation produced during the evaporation process,  by integrating the entropy flux at $\sI^+$.  More precisely, assuming that the black hole is created with initial mass $M_0$ via a rapid gravitational collapse (we will take $J=0=Q$ for simplicity), one has that the intensity of the Hawking radiation at $\sI^+$ that follows from Hawking spectrum is (Stefan-Boltzman law)  
\be
I_{E}=\sigma A T^4,
\ee
  where   $T$  stands for the Hawking temperature  $ \kappa/2 \pi $  (with $ \kappa$, the surface gravity that  for a Schwarzschild  black hole is $ \kappa =   \frac{1}{4M}$) and $\sigma$  is a constant  that takes into account the grey-body factors,  as well as the contributions from the different species and number of degrees of freedom of each of them respectively. The entropy flux per unit retarded time in the radiation is $\dot S=I_{E}T^{-1}$. Energy conservation requires $\dot M=I_{E}$ from where we get $dS/dM=T^{-1}$. Integrating one obtains
\be\label{entro-pia}
\Delta S=\int_{M_0}^{0} \frac{8\pi}{\ell_p^2} M dM=\frac{A_0}{4\ell_p^2},
\ee
where in the integration we assumed that there was no radiation initially at $\sI^+$. The previous argument assumes that the radiation process is quasi-stationary in the sense of thermodynamics. In general, one has the inequality
\be\label{gsl}
\Delta S\ge \frac{A_0}{4\ell_p^2}.
\ee  
One way of altering the quasi-stationarity consists of feeding the BH from the outside during evaporation. A process that can increase the final entropy radiated $\Delta S$ arbitrarily. The previous analysis generalizes to the rotating case if, in addition, we use conservation of angular momentum (but the calculation is more involved as it requires the precise angular momentum dependence of the grey-body factors). 

\begin{figure}[h] \centerline{\hspace{0.5cm} \(
\begin{array}{c}
\includegraphics[width=8cm]{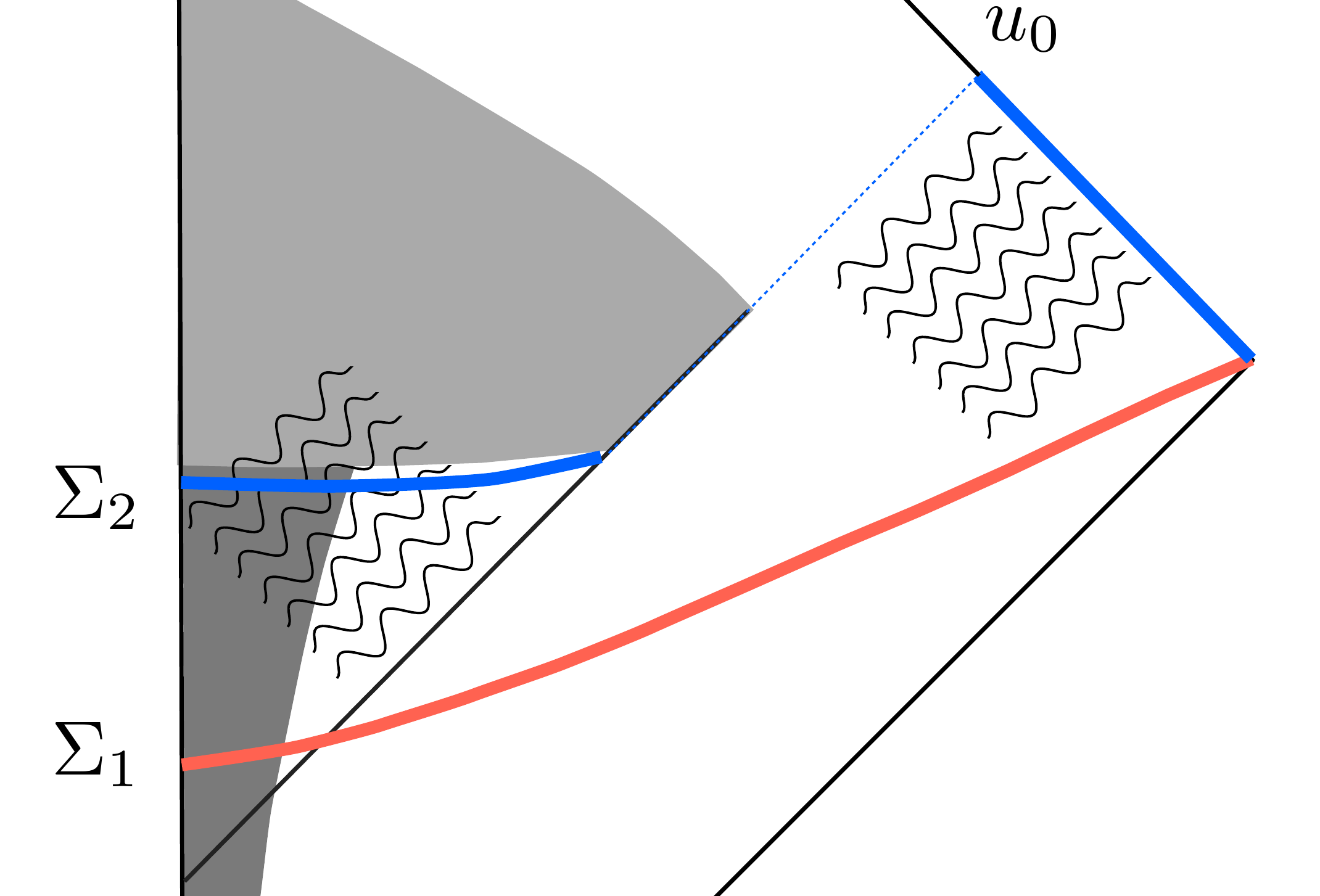}
\end{array}
\)}
\caption{Two  `instants' represented by suitable Cauchy surfaces relevant for the discussion of entropy and information.}
\label{td4}
\end{figure}

The previous discussion provides a clear picture of the nature of the entropy of the outside world (the one entering the GSL) at the end of black hole evaporation. On the one hand it is the entropy contained in the emitted Hawking radiation at $\sI^+$ which (in such quasi-stationary case, adiabatic from the perspective of the GSL, \eqref{gsl}) it matches the initial  black hole entropy $A_0/(4\ell^2_p)$. On the other hand, this entropy is also entanglement entropy, namely 
\be
\Delta S=-{\rm Tr}[\rho_{\rm out} \log(\rho_{\rm out})]=\frac{A_0}{4\ell^2_p}
\ee  
where $\rho_{\rm out}$ is the reduced density matrix  characterizing  the state  of the quantum fileds on $\Sigma_2$ in Figure \ref{td4} (relevant for observers in the outside region of $\Sigma_2$ idealized by  the portion of $\sI^+$ for $u\le u_0$) once the  degrees of freedom corresponding to the inside are traced out. Here, the inside portion of $\Sigma_2$  is given  by the `latest possible'  instant before the quantum region where spacetime representations are compromised due to Planckian effects (here, we are pushing the semiclassical approximation as far as we can but the same was done in the previous calculation of the entropy of the radiation \eqref{entro-pia}). The last equality also assumes quasi-stationarity of the evaporation process and follows from \eqref{entro-pia}. 
between the outside and the inside of the black hole (entanglement entropy between the two thicken portions of  $\Sigma_2$ in Figure \ref{td4}).  

The previous equation holds due to the fact that the Hawking radiation is produced (according to the rules of quantum field theory) in perfectly correlated pairs of outgoing particles, with one particle radiated to infinity and the other into the black hole singularity. This is only a special case of a very general feature of particle creation in curved spacetimes \cite{Wald:1975kc} and particle creation by external potentials.   It is precisely these correlations---established according to the rules of standard quantum field theory applied in a regime where we expect it to be a valid approximation---that grant the unitarity of the evolution among Cauchy surfaces to the past of the quantum gravity region (like  $\Sigma_1$ in Figure \ref{td4}). Moreover, unitarity is expected to hold when extrapolating this evolution to the  `latest possible time' $\Sigma_2$. There is  a null component of $\Sigma_2$, represented by a dotted line, that we assume not to play an important role (this assumption is expected to hold as the degrees of freedom in the radiation near the end of evaporation are ultrarelativistic  outgoing modes; quantum gravity effects might also play a role, but we expect them not to change the situation drastically to the point to invalidate the qualitative features of the scenario). The unitarity is precisely preserved by the entanglement between the outside and inside portions of these Cauchy surfaces and, thus, the radiation entropy is concurrently  {\em standard thermodynamical entropy} and {\em entanglement entropy} between the outside and the inside of the black hole.

The perspective adopted here is that the outside entropy, the one satisfying the GSL \eqref{gsl}, is simply entanglement entropy, not only at the `instant' $\Sigma_2$, but for all the Cauchy surfaces with the quantum gravity region to their future (as for example $\Sigma_1$ in Figure \ref{td4}).
Such perspective opens the way to a fundamental definition of black hole entropy as a measure of the (local) entanglement between the degrees of freedom right outside with those right inside the event horizon of a black hole \cite{Bombelli:1986rw}. More precisely,  given a pure state $\rho=\ket{0}\bra{0}$ of a quantum field on a Cauchy surface, such as $\Sigma_1$ in Figure \ref{entro-pia}, the entanglement entropy between the inside and the outside of $\Sigma_1$, defined with respect to the black hole horizon, is divergent due to the contribution of ultra-local entanglement across the horizon in local quantum field theory. If one introduces a cut-off  length scale $\epsilon$ to regularize such `high-energy' or UV, contributions one finds
\ba\label{entro}
S_{\rm ent}&=&-{\rm Tr}[\rho_{\rm out} \log(\rho_{\rm out})]\n \\
&=& f_1 \frac{A}{\epsilon^2}+f_2 \log(A \epsilon^{-2})+S_{\rm rad},
\ea 
where $S_{\rm rad}$ is a finite contribution representing the entropy of the radiation outside of the black hole. This term is expected to vanish in a suitable vacuum state if $\Sigma_1$ is sufficiently early, so that it does not contain any Hawking particles yet. One would like to associate with the entropy of the black holeI  the term proportional to the black hole area $A$ which diverges as $\epsilon^{-2}$. However, this term is not only divergent, but also ambiguous in quantum field theory, as it depends, for instance on the details of the regularization procedure used, as well as on the number of degrees of freedom involved (the so-called species problem).

In addition to having the correct formal dependence on $A$, there are reasons to believe that it is precisely that UV divergent term that---once made finite
by the inclusion of the missing quantum gravitational degrees of freedom---encodes the correct notion of black hole entropy that enters the laws of black hole thermodynamics. We can express this conjecture as
\be
 f_1 \frac{A}{\epsilon^2} \ \ \ \ \ \ \ \ {\begin{array} {ccc} {\longrightarrow}\\ \rm \van in\  quantum\\ \rm \van gravity\end{array}} \ \ \ \ \ \ \ \ \frac{A}{4\ell_p^2}.
\ee
One can find support for this conjecture by studying perturbations of  the Hartle-Hawking state of an eternal Schwarzschild black hole spacetime \cite{Perez:2014ura}\footnote{This is motivated by a simpler calculation in flat spacetime given in \cite{Bianchi:2012br}.} where, perturbing the definition \eqref{entro}, one shows that 
\be \delta S_{\rm ent}= \frac{\delta A}{4\ell_p^2}+\delta S_{\rm rad},
\ee
where 
\be
\delta S_{\rm rad}=\frac{1}{T} \int_{\sI^+} \braket{\delta T_{\mu\nu}} \xi^\mu dS^\nu=\frac{\delta E_{\rm rad}}{T},
\ee
$\xi^a$ is the stationarity Killing field of the background geometry, and expectation values are taken in the Hartle-Hawking state. The last equality involves the change of the radiated energy at infinity due to the perturbation $\delta E_{\rm rad}$, and the Hawking temperature $T$ (heat over temperature: one finds again the direct link between entanglement and thermodynamics when black holes are involved). Note that the argument leads to the desired Bekenstein-Hawking horizon area contribution required by the previous conjecture (independently of any cut-off and number of species in the quantum field) because the area contribution arises explicitly from the semiclassical Einstein's equations via the Raychaudhuri equation describing perturbation of the generators of the black hole horizon 
sourced by $ \braket{\delta T_{\mu\nu}}$. Here we see how the inclusion of gravity actually resolves simultaneously the divergence and the ambiguity issue (including the species problem).  It  is  worth noting that the issue of  possible  dependence on  the   number of  particle  species  drops  out of the  analysis  as their number affect in  the same  manner   both the calculation of the entropy and the change in the area. 

More generally,  one can argue for the validity of the conjecture from the UV structure of entanglement in any arbitrary Hadamard state (states whose $2$-point function mimics the UV structure of the Minkowski vacuum and used to renormalize the energy momentum tensor in semiclassical gravity) in quantum field theory combined with a suitable hypothesis on the granularity at the Planck scale (e.g.  loop quantum gravity provides one). 
More precisely, assuming that the background geometry is well approximated by that of a stationary black hole (a Kerr-Newman black hole), one has that there are (uniquely defined) local observers in a neighbourhood outside, and  very close to the horizon, that see the horizon at rest (a sort of `rest frame of the horizon'). These are the observers with four velocities $u^a=\chi^a/\sqrt{|\chi\cdot\chi|}$, where $\chi^a$ are the Killing generators of the horizon given by
\be
\chi^a=\xi^a+\Omega  \psi^a
\ee  
where $\xi^a$ is the previously evoked time translational Killing field and $\psi^a$ is the axial symmetry Killing field ($\Omega$ is the angular velocity of the horizon entering the first law \eqref{1st}). The rules of local quantum field theory imply that such stationary observers detect around them a thermal environment with local temperature $T_{\rm local}=2\pi/\ell +\sO(\ell)$, for $\ell$ is the space-like distance to the horizon in a background geometry assumed to be stationary to the past (this is hence the distance to the bifurcating surface of a fiducial \cite{wald paper} bifurcating Killing horizon). Thus, as the temperature grows as $\ell\to 0$, the UV structure of quantum states in quantum field theory implies that, in this limit, all degrees of freedom available must look like as if they were excited at infinite temperature in order for the state to look just like the vacuum locally, say, from the point of view of freely falling observers. Thus, the state must look like an infinite temperature state for the special observers selected by the time translational symmetry of the local geometry near the horizon (the natural rest frame of the system)  in order not to have something like a fire wall at the horizon. Consequently, in the limit $\ell\to 0$, all degrees of freedom must be equally likely---in the density matrix defined in the frame of such preferred observers---and be maximally correlated with partner modes on the other side.  Under such conditions the UV contribution to the entanglement entropy (the divergent piece) is given by 
\be
 f_1 \frac{A}{\epsilon^2} \ \ \ \ \ \ \ \ {\begin{array} {ccc} {\longrightarrow}\\ \rm \van   Hadamard\end{array}} \ \ \ \ \ \ \ \ \log(N)
\ee
where $N$ is the number of degrees of freedom characterizing the the infinitely-thin atmosphere {\it close to the horizon} (the UV modes) that we can regard as horizon states:
$N$ corresponds to the dimension of the Hilbert space of horizon states.

The number $N$ is divergent in an effective quantum field theory description simply because it does not take into account the fine grained structure of quantum geometry. We expect that this apparently divergent number should actually be bounded by the area of the black hole (the entanglement surface) in a suitable theory of quantum gravity. This is actually the case in models of quantum geometry emerging from loop quantum gravity \cite{G.:2015sda} where the dimension of the allowed piece of the Hilbert space---obtained from constraining the area of the horizon to be a macroscopic number $A$---goes like $N\approx \exp{\eta A/\ell_p^2}$ for a dimensionless number $\eta$ whose value is subjected to ambiguities associated with the precise definition of the model. One gets then
\be\label{lqg}
 f_1 \frac{A}{\epsilon^2} \ \ \ \ \ \ \ \ {\begin{array} {ccc} {\longrightarrow}\\ \rm \van loop\  quantum\\ \rm \van gravity\end{array}} \ \ \ \ \ \ \ \ \frac{\eta A}{\ell_p^2}
\ee
More precisely, in loop quantum gravity, one counts number of microscopic states of the spacetime geometry containing a black horizon of a fixed macroscopic area. Such number is finite and its finiteness is directly related to the fact that the underlying theory predicts a fundamental granular structure where spacetime emerges from the contributions of fundamental geometric building blocks. Under such circumstances, the macroscopic constraint that the horizon area must be $A$ cuts off the number of available combinations \footnote{It is important to point out that, despite the insightful value of these results, the meaning and computation of black hole entropy in loop quantum gravity remains open to an important extent. 
The main limitation of these calculations is the lack of sufficient control in the description of the physical states in the Hilbert space of quantum gravity  corresponding to a given macroscopic black hole including all the degrees of freedom expected to play a role.}.

This should happen, generically, if the smooth spacetime geometry and the smooth fields of general relativity emerge from a more fundamental Planckian physics, which is discrete at the most basic level. Loop quantum gravity \cite{rovelli_2004, thiemann_2007} and causal set theory \cite{Bombelli:1987aa} are examples of mathematical models where this idea is partially realized. In such approaches, smooth fields and smooth geometry are expected to arise only as effective description of the collective behaviour of microscopic building blocks under suitable circumstances. 

\subsubsection{A key hypothesis about quantum gravity}

Even though the goal expressed in the previous paragraph remains an open problem at the moment, a feature that appears to be common to approaches of this type is that the data encoded in a classical solution---a smooth geometry solving  Einstein's equations with smooth fields defined on it, and satisfying suitable field equations---would not identify uniquely a microscopic state in the fundamental quantum theory. More precisely, it seems clear that the presence of a non trivial microscopic structure implies  the existence of new microscopic degrees of freedom with no smooth effective field theory description.   For instance, the quantum state representing flat spacetime with matter in the Minkowski vacuum is to arise from the suitable superposition of  fundamental building blocks of Planck size where the microscopic arrangement of individual pieces is not completely fixed by the macroscopic flatness requirement. In particular, what I am emphasizing here is the possibility that, in such theories, what one would loosely call  the  `ground state' would be, in contrast with the quantum field theoretic intuition,  highly degenerate. Such fundamental degeneracy is precisely the source of black hole entropy in the counting that leads to \eqref{lqg} in loop quantum gravity,  where a single classical black hole corresponds in the quantum theory to a large multiplicity of microscopic states.   For reasons that will become clearer as I develop my argument, such degeneracy must also play a central role in the discussion of unitarity in black hole evaporation.

Unfortunately, the status of loop quantum gravity (and to my knowledge, the status of any other approach to quantum gravity) does not allow for a detailed dynamical analysis of the black hole formation and evaporation process. However, the consideration of the question from the loop quantum gravity perspective suggests a natural scenario where the question of the fate of information takes a new form.  In order to be as concrete as possible, I will write a series of formal equations whose purpose is to illustrate an idea more than being quantitatively correct or predictive. These equations are to be taken with   {\it a grain of salt} in the sense that, in some situations, they are, strictly speaking,  incorrect (like, for instance, when I will represent the final state of the Hawking radiation after evaporation as a thermal state at a given temperature, disregarding the fact that temperature of the emitted  radiation  changes as the black hole shrinks), or, in other cases,  they express some expected property of quantum gravity realized (by the concrete mathematical translation) necessarily in a rigid manner; a manner that must be considered (if the message is communicated as I expect it) with the necessary indulgence and flexibility of a concept that can only be realized precisely if the missing formalism of quantum gravity was available.  Along these lines, I assume, for instance, that the Hilbert space of quantum gravity is the tensor product of the quantum geometry Hilbert space times the matter Hilbert space, namely 
\be
\sH=\sH_{\rm geo}\otimes\sH_{\rm matter}.
\ee
 This assumption is probably wrong at the fundamental level where it is not unreasonable to imagine that the correct description could be embodied in a sort of unified structure from which both matter and geometry emerge at low energies. However, here I make this choice just in order to simplify the argumentation and the notation in formal equations. As suggested by the structure of loop quantum gravity, 
there are infinitely many quantum geometry states corresponding to a single classical solution differing in degrees of freedom at the Planck scale to which smooth field observables are not sensitive.  Such degrees of freedom could be pictured as defects in the Planckian fabric of quantum gravity to which one is simply not sensitive when using the coarse low energy probes of macroscopic agents (like the molecular structure that escapes the smooth characterization of the Navier-Stokes effective theory of fluids).
For instance, when it comes to flat Minkowski spacetime, we assume that the different quantum geometry states, which are seen as flat Minkowski states, are members of a (for simplicity assumed to be) countable set, so that we can write them as
\be
\ket{\rm flat,n} \in \sH_{\rm geo},
\ee  
where the label $n\in \N$. For each such quantum geometry state there is a corresponding matter state (vacuum state) of the matter ${\ket{0,n}}\in \sH_{\rm matter}$. The states ${\ket{\Psi_{0,n}}}={\ket{\rm flat,n}} \otimes{\ket{0,n}}$ are physical states solving all the dynamical equations of quantum gravity (all the constraints in the canonical language). These states are vacuum states with not wavelike matter excitation (no infrared photons or gravitons) and labelled `flat' as for all $n\in \N$ they satisfy
\be\label{adm}
\widehat M_{\rm ADM} {\ket{\rm flat,n}} \otimes{\ket{0,n}} =0,
\ee
 where $\widehat M_{\rm ADM}$ is the ADM mas or ADM Hamiltonian equation.
The states \be {\ket{\Psi_{0n}}}={\ket{\rm flat,n}} \otimes{\ket{0,n}}\ee represent the degenerate `vacua' of quantum gravity.

Motivation for the existence of a large number of states satisfying the low curvature requirement embodied in equation \eqref{adm}  comes from experience in loop quantum gravity, where the theoretical framework offers concrete insights of its validity in various ways. The first indication comes directly from the study  of the nature of quantum gravitational degrees of freedom at the Planck scale (quantum geometry), where, due to the discrete structure of quantum states,  there are new degrees of freedom at the Planck scale with no fields theoretic semiclassical analog.  More indications  arise 
from the analysis of symmetry reduced models where their simplicity allows for rigorously proving the analog of \eqref{adm}. In the first case, one observes that, due to their singular nature, these Planckian degrees of freedom are to be pictured as defects in the fabric of geometry rather than propagating excitations. A possible candidate for such Planckian microscopic degree of freedom is the one captured (in loop quantum gravity) by the degeneracy of the volume eigenstates  where each volume eigenvalue (defining an elementary volume of quantum space) 
is doubly degenerate producing a $q$-bit of degeneracy for each such building block from which continuous geometry is to be built. Another feature suggesting the existence of these hidden defect-like degrees of freedom comes from the study of the spectrum of the area in loop quantum gravity. 
Such underlying degeneracy of coarse grained geometry advocated in \eqref{adm} actually plays a key role in the computation of black hole entropy.  
Here, a single classical isolated horizon labelled by its macroscopic classical area admits a number $N$ of  microscopic eigenstates of the area $A$ with eigenvalues $a$ in the microscopic range $a\pm \ell_p^2$ grows exponentially with $a$ as $\exp(\eta a/\ell_p)$ (\cite{Perez:2017cmj, BarberoG:2015xcq} and references therein). 
Now, such abundance of microscopic states cannot be associated to degrees of freedom with a  classical interpretation. Indeed, the isolated horizon phase space eliminates, by  the very definition of the boundary condition defining the system, the possibility of local excitations on the horizon by definition \cite{Ashtekar:1997yu}. Thus, these new degrees of freedom are like quantum features for these black holes, a single macroscopic black hole corresponds to any of these huge number of
 microstates differing by ultralocal details (defects) and not by wavy modes like gravitons in the perturbative regime.

Additional motivation that resonates with the previous example comes from the analysis of certain mini-superspace models. Indeed, the exact analog of equation \eqref{adm} arises in the case of symmetry reduced models for quantum cosmology and black holes where the infinitely many degrees of freedom of quantum gravity are reduced to finitely many via symmetry assumptions.  Despite the simplicity of these models---which are toy models based on the same type of representation of the  algebra of observables, as in the full theory of loop quantum gravity,   is  taken  to  underlie  the definition of the quantum theory---one finds two remarkable properties which cannot be separated from each other: On the one hand, the structure of the Hilbert space is such that the big-bang singularity (or the interior singularity in the case of black holes) is resolved with a notion of quantum dynamics that is well defined across a quantum gravity region as in Figure \ref{td4}, while, on the other hand, the loop quantization produces a drastically larger Hilbert space that includes new hidden degrees of freedom with no low energy interpretation.  In the case of cosmology, this implies, for instance that the solutions of the Hamiltonian constraint are infinitely degenerate.  As there is no well defined notion of energy in cosmology, one cannot write the analog of \eqref{adm}.  However,  there is a perfectly well defined notion of energy in a closely related formulation, known as unimodular gravity \cite{Unruh:1988in, Smolin:2010iq, Smolin:2009ti, Chiou:2010ne}, which is simply given by a conserved quantity playing the role of the cosmological constant. It is easy to see that in the singularity avoiding quantization mentioned above \cite{Amadei:2019ssp,  Amadei:2019wjp} the eigenstates of the cosmological constant are infinitely degenerate in the spirit of \eqref{adm}. The ADM mass can be loop-quantized in the case of spherically symmetric models of black holes and show that it is, indeed, infinitely degenerate with a degeneracy space labelled by quantum numbers with no classical correspondence: microscopic defects.

Finally, I must admit that the previous are facts in the context of models describing black holes or cosmology in loop quantum gravity. At present, one cannot be certain that the property that I formally postulate in \eqref{adm} is a feature of the correct quantum gravity description of black holes. This is why the previous are only indications motivating a perspective. 
One should be aware of the risks associated with overemphasizing  insights produced only by simple models in the context of a theory for which the precise mathematical way the continuum limit arises remains unclear. The reader (as I do) must bear this caveat in mind.

My first task in the discussion of black hole formation and evaporation process is to express the form of the initial state.  As in Daniel's case, I will assume that the background spacetime is basically flat in the far past.   I will also assume that the initial state is a tensor product state of the geometry and matter degrees of freedom. This assumption is certainly not necessary (see \cite{Perez:2014ura} for more details). However, as the main ideas are not affected by it, we make it in order to simplify the notation. Nevertheless, as we will discuss later, such assumption is an extreme idealization of the fact that the initial state must be special and `low entropy' in a sense. Finally, without lost of generality, we can assume that the initial quantum geometry state is given by $\ket{\rm flat,0}$ in the decomposition of \eqref{adm}, as we can always define the basis of macroscopically-equivalent vacua in order to satisfy this requirement. With all this we can write\footnote{Here we note  that, as there are hidden variables for macroscopic observers, it is also (in practice) natural to think of the initial state as a density matrix  ( of the proper kind, and  thus,  understood as reflecting epistemic features)} . Information in that case is degraded (as my scenario implies) in accordance with the cartoon of Figure \ref{second}. 
\be\label{iniini}
\ket{\Psi_{\rm initial}}=
\underbrace{\ket{\rm flat,0}_{\ {} }}_{\rm quantum \ geometry}\!\!\!\!\!\!\! \otimes\ \    \underbrace{\hat C(f) \ket{0_{\rm in},0}}_{\rm matter\ fields }
\ee
representing an initial geometry well approximated by a low curvature (zero curvature) background $\ket{\rm flat,0}_{\ {} }$, with matter  encoded in the state $\hat C(f) \ket{0_{\rm in},0}$  representing an initially diluted semiclassical matter distribution with profile $f$  (that will eventually collapse to form the black hole), here denoted by the action of the coherent-state operator $\hat C(f)$ acting on the vacuum compatible with the quantum geometry state.   
 This is a very special state and reality need not be described exactly by such a simple product state, but we must keep in mind that gravitational collapse is expected to be a highly time asymmetric process where the past is extremely special (and low entropy) \cite{penrose-roadtoreality-2005}. The very special nature of the past embodied in such initial condition captures a central physical feature of the system (we admit) in a highly idealized manner. 
 
The spacetime history of black hole formation and evaporation is well represented by Penrose's cartoon
of irreversibility reproduced in Figure \ref{second}. Before  the formation of a black hole,  the story of our system exploring larger and larger portions of the available phase space is the usual and standard one involving, mostly, the non gravitational degrees of freedom in the semiclassical cloud $f$,  its self interactions,  and those with gravity (producing gravitational waves and matter radiation and the microscopic details of the evolution of the cloud becoming a star that eventually will collapse). The same tale that we apply to the phenomenological description  of systems involving many degrees of freedom like molecules, atoms,  and fundamental particles. A key point of my perspective is that the (well motivated yet here postulated) existence of the underlying microscopic Planckian  granularity necessarily implies that the story continues along very similar general lines after the black hole forms and evaporates, as far as the fate of information is concerned. 

What changes in the later phase of black hole formation and evaporation is that, as the singularity forms inside the collapsing body,  a new and huge new portion of phase space becomes available for the relevant degrees of freedom to be explored (as when the bottle breaks and the evaporation of the perfume inside, forbidden so far,  can finally take place).  The  description  of  gravitational collapse and the formation of what we classically call a singularity, must  be followed by  al  characterization of what takes place  beyond the event horizon reaching the  end of what is  susceptible  of  a classical description  (this is of course  not a singularity, but rather a quantum gravity region across which dynamics ought to  remain  well defined in the appropriate fundamental description).  The singularity is a quantum gravity region where tidal effects and densities become Planckian, bringing the system in contact with the Planck scale and its microscopic degrees of freedom (like the lighter setting a newspaper on fire in our emblematic introductory example).
The gravitational collapse ignites interactions with the Planckian regime inside the black hole horizon that cannot be ignored in seeking a unitary description of black hole formation and evaporation. 

\subsubsection{Quantum dynamics near the singularity is central for the discussion of the information issue}

In order to understand the fate of information in black hole evaporation, one must take into account the existence of such Planckian degrees of freedom. This is fairly obvious for the degrees of freedom of the initially diluted semiclassical cloud becoming later a collapsing star that evolves right into the quantum gravity region where general relativity predicts a singularity. However, this is also true for the dynamics of the vacuum $\ket{0_{in}, 0}$,  which involves (via the interaction with the low curvature gravitational field) the Hawking pairs and the Hawking radiation emitted to $\sI^{+}$. Black hole evaporation implies that the energy content in the initial cloud has to be somehow annihilated by the back reaction of the Hawking radiation. The local characterization of this process requires a full quantum gravitational treatment, as well as the interaction between the matter forming the black hole and the Hawking particle can only take place inside the quantum gravity region. This is clear from the fact that the gravitational collapse is a fast process where the matter sources reach the singularity in time scales of the order of the initial  black hole mass $M_0$, while Hawking radiation is a slow process with characteristic times of the order of $M_0^3$. This is also apparent from the causal structure represented in Figure \ref{PD-nous1}. Non local aspects of quantum field theory should not be neglected. For instance, one can check that vacuum polarization contributions to  $\braket{T_{ab}}$ produce `negative energy densities' (as measured by observers falling with the collapsing matter) close to where the sources hit the singularity \cite{Perez:2014xca}.  However, these effects are all important close to the singularity where  quantum field theory on curved spacetimes cannot be trusted. A full quantum gravity description of such high curvature regime is mandatory. 

Focusing on the Hawking radiation, there are two aspects that are central for our discussion, that are well established from semiclassical physics: on the one hand Hawking radiation is produced by the particle creation generated via the interaction of the matter fields with the curved spacetime background, on the other hand, it is the back reaction of this radiation that is responsible for the evolution of the black hole geometry and its eventual evaporation. We know, from the general theory of particle production by the gravitational field that particles are produced in maximally entangled pairs (see for instance \cite{Wald:1975kc}). In the present particular case, this implies that for each outgoing Hawking particle emitted as Hawking radiation to $\sI^{+}$ there is an outgoing particle falling into the singularity which is maximally entangled with its Hawking partner. Moreover, even when it takes an infinite affine parameter flight for the Hawking particle to reach $\sI^+$, the inside partner hits the singularity after a finite change of the affine parameter. Therefore, any question about unitarity---which precisely concerns the fate of the entanglement between these two particles---necessarily involves a quantum gravity description of the dynamics inside the quantum gravity region. Once more, we conclude that the question of unitarity is a strong quantum gravity question \footnote{There are a series of analysis in the literature of the question of information that systematically avoid dynamical considerations near the singularity. For the reasons evoked so far, such discussions cannot offer, in my view,  any serious resolution of the apparent paradoxes involved. We will discuss this important point further in Section \ref{page-curve}.}.

It is possible to argue that the Hawking particles falling into the singularity must interact strongly with the microscopic Planckian structure of the spacetime as they become infinitely blue shifted during their evolution toward the singularity. In order to do this, we consider one of these particles and neglect the back reaction of the Hawking radiation---recall that, as mentioned in the previous paragraph, the trip of the Partner hitting the singularity is very short---and describe the particle as propagating in the black hole geometry freely along a geodesic. For simplicity, we consider a massless particle (like a photon or a graviton) and simplify the discussion further by assuming  a Schwarzschild background. If $k^a$ denotes the four momentum of the particle, then the symmetries of the background imply the conservation of the angular momentum $\ell\equiv k^a\psi_a$ and the Killing energy $E\equiv k^a\xi_a$ , where  $\psi^a$ and $\xi^a$ are the rotational and stationarity Killing fields of Schwarzschild. One can object that these are only approximately conserved quantities as the black hole is actually evaporating, so that the previous are only approximate symmetries. However, as mentioned before, for macroscopic black holes the time scales of evaporation are enormously longer than the dynamical scales involved in the particle fall. Thus, this approximation should be accurate for sufficiently massive black holes. Now, it is easy to use the previous constants of motion in order to calculate  the frequency measured by a radially freely falling observer toward the singularity (for instance, those that are normal to the $r=$constant hypersurfaces). The result is
\be\label{jazz}
\omega^2(r)=\frac{\ell^2}{r^2}+\frac{r}{2M} { E}^2 +\sO\left(\frac{r^2}{M^2}\right).
\ee
%
%
 The divergence approaching $r=0$ would be the same for any regular observers measuring $\omega$. The key point here is that the Hawking spectrum contains particles with non vanishing $\ell$ \cite{Page:1976df}. Thus, the frequency of such particles is infinitely blue-shifted as observed by the local freely falling observers (this is to be expected from angular momentum conservation and the fact that the length of the orbits of $\psi$ are shrinking to zero), as well as by any other observer related to it via a finite Lorentz transformation.  Only exactly spherically symmetric modes with $\ell=0$ would become IR at the singularity. However, even this last conclusion is not expected to be significative, if the black hole rotates or if we consider that, at the fundamental level, background geometries with exact spherical symmetry inside the black hole  are of measure zero.  Thus, the partners falling into the black hole develop Planckian wave-lengths close to the quantum gravity region and quantum gravitational interactions between the Hawking particle and the Planckian microscopic structure cannot be avoided.

A clear indication that the back reaction of the background to the presence of Hawking radiation becomes non-negligible near the singularity---independently of how semiclassical the blackhole might look from the outside---comes from the behaviour of the renormalized expectation value of the energy-momentum tensor in the quantum state of matter associated to the Hawking radiation. This can be analytically computed in an effective 2d model if one concentrates on $\ell=0$ modes and neglects backscattering \cite{Fabbri:2005mw}. The result is that energy densities and pressure contributions from the quantum state diverge as one approaches the singularity as $r\to 0$. Such divergences are not a consequence of the approximations used and are known to be present in the 4d case even though it is not possible to give analytic expressions for it. This divergence in the source of the semiclassical gravity description implies that a full quantum gravitational treatment is necessary.     

Note that, despite appearances, the previous discussion of wavelengths and scales of particles is indeed a Lorentz invariant discussion. 
The concrete quantitative expression \eqref{jazz} was obtained by using freely falling radial observers which are the preferred (at rest with the black hole) observers
in the interior of a Schwarzschild black hole.  Even when this quantitative expression will change by changing the observers, the important fact is not the dependence of $\omega$ with $r$, but simply the observer independent statement that $\omega$ will diverge for any arbitrary observer (with the exception of the singular non-significant case of an infinitely boosted observer in the outer direction). This is why the insights drawn from the present discussion are reliable as they do not violate the expected general covariance of the fundamental theory.  
 
 \subsubsection{The concrete scenario}
 
\begin{figure}[h] \centerline{\hspace{0.5cm} \(
\begin{array}{c}
\includegraphics[width=8cm]{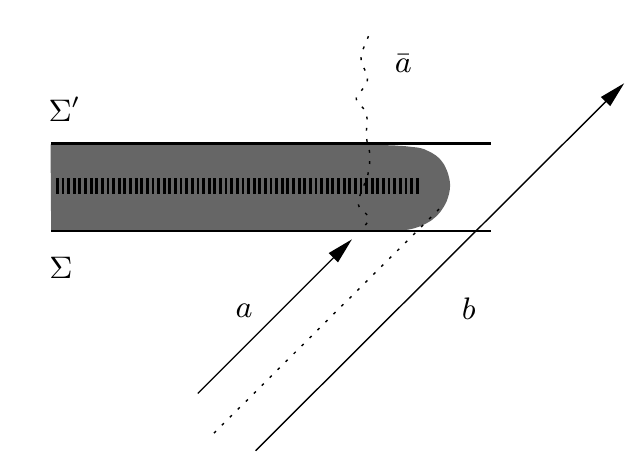} \end{array}\ \ \ \ \ \ \ \ \ \ \  \begin{array}{c}
 \includegraphics[width=4.5cm]{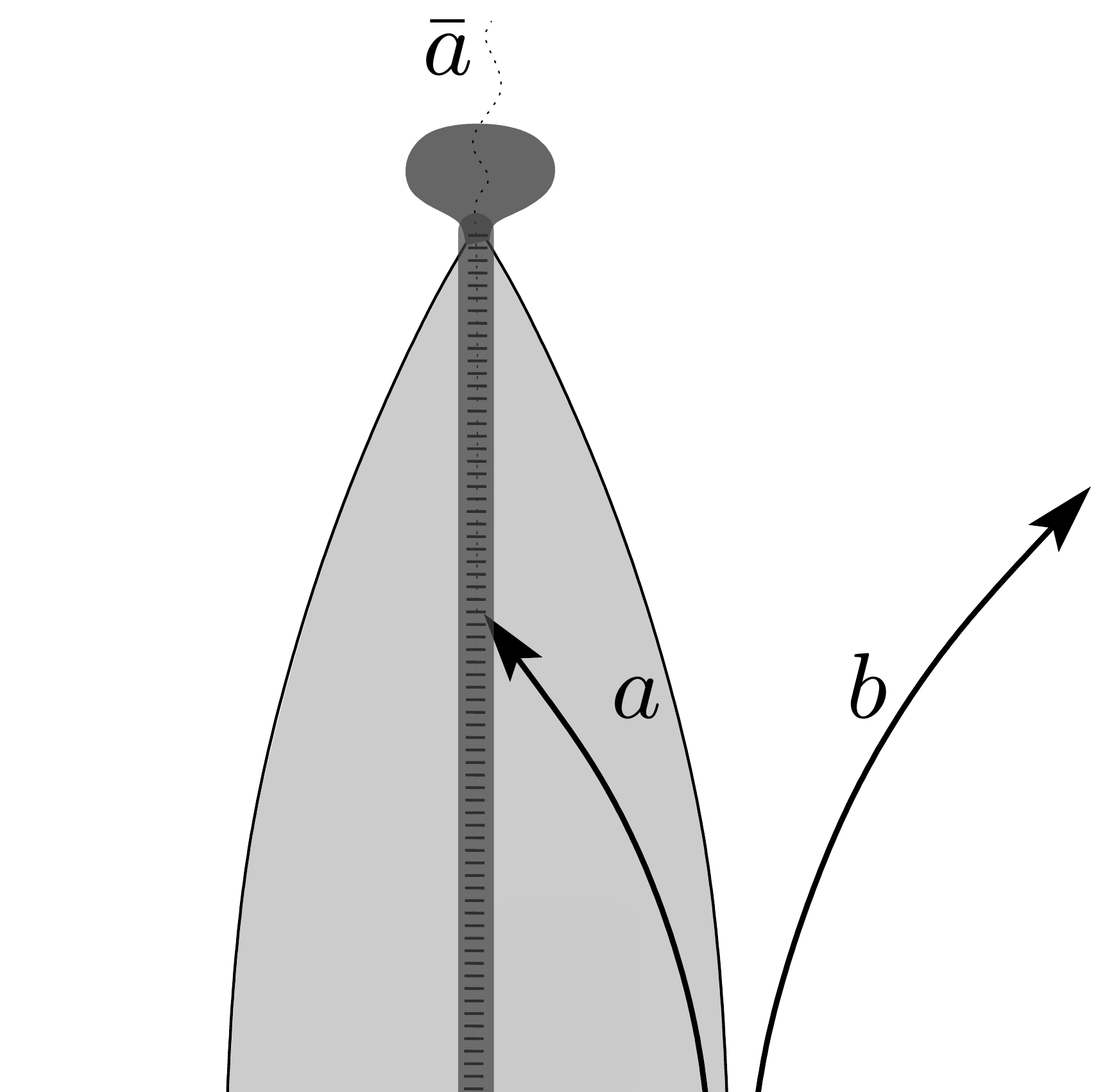}
\end{array}
\)}
\caption{ A maximally entangled Hawking pair $(a,b)$ is created via the interaction of the gravitational field and the vacuum. Interactions in the quantum gravity region transfer the quantum entanglement between $a$ and $b$ to entanglement between a Planckian defect $\bar a$ and $b$. The defect $\bar a$ emerges to the future of the quantum region into a macroscopically flat vacuum environment. This transfer is granted by monogamy of entanglement while the energy-less nature of the final degree of freedom is a consequence of the quantum gravity infinite degeneracy of the flat vacuum. The picture on the left shows the process in a conformal representation where light rays are at $\pi/4$ radians. The right picture is a representation of the process where the area of the black hole horizon is more faithfully represented. }
\label{percol}
\end{figure}

With all this preliminary discussion presented, we are now ready to describe the scenario that I proposed in \cite{Perez:2014xca}. 
As mentioned at the beginning of Section \ref{bulb},  when describing the analogy with hydrodynamics of ships with bulbous bows,  the most reliable description of the black hole formation and evaporation is understood from the  perspective of observers at infinity.  This is why the Penrose diagram of Figure \ref{funo} will be the framework of the present discussion. 
The first assumption is that there is  well defined evolution across the quantum region inside the black hole so that we can represent the situation as in Figure \ref{funo} \cite{Bojowald:2001xe, Ashtekar:2006wn, Ashtekar:2005cj}. In such a context, a `scattering theory' representation (where an in-state evolves into an out-state) is possible even though the result (as we will argue) cannot be translated into the language of effective quantum field theory. A maximally entangled Hawking pair $(a,b)$ is created via the interaction of the gravitational field and the vacuum (see Figure \ref{percol}). Particle $b$ goes out to $\sI^+$ as Hawking radiation carrying positive Killing energy. The entangled partner $a$ falls into the singularity with negative Killing energy and (assuming conservation of energy at infinity) reduces the Bondi mass of the black hole in an interaction whose precise local nature can only be understood in a full quantum gravity dynamical calculation. Conservation of energy at $\sI^+$ implies, however, that,  as a consequence of this interaction, the particle $a$ `annihilates a part of the mass' contributed by the (matter) source responsible for the formation of the black hole (the initial cloud $f$ in \eqref{iniini}).  The maximal entanglement between the pair $(a,b)$ cannot be maintained because of the very generic property of entanglement known as monogamy. More precisely, as a consequence of the quantum gravity interaction between $a$ and the Planckian degrees of freedom in the quantum gravity region (made unavoidable by the infinite blue shift previously described near the would-be-singularity), and monogamy of entanglement, the latter must be transferred to the microscopic degrees of freedom. In a theory where the property \eqref{adm} and conservation of energy at $\sI^+$ hold, there is only about one  Planck mass available at the end of the evaporation process.  This implies that the entanglement must be transferred to extremely low energy excitations. Now, due to the overwhelming abundance of `defect-like' degrees of freedom at the Planck scale, it is entropically clear that these correlations must be transferred to the defects that (according to \eqref{adm} and the discussion that motivates it) define the Planckian microscopic degeneracy of the vacuum.

Let me try to illustrate the previous argument in terms of a few formal equations.  
First recall the  structure of the Minkowski vacuum state $\ket{0}$ of a quantum field when written in terms of the modes corresponding to Rindler accelerated observers with their intrinsic positive frequency notion, namely 
\be
\ket{0}=\prod_{k}\left( \sum_n \exp\left({-n\frac{\pi \omega_k}{a}}\right) \ket{n,k}_{R}\otimes\ket{n,k}_{L}\right),
\ee
where $\ket{n,k}_{L}$ and $\ket{n,k}_{R}$ define the particle modes---as viewed by an accelerated observer with uniform acceleration $a$---on the left and the right of the Rindler wedge with $n$ particles in the mode with wave number $k$
\cite{Wald:1995yp}.
Here, we see, from the form of the previous expansion, that, even when we are dealing with a pure state (if we define the density matrix $\ket{0}\bra{0}$), the reduced density obtained by tracing over one of the two wedges would produce a thermal state with $T=a/(2\pi)$. 
 
Building on the analogy with the previous expression, the statement in the perspective we propose on the purification of information in   
BH evaporation  can we schematically represented (the following is certainly not a precise equation) by  
\be\label{final-state}
\ket{\Psi_{\rm final}}=\mathbf{U}
{\ket{\rm flat,0}} \otimes \hat C(f) \ket{0_{\rm in},0}
=\prod_{k}\left( \sum_n \exp\left({-\frac{\beta}{2} n \omega_k}\right) \ket{\rm flat,n} \otimes a^\dagger_k\ket{0_{\rm out},n} \right),
\ee
where the unitary evolution operator $\mathbf{U}$ acts on the initial state \eqref{iniini}  (for instance, defined on $\Sigma_1$ of Figure \ref{PD-nous1}), the $\ket{k,n}\equiv a^\dagger_k\ket{0_{\rm out},n}$ denotes a Hawking particle on the out-vacuum $\ket{0_{\rm out},n}$ compatible with the flat quantum geometry state $\ket{\rm flat,n}$.   The final state on the right is a pure state where the initial correlations between Hawking partners (which are basically vacuum quantum field theoretic correlations in $\ket{0_{\rm in},0}$) have been transferred to correlations between the Hawking quanta $\ket{k,n}$ and the many quantum flat geometries $\ket{\rm flat,n}$  (this final state can be assumed to reflect the state on $\Sigma_2$ in Figure \ref{PD-nous1}). At the end of the evaporation, there are no localized remnant hiding the huge degeneracy inside; there is only a large superposition  of states that are inequivalent in the fundamental quantum gravity Hilbert space,  but seem all the same for low energy agents (see cartoon representation of Figure \ref{funo}). Such degrees of freedom cannot be captured by any effective description in terms of smooth fields (EQFT) for the simple reason that they are discrete in their fundamental nature.

 \begin{figure}[h] \centerline{\hspace{0.5cm} \(
\begin{array}{c}
\includegraphics[width=14cm]{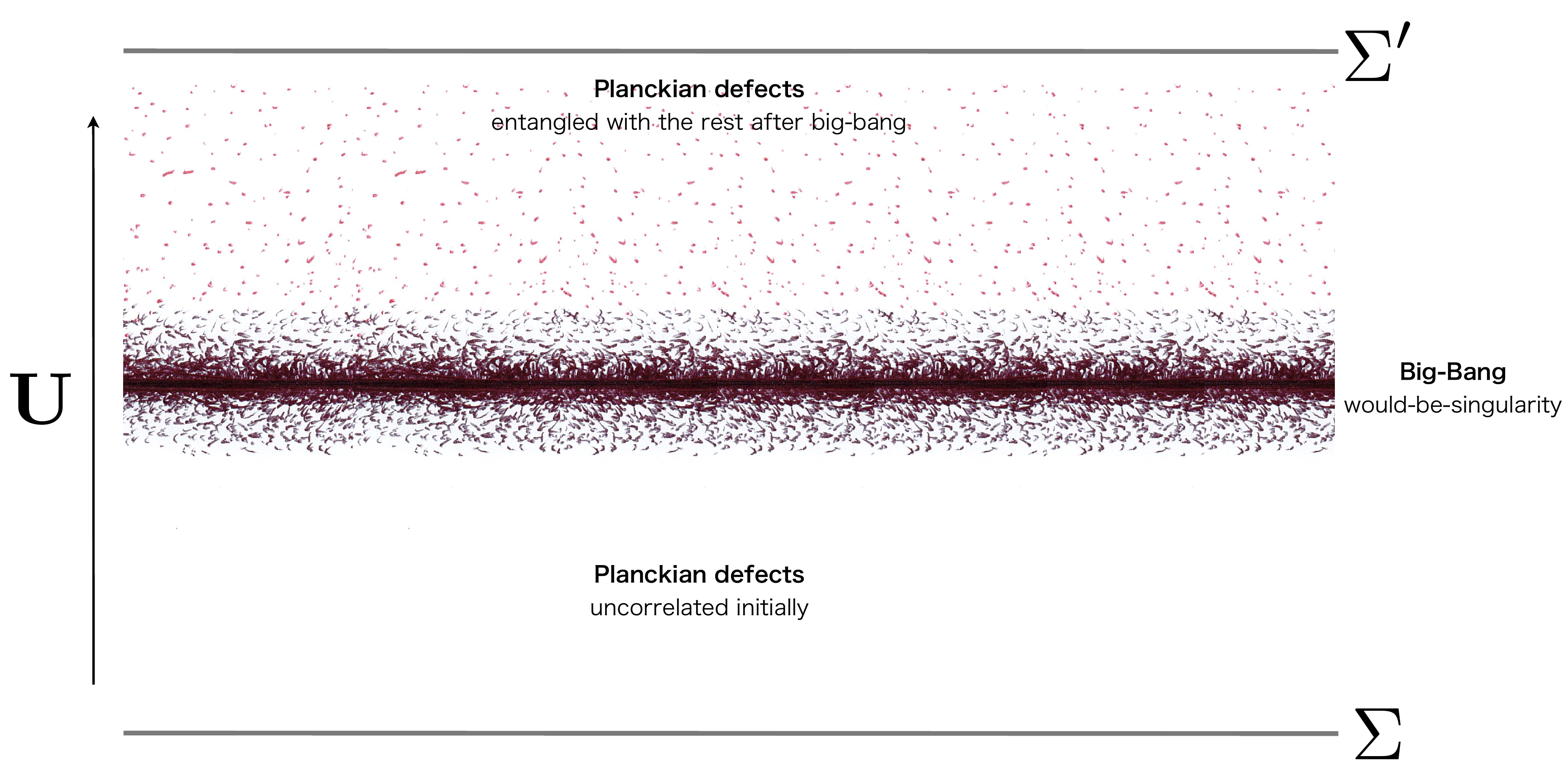}
\end{array}
\)}
\caption{Diagram illustrating (effectively) the natural scenario, suggested by the fundamental features of LQG, for the resolution of the information puzzle in black hole evaporation \cite{Perez:2014xca}. As in Figure \ref{funo}, one should keep in mind the limitations of such spacetime representation of a process that is fundamentally quantum, and, hence, only understandable in terms of superpositions of different spacetime geometries.}
\label{fdos}
\end{figure}

{ It is important to reiterate a key (and possibly not so intuitive) feature of the degrees of freedom with which the Hawking radiation is entangled. This is the fact that they carry nearly zero or exactly zero energy as required by the statement in equation \eqref{adm}. However, they cannot be thought of as low energy particle like excitations that would consequently be associated with very long wave lengths. These are `exotic' degrees of freedom from the perspective of effective field theories (as emphasized before, they do not admit a characterization in such terms) that do not satisfy the usual Einstein-Planck relationship $E=\hbar \omega$ with some wavelength $\lambda\sim 1/E$. \footnote{ This is not so exotic after all if we thing of the familiar case of  a non relativistic charged particle in a two dimensional infinite perfect conductor in a uniform magnetic field normal to the conducting plane. The energy eigenvalues are given by the Landau levels $E_n=\hbar \omega_{\rm B} (n+1/2)$ where $\omega_{\rm B}=qB/(mc)$ is the Bohr magneton frequency, but they are infinitely degenerate. There are canonically conjugated variables $(P,Q)$ associated to the particle that are cyclic, i.e., do not appear in the Hamiltonian. In this case, one can produce wave packets that are as `localized' as wanted in the variable $Q$ without changing the energy of the system. Interestingly, this is a perfect example of a system where one could have an apparent loss of information of the type we are proposing here (for a more realistic analog gravity model discussing the information paradox along the lines of the present scenario see \cite{Liberati:2019fse}). If one scatters a second particle interacting softly with the charged particle on the plate so that the interaction does not produce a jump between different Landau levels, then correlations with the cyclic variables would be established without changing the energy of the system. This is the perfect model to illustrate the possibility of decoherence without dissipation.} They are described as Planckian defects, nevertheless, they do not carry Planckian energy. The point is that such relationship only applies under suitable conditions which happen to be met in many cases but need not be always valid.
One case is the one of degrees of freedom that can be thought of as waves moving on a preexistent spacetime. This is the case of particle excitations in the Fock space of quantum field theory, or effective quantum field theories which are defined in terms of a preexistent spacetime geometry. There is no clear meaning to the above intuitions in the full quantum gravity realm where the present discussion is framed. Indeed, the defects that we invoke appear when a background geometry is built from elementary Planckian units, the degeneracy postulated in the construction of a flat background reveals the existence of degrees of freedom with no weight.  Even when such relations (linked to the usual uncertainty principle of quantum mechanics) should hold in a suitable sense for suitable emerging degrees of freedom, there is convincing evidence in approaches to quantum gravity like loop quantum gravity that this is not the case for all the degrees of freedom involved. }

At the present stage of development of different approaches to quantum gravity, it is impossible to test the scenario that I propose in a fully realistic gravitational collapse situation. 
However, one can actually show that the key ingredients that are needed are actually realized in existing models. We have mentioned the great degeneracy of microscopic states compatible with a single macroscopic black hole geometry in the black hole entropy accounts based on the isolated horizon boundary condition (see \cite{Ashtekar:2004cn} and references therein), and the fact that a similar degeneracy of micro-states is also present in symmetry reduced models of black holes and cosmology \cite{Olmedo:2016ddn}. In the later case, one can show that the dynamical decoherence process evoked above actually takes place during evolution across the would-be-singularity: low energy macroscopic degrees of freedom interact in the quantum gravity region with the microscopic `defect-like' degrees of freedom. As in the context of the previous general discussion, one must start with an initial state suitably chosen so that macroscopic features are not entangled with the microscopic defects. Quantitatively, one choses a `special' initial state so that the entanglement entropy---defined by the Von Neumann entropy of the reduced density matrix where one traces over the defects---is zero. Then, one  shows that, generically (independently of the initial state and of the matter couplings involved), entanglement with the microscopic degrees of freedom is established when evolving across the big-bang quantum region and the entanglement entropy grows: if we declare the macroscopic observers not to be sensitive to the microscopic defects, then pure states evolve into mixed states for them. If all the degrees of freedom are taken into account, evolution remains unitary across the quantum gravity region. 

One can generalize this to the spherically symmetric and stationary black hole models of the type considered in the loop quantum gravity \cite{Gambini:2013hna}. These models share with the cosmological ones the existence of the microscopic degrees of freedom and (even when they do not include black hole evaporation) they allow for the consideration of Hawking radiation of a test field making the analogy with the real physical case clearer. At the technical level, the situation is very much like the cosmological case due to the fact that for, such black hole models, the interior is isomorphic to a homogeneous (but not isotropic) cosmological model also called a Kantowski-Sachs model. Even when the homogeneity is broken by the matter falling into the black hole the model is still useful as an approximation. This is so if we concentrate on  the spherically symmetric  Hawking partner $a$ which can be physically approximated by a stationary matter contribution (compatible with the homogeneity assumption) due to the infinite redshift behaviour of the $\ell=0$ modes captured by \eqref{jazz}.  Thus, the quantum dynamics of both the geometry, as well as the falling partner $a$, can be suitably represented by the existent  Kantowski-Sachs models. Monogamy of entanglement, as well as a non trivial quantum interaction between $a$ and the microscopic degrees of freedom in the quantum gravity region, imply the type of decoherence that is at the heart of my scenario.


\subsubsection{The future of the quantum gravity region}

Assuming that the defects evoked in the previous discussion are associated to a microscopic geometric quantity, one could hope to be able to estimate the spacetime region that would be necessary to carry enough bits to purify the Hawking radiation. Arguments of this type cannot be taken too literally, as the only clear geometric picture must come from a detailed dynamical description using a fully developed quantum gravity framework. 

\begin{figure}[h] \centerline{\hspace{0.5cm} \(
\begin{array}{c}
\includegraphics[width=6cm]{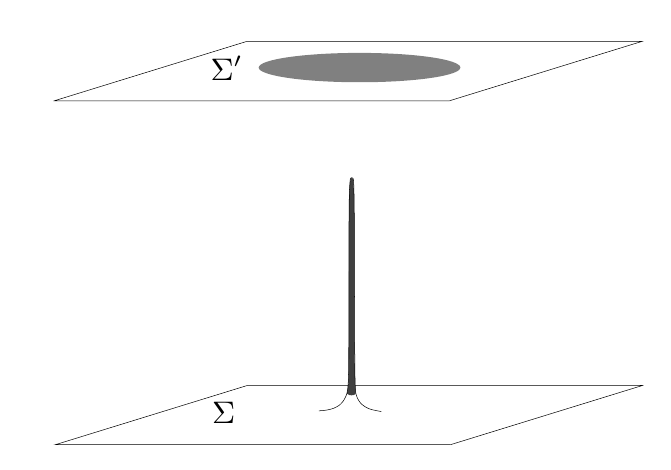}
\end{array}
\)}
\caption{A qualitative representation of the Riemannian geometry of $\Sigma$ and $\Sigma^{\prime}$ of Fig \ref{slicing}. The shaded regions are those `touching' the quantum region.} \label{slicing2}
\end{figure}

Lacking a clear fundamental account, one could aim at an heuristic estimate based, for example, on the consideration of the volume quantum numbers and their $2$-fold degeneracy in loop quantum gravity as candidates for the defect degrees of freedom. In such case, we can obtain some geometric information about the transition from the past of the quantum gravity region---as characterized in a time slice $\Sigma$ (Figure \ref{slicing2}) to its future in the region where the black hole has already evaporated--.  In such future region, the quantum gravity region becomes `visible' to future exterior observers, and represents, from the classical perspective, a sort of   `naked' singularity. How big needs this `naked'  singularity region be as measured on a time slice $\Sigma^\prime$ (Figure \ref{slicing2}) just at the time where spacetime geometry becomes applicable again?
In order to give a tentative answer to this question, we first notice that the volume of the shaded region, as measured in the past slice $\Sigma$ in Figure \ref{slicing2}, depends on the details of the evaporation process and on quantum gravity effects in the non-semiclassical regime. Nevertheless, dimensional analysis alone---combined with the physical input that this volume should diverge in the $\hbar\to 0$ limit (classical limit)---implies that, in the large mass (initial mass) $M_0$ regime, it should go like
\be V(\Sigma)\propto M_0^3 ({M_0}/{\ell_p})^{\alpha},\ee  
where the missing proportionality constant, and $\alpha>0$ depend on the interior dynamics (e.g. $\alpha=5/2$ when modeling the evaporation process with an advanced Vaidya metric \cite{Perez:2014xca}).  
We can now estimate the scaling of the volume of the shaded region in the flat hypersurface $\Sigma^{\prime}$ by requiring that there are at least as many volume `bits' as necessary to purify the radiation emitted during the Hawking evaporation process. For a `simple' black hole---one formed quickly and let alone and isolated during the whole evaporation process---the equality \eqref{entro-pia} holds, and thus the number is given by the initial BH entropy $A(M_0)/(4\ell_p^2)$, where $A(M_0)$ is the area of the BH at retarded time $u_0$ (see Figure  \ref{slicing2}). From this one gets
\be V(\Sigma')\propto \ell_p {M_0}^{2}.\ee  
From which a characteristic size $L=V(\Sigma')^{1/3}\approx 10^{-11} (M_0/M_{\odot})^{2/3}m$ of the naked {\em would-be-singularity} follows. For a BH  with $M_0=10^{15}g$ (e.g.  primordial BHs completing evaporation at present) this gives a size of about $10^{-2} \ell_{LHC}$ where $\ell_{LHC}$ is the shortest scale to which the {\em large hadron collider} is sensitive today.
The life time $\tau=u_2-u_1$ of the naked region as measured from $\sI^+$ would be of the same order in geometric units. So, even when the {\em would-be-singularity} region is macroscopic in comparison to the fundamental Planck scale, the size of the naked event can be quite small in cosmological or even particle physics scales. Thus, the final quantum gravity event where purification takes place could be a very small, and very short process. The previous estimate of the size of the quantum region assumed the black hole to  have a simple formation history. The extent of this region is unbounded in our prescription and will depend on that history (of course this is not surprising in a theory where information is not destroyed).

\subsubsection{Vacuum-geometry state {\em `ages' }when black holes evaporate (the irreversibility of the evaporating process)} \label{vgs}

When considering the final pure state $\ket{\Psi_{\rm final}}$ as given in \eqref{final-state}, one can define the pure density matrix 
\ba
\rho_{\rm final}&=&\ket{\Psi_{\rm final}}\bra{\Psi_{\rm final}}\n \\
&=&\mathbf{U} \ket{\Psi_{\rm initial}}\bra{\Psi_{\rm initial}}\mathbf{U}^\dagger,
\ea
which is simply the unitary evolution of the initial density matrix constructed from $\ket{\Psi_{\rm initial}}$ defined in \eqref{iniini}. If one defines the reduced density matrix by tracing out the defect degrees of freedom in the geometry, one finds $\rho_{\rm final}^{\rm reduced}(\rm matter)$ is the mixed  state representing  the Hawking radiation when one ignores the geometry purifying excitations (which is a natural thing to do for low energy observers). The Von Newman entropy $S(\rho_{\rm final}^{\rm reduced}(\rm matter))$ represents the (thermal) entropy of the Hawking radiation satisfying the generalized second law \eqref{gsl}.

Alternatively, starting from the final pure state $\rho_{\rm final}$, one can trace out the matter degrees of freedom and obtain $\rho_{\rm final}^{\rm reduced}(\rm flat)$. The physical justification for this is also natural: late observers at $\sI^+$ living in retarded time $u>u_0$ might not even be aware of the previous existence and evaporation of a black hole. They are not sensitive to the Hawking particles emitted before they even learned about doing physics. The physics of their interest will be well described by either the pure density matrix $\rho_{\rm final}$ or the mixed one $\rho_{\rm final}^{\rm reduced}(\rm flat)$, the entropy of which is (by the purity of the original state) 
\be\label{entro-lola} S(\rho_{\rm final}^{\rm reduced}(\rm flat))=S(\rho_{\rm final}^{\rm reduced}(\rm matter)).\ee
 From their perspective, and assuming they have access to the fundamental degrees of freedom, the geometry state representing flat spacetime is mixed and contains (due to the previous history that involved the evaporation of a black hole) a certain entropy. From this, one concludes that, as black holes form and evaporate in our universe, the state of spacetime grows old with a measure of its age given by the value of entropy \eqref{entro-lola}.

\subsubsection{On the relation with traditional remnant scenarios} One possibility is to assume that such purifying degrees of freedom are particle excitations coming from what is left of the BH (a remnant). Now, given that these particles must be extremely infrared as only 
 $E_{\rm \va late}\approx m_p$ is available for purification, then a simple estimate of the time (denoted $\tau_p$) that the process would have to last if this is the main channel for purification yields $\tau_p\approx (M/m_p)^4$. This is the scenario of an extremely long lasting point-particle-like remnant with a huge internal degeneracy which is sometime claimed to be problematic from the point of view of effective quantum field theory due to the apparent large probability for pair production that this objects would have. Although, for reasons related to the following subsection it is by no means clear to me why the correct effective field theoretical description would actually lead to such inconsistencies (for a standard account of this problem see \cite{Chen:2014jwq} and references therein).   Of course, such problems with large production probability rates arise only of one assumes that it is legal to treat the remnant as a point particle in effective field theory. This assumption seems of questionable validity from the features of the black hole internal geometry evoked before: if remnants exist they would enclose  a huge internal extension with a non-trivial probably non-classical geometry inside. 

Nevertheless, I would like to emphasize that the scenario that I  am describing here is not that of a traditional remnant. The purifying degrees of freedom are not localized within a single point-like object, but rather scattered to the future of the quantum gravity region in the microscopic structure of the final state. These degrees of freedom escape the description in terms of effective field theory, they carry information but (close to) zero energy---recall \eqref{adm}---their existence is expected  in loop quantum gravity (in particular),  as well as in any approach to quantum gravity where low energy spacetime geometry and fields arise from the collective contribution of Planckian granular structures.  The purifying action of such defects corresponds to a concrete quantum gravity realization of decoherence with negligible dissipation \cite{Unruh:2012vd}. 

\subsubsection{Black holes might be  able  to  hide an unbounded amount of information}
 \label{page-curve}

A point that should be explicit already in the previous presentation of my picture is that, according to it, black hole entropy (that we see as the UV contribution of entanglement entropy as discussed in Section \ref{uv-ir}) is not a measure of the number of internal degrees of freedom of black holes. Indeed, as these degrees of freedom are correlated with the Hawking radiation outside, and the later can be arbitrarily large depending on the past history of the given black hole, they are necessarily unbounded in the fundamental theory. This is perfectly consistent with the semiclassical description as arbitrarily old black holes with a long history (like the emblematic example of the black hole fed from the outside to compensate for its evaporation an arbitrarily long time) hide behind their horizon an arbitrarily large volume Cauchy surface \footnote{For explicit analyses that formalize this point see for instance \cite{Christodoulou:2016tuu}.}.  
The popular intuition that black hole entropy is a measure of the number of internal states is simply not correct in our description \footnote{In relation of the discussion of the previous section, note that these internal degrees of freedom remain causally disconnected from outside observers prior to retarded time $u_0$ in Figure \ref{PD}, where we assume that the black hole stops evaporating by leaving behind a stable (or very long lived) black hole remnant ( which might or not  take  the   form of a `particle like' object (as judged from the outside) at the origin of the (almost) Minkowski part of the diagram for $u\ge u_0$). The gravitational non-trivial nature of the  structure of these objects that, on the one hand,  might  seem  particle-like  from the  asymptotic  perspective, yet hide  arbitrarily large universes with non trivial geometric and causal features inside, suggests that these ``defects"   do not admit an effective field theory formulation, where for instance,  pairs  can be created and destroyed from quantum fluctuations of a   `vacuum state' on a `flat background geometry'.  In fact  it  might well be   that such   ``defects"   can only be   characterized   as inherently delocalized, when  one  attempts to  talk  about them in  the standard  language of   space-time   geometry,  perhaps something analogous  as  what what occurs  when   talking  about the  entanglement in a  EPR-B  setup.  }.  
Even when this might be a good intuition for composite non-relativistic systems the intuition does not hold in the context of strongly gravitating systems. 
This is a point of strong conflict with certain naive formulations of the so-called holographic principle.

 \begin{figure}[h] \centerline{\hspace{0.5cm} \(
\begin{array}{c}
\includegraphics[width=9cm]{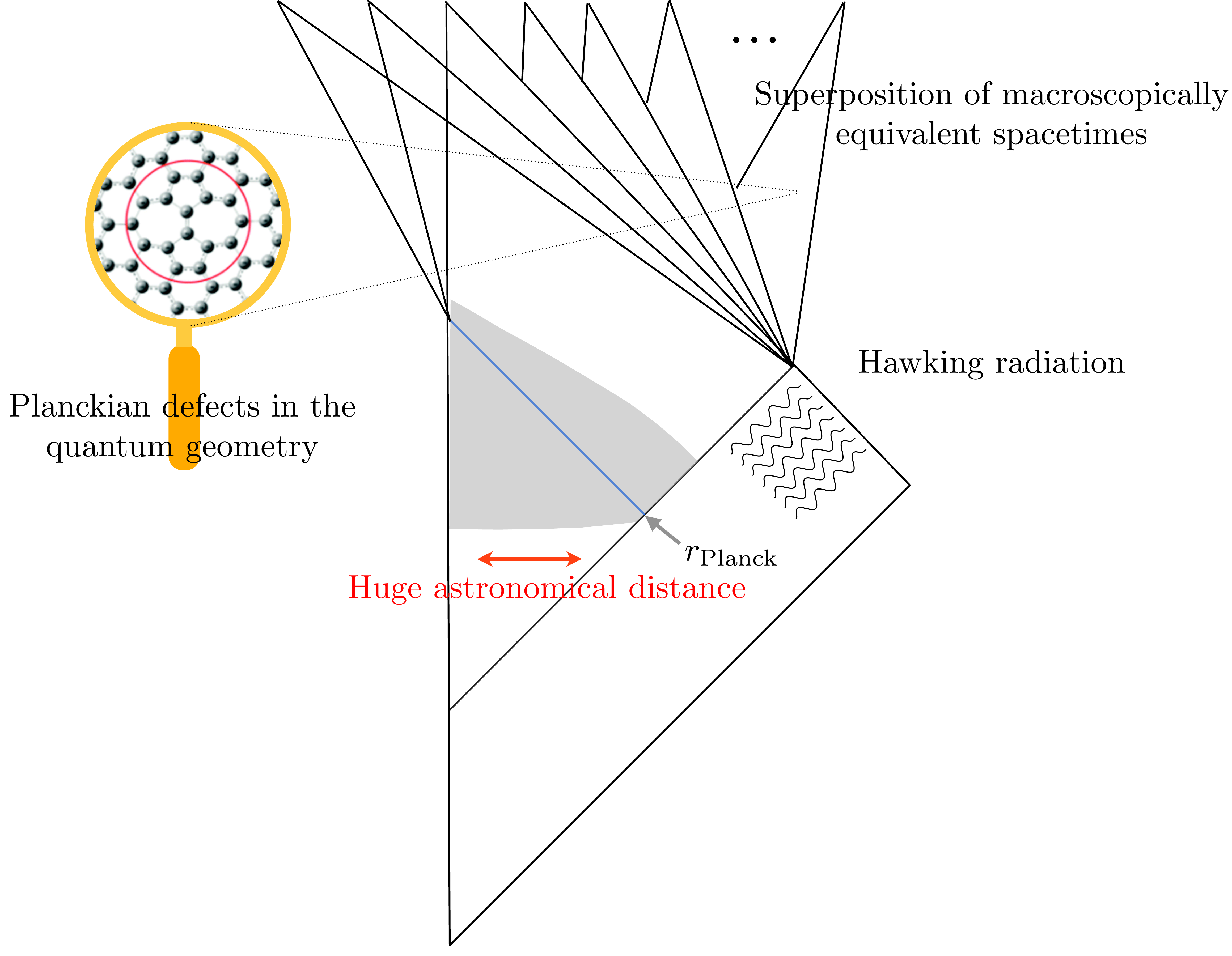} \end{array}\ \ \begin{array}{c}
 \includegraphics[width=9cm]{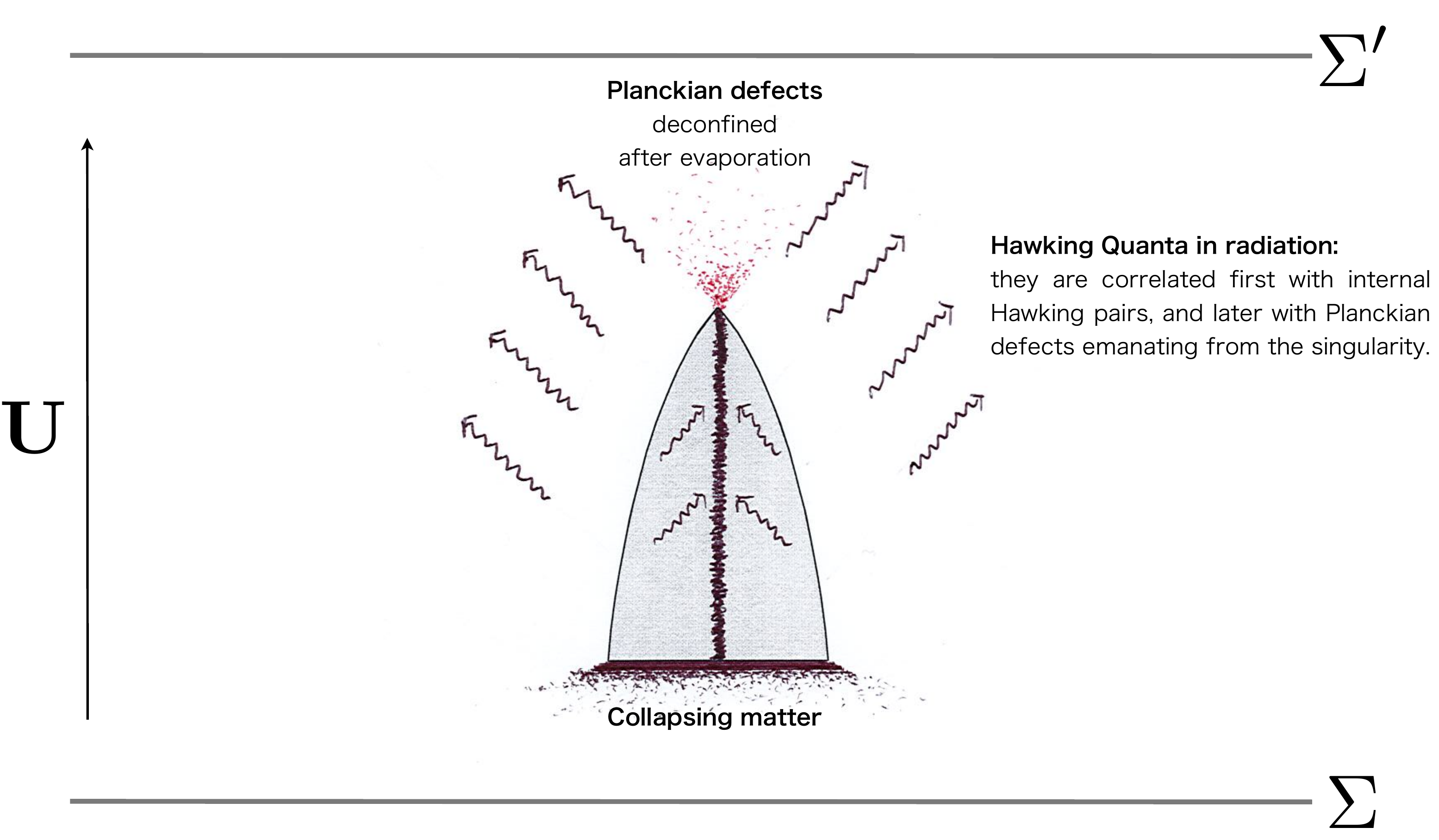}
\end{array}
\)}
\caption{{\bf Left panel:} Penrose diagram illustrating (effectively) the natural scenario, suggested by the fundamental features of LQG, for the resolution of the information puzzle in black hole evaporation \cite{Perez:2014xca}. The shaded region represents the {\em would-be-singularity} where high fluctuations in geometry and fields are present and where the low energy degrees of freedom of the Hawking pairs are forced to interact with the fundamental Planckian degrees of freedom. {\bf Right panel:} Same situation as a scattering process from an initial to a final Cauchy hypesurface. This figure contains basically the same information as the Penrose diagram. Its additional merit, if any, is the more intuitive representation of the shrinking black hole, as well as the time scales involved (collapse time is very short with respect to the evaporation time). Both these features are absent in the conformal representation on the left. There are limitations of such a spacetime representation of a process that is fundamentally quantum and, hence, only understandable in terms of superpositions of different spacetime geometries.}
\label{funo}
\end{figure}



\section{DANIEL'S PERSPECTIVE}
\label{Section-Daniel}
\subsection{Breakdown of unitarity}
\label{BU}
 I  should  start  by  clarifying that the perspective  I  will  discuss  here  has  been  developed  though the  years  with several colleagues,  in works  which  will be  citing  throughout the discussion.
 
In this  section,  we  will be  concerned  with the option  of  negating  the  7-th  
 premise. In fact, such   possibility was considered  ever   since  Hawking  's  discovery (starting  with Hawking  himself),   but apparently rapidly dismissed 
  by him. Early analysis of the possibility  by  \cite{Banks:1983by}  indicated  that  evolution from pure states to mixed states would give rise to unacceptable violations of either causality or energy-momentum conservation, not only in the context of black hole evaporation, but also in ordinary laboratory situations. Those  arguments   were then revisited in \cite{Unruh:1995gn}  and   shown to  be overstated.

 
Recent  detailed studies   indicate  that   models  which effectively incorporate  that feature   do  not only  offer  a  simple  path  for  diffusing the black hole information puzzle\cite{Modak:2014vya, Modak:2014qja, Bedingham:2016aus, Okon:2016qlh, Modak:2017yth, Okon:2017pvc}, 
 but  also provide tools with  the potential to  solve a number of open problems in theoretical physics,  ranging  from the  emergence of the seeds  of cosmic structure out of filed uncertainties  in the inflationary epoch \cite{Perez:2005gh, DeUnanue:2008fw, Leon:2010fi, Leon:2010wv, Leon:2011hs, Landau:2011aa, Diez-Tejedor:2011plw, Landau:2011ljv, Canate:2012nwv,  Leon:2017yna, Leon:2016ysi},  accounting for the value of the cosmological constant \cite{ 
Josset:2016vrq, Perez:2018wlo, Perez:2017krv} dealing with the  problem of time  in quantum gravity 
\cite{Okon:2013lsa, Okon:2016pty}  resurrecting the possibility of   Higgs  inflation \cite{ Rodriguez:2017rjh},  or addressing the so called    $H_0$ tension \cite{ Perez:2020cwa, Perez:2019gyd}. It  should  be noted that  the  latest   proposals  regarding the  cosmological constant and  the  $ H_0$  tension have  been  framed  in the context of an hypothetical QG   granularity, or spacetime defects  but,  in my view, and,  as argued in \cite{Sudarsky:2020wts},  those,   and the  collapse   events of   spontaneous collapse theories, might be, in  the ultimate instance, different  manifestations of the  same physics.    

 In fact, the idea {of} explicitly  modifying quantum  theory, replacing the  unitary  deterministic  evolution  provided  by Schr\"odinger's  equation,  has  its own independent history  and    is a highly active  area  of theoretical  and    experimental research\cite{Bassi:2012bg}.  
 The  so called  spontaneous collapse theories  are  modifications of quantum mechanics  that   were  designed to   deal  with the conceptual problems faced  by   text  book versions of quantum theory and  specifically   in order  to  remove the measurement problem.  The  earliest,  but slightly incomplete  attempt   appears to  go all the  way to  \cite{Pearle:1979vm}.

The main complete versions  include the   original GRW  one, \cite{Ghirardi:1985mt},  in which the   collapses  are discrete   events  involving sudden stochastic   changes in the quantum state of a system  occurring   spontaneously,  i.e.,  without the need of anything like an observer or  a   measurement device. The other  one is CSL\cite{Pearle:1988uh},  a version of the theory in   which the collapse  dynamics is  incorporated directly into the continuous evolution through the use of a  Weiner-like  processes. 

The proposal  of relating  these ideas to the  black hole  evaporation problem  might  be traced back  to  an  article  by R.  Penrose  \cite{Penrose:1996cv}, who   offered an  argument  based on  the  consideration of the possibility of black holes reaching  equilibrium  with some thermal environment in a finite box. There he  argues  that the   feasibility os  such  equilibrium   would  require  that, together    with  what  he  viewed as the {\it  many to one paths}  involving   black hole  formation and  subsequent evaporation,  there  should  exist  processes  that might be  characterized  
as  involving  {\it  many paths  to one path},  and   which  he   described  as  {\it self-measurements or   self observations}, and which  seem    very close to   what we today call  spontaneous  collapses (see  figure \ref{second} bellow).

 \begin{figure}[h] \centerline{\hspace{0.5cm} \(
\begin{array}{c}
\includegraphics[width=8cm]{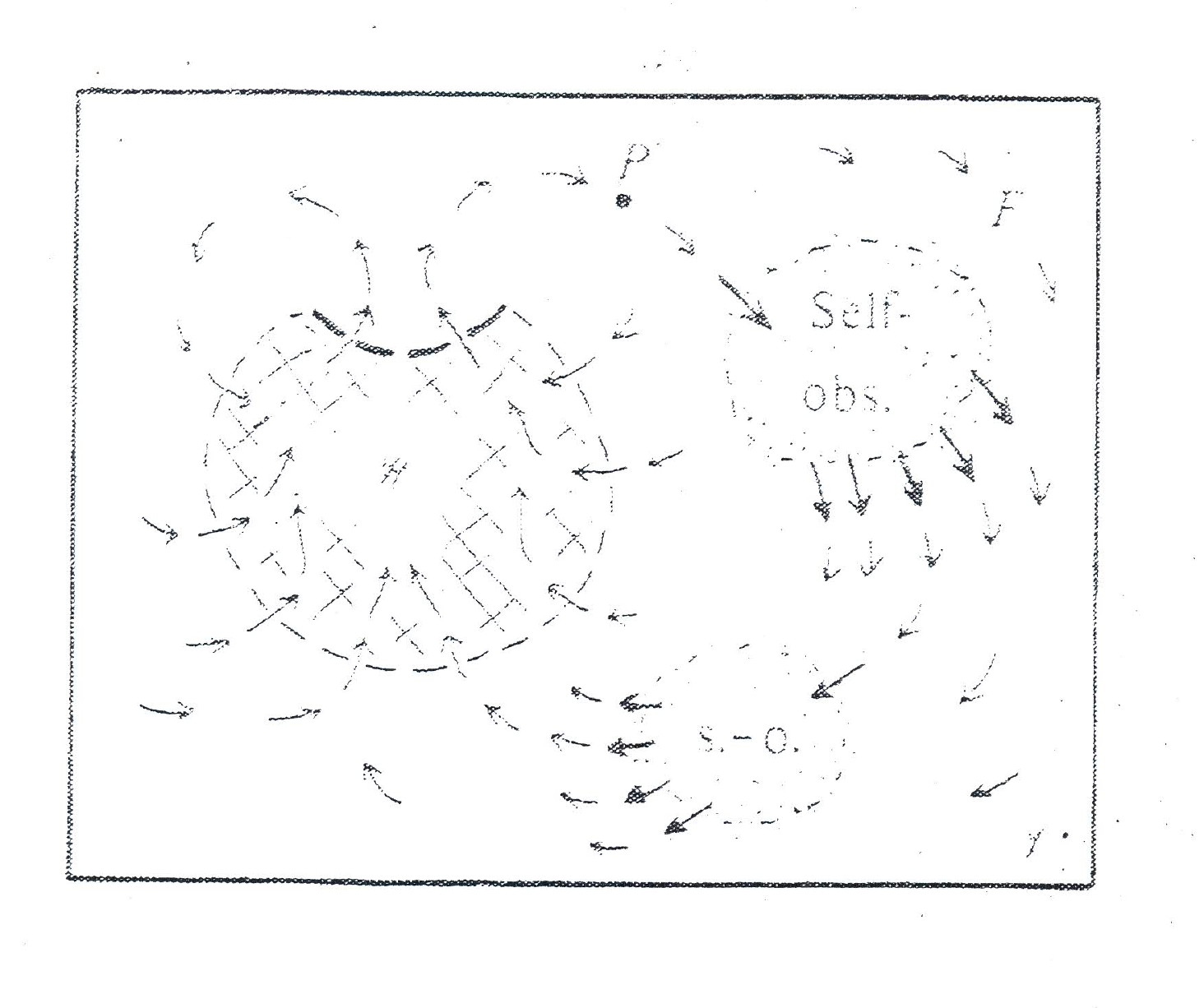}
\end{array}
\)}
\caption{The phase space characterization of a black  hole in   thermal  equilibrium  with  a  thermal  environment. Borrowed  from Penrose.}
\label{second}
\end{figure}

 Recently, the proposal has  been taken further  arguing   that    the  theory might account  for the  black hole  information   puzzle,  in particular  the  loss of unitary relation  between the   state of matter fields  previous to the formation of a  black hole  and  that   of the  same matter fields     after  a complete evaporation   of   the black hole via   the Hawking process. Here,  we  will  offer a brief   review  of   the basic   ideas  behind  the  implementation of   such  account   and     argue,    furthermore, that once   one  accepts the loss of unitarity in the  black hole evaporation, one  is   strongly compelled to  accept   that  non-unitary aspects in the   evolution of  quantum systems   is a  general  occurrence that might  be, in fact,  what is codified  at the phenomenological level in theories  such as GRW  and CSL.

 There are   several issues that must be  dealt  with in describing a coherent and  self consistent theory of this kind in all detail,  and  what we have at the present is  an  ongoing process  along the lines  that  we  will  sketch   in the following.  
 Some  of them  will be   made completely clear and transparent in the   discussion, while   others  are  still the  subject of  ongoing research,  and the  program must   deal  with   substantial   and  rather   nontrivial  technical  and   conceptual  problems.
 
We  will  first argue  that  accepting  information loss in the context of an evaporating black hole,  together with a quantum gravity outlook regarding the non-fundamentality of classical spacetime, leads one  to the   expectation of information being lost   in  general. {This  kind of  argument  seems to have  been considered already by   S. Hawking  at the time of  his  discovery and  taken up  by others.}

In other words,  if unitary evolution at the fundamental level is not  universally valid  in nature, then it wold  
be  unlikely to be  valid  just  when  black holes are involved. So, if we   
 take the view  that it  fails to hold in the context of  a black hole evaporation, then it would   seem  that such departure from unitary 
 evolution must be rather  generic aspect of nature  (although perhaps in a less
  important qualitative level than in those situations  in which  a black hole is involved). After all   we know  that   quantum theory 
   might be represented as a  path integral,  so that 
 every physical  process  receives contributions  from all    available  evolutionary paths\footnote {In the  context of,  say,  an effective  interacting quantum field theory, this is   manifest  in the fact that   the amplitude  for  every process  receives contributions  of an  infinite  series of terms   usually represented by the so called Feynman  diagrams. Note that  in  such   sum,  one must, in principle,  include  the    non-perturbative  type  of  contributions  involving   all conceivable intermediate  states often contributing  ``off  shell", for instance,  {\it virtual  sphalerons} in the electro-weak interactions, as well as  effective   summations  of non-perturbative effects  which   are  represented by  {\it instantons} in  euclidean    versions of  space-time processes.   Thus,  following   our analogy,    we  should expect that   in  every physical process  there would   be  contributions involving 
 {\it  ``virtual black holes" } of various  kinds,  and,  if  their occurrence   is associated  with a breakdown of unitarity,   such breakdown can be expected to percolate to the  original process under consideration.  
According to  the   emergent  perspective  regarding the  nature  of   classical space-time, the issue  might be  expressed  as  indicating that   the    excitations   of  quantum geometrical nature  that might   be   characterized  as    {\it virtual   black holes}   can be  expected to  occur in  all ordinary situations, and,   if   macroscopic  black holes led  to   departures from unitarity,  so  must these  ever-present  ``microscopic " \& virtual black holes.  Departures from  unitary  evolution   must,  thus be   a common  occurrence (in  fact,   such observation was raised  early on by Hawking himself in the very paper reporting his  dramatic  discovery 
 \cite{Hawking:1976ra},  
 although  it  seems  hiss reaction at the time  was to use   such conclusion in order   to cast doubts on the overall     possibility itself).}.  In fact, those non-geometric   excitations connected  with departures from   deterministic unitary evolution  can be  expected  to be even  more generic  than  classical  black holes, as they need not evolve towards  stationary stages,  nor   satisfy the  laws  of  classical  general relativity, since   they   need  not be  {\it ``on shell"} (just  as   ordinary    QFT  forces  us to consider  radiative,  off   shell,   contributions,  once non-trivial interactions are involved)  .

 \subsection{Relativistic collapse}
\label{CovColl}

  Spontaneous  collapse theories   have been  developed    and  studied  for  about   four decades now  \cite{Ghirardi:1985mt,  Pearle:1988uh} in  order to   deal  with  the conceptual  difficulties that plague   the    textbook versions\footnote{Including the  so called  Copenhaguen interpretation, the   Von Newman interpretation and so on.} of quantum theory, and, in particular,  the   measurement  problem (see for instance \cite{maudlin1995three, Schlosshauer:2003zy, Adler:2001us, Okon:2015fgr}).
   
     Most work has, so far,  been  focused  on applications    to laboratory  situations  aiming to set   bounds on the   theories',  parameters,  or  to   observe  actual  deviations  from  {\it standard  quantum theory}.
     
     In recent  years  various  proposals  for  relativistic    versions   of that kind of  theories  have  been put forward, at least in  schematic   ways \cite{Bedingham:2010hz, Tumulka:2006, Pearle:2014tda}.  We  will not discuss    any of those   proposals in  much detail,  but  rather, we  will   consider   what  we    envision as a  generic   relativistic collapse model which    might  be  applied to   quantum field theory,   and then   briefly   explore  their application   to the  black hole  evaporation  situation.

     We  will  describe such -- yet to be fully formulated--  theory in  terms of  
the interaction picture 
 in which the quantum state of matter $|\Psi_{\Sigma}\rangle$ is assigned to a space-like hypersurface $\Sigma$, while the
  quantum fields   carry  the standard- 
   non-collapse part of the evolution. This  might be
      expressed in  the Tomonaga-Schwinger  formal  language  as  stating that   the change   in  the  state   corresponding to   the  hypersurfaces $\Sigma $ to  the   one   corresponding to an  hypersurface   that has  been deformed infinitesimally to the future  of $ \Sigma $  in  an arbitrary way.   In the standard   situations  involving  ordinary interactions, the  corresponding state changes according to 
\begin{align}
i\delta|\Psi_{\Sigma}\rangle
=  \Delta V \hat{\cal H}^{(Int)} (x)|\Psi_{\Sigma}\rangle,
\label{TOM0}
\end{align}
where ${\cal H}_{\rm int}$ is the interaction Hamiltonian density.  A   similar  kind of evolution, might be thought as  represent  the   spontaneous collapse dynamics,  although with  important differences, including the fact that the  evolution is not unitary, nd that contains    stochastic  elements.  
  
Here, $\Delta V$ is the spacetime volume enclosed between $\Sigma$ and $\Sigma'$   assuming  that no point in $\Sigma$ is to the future of $\Sigma'$.  The covariance of the  scheme,  as   in  ordinary  QFT,   involves   requiring     $[\hat{\cal H}_{\rm int}(x),\hat{\cal H}_{\rm int}(y)] = 0$ for spacelike separated   events $x$ and $y$, thus,  ensuring that the state corresponding to  the  evolving   of the corresponding  state  as the associated   hypersurface  {\it "moves"}  across the points $x$ and $y$ is independent of the order in which this is done.  This  requirement ensures,  more generally,   the foliation independence of the  general - non infinitesimal- state development.


 For simplicity, we  will illustrate  the    basic  ideas   using    a  theory  involving  discrete collapses. The basic view  is that, when   looked as a  block,  space-time  will be  sprinkled   (in a manner  to be precisely determined  by the specific theory)  with  spontaneous  {\it collapse events} $\lbrace x_i \rbrace$.  
 Associated with each one of such  collapse events tied to  space-time point $x_i$,  the state  will undergo  a  discrete
   change (the  details    of which must also be specified  by  the  collapse theory and the  actual  physical  situation).   That  is,   when   we   consider the characterization of the  system  on the hypersurface $\Sigma$ as it  crosses the   collapse  event $x$, the state ceases to  evolve according  to   Schr\"oedinger's  equation, and the   dynamics   is replaced  by  a   discrete  spontaneous  change,  that might be characterized  by  a  sharply peaked contribution   of compact support  to ${\cal H}_{\rm int}$, 
  which might be  idealized,   for the time  being, as  represented   by a  4-dimensional  $\delta $- function.  In that case the change  in the state  as  such    collapse event   is  crossed  would  be  given by: 
\begin{align}
|\Psi_{\Sigma}\rangle \rightarrow |\Psi_{\Sigma^+}\rangle  = \hat{L}_x(Z_x) |\Psi_{\Sigma}\rangle,
\end{align}
where $\hat{L}_x$ is the collapse operator at $x$ and $Z_x$ is a random variable which corresponds to the collapse outcome\footnote{One  normally  assumes that there  is  a fixed probability of a collapse event  occurring in any incremental space-time region of invariant volume,  but this  need not be so,  and  the  rate might  be controlled by the  actual physical situation, as  in the  proposal  we  will consider later.}. This results in collapse events which have a Poisson distribution  with  density  $\mu$, (in terms of  space-time  4-volume).
 This distribution of collapse events in spacetime ought to be  covariantly defined and should  make no reference to any  {\it apriori} preferred foliation \footnote{Although, in principle,  it is possible  to consider  theories in which  the contingent aspects  of  the physics itself determines a preferred  foliation, as proposed in \cite{Durr:2013asa}.}.

The collapse operators  are taken   to   satisfy the following {\it completeness} condition: 
\begin{align}
\int dZ |\hat{L}(Z)|^2 = 1.
\label{COMP}
\end{align}
 with the  assignment of the probability density for the outcome  $Z_x$, for a collapse event on the state $|\Psi_{\Sigma}\rangle$ at point $x$,   given by 
\begin{align}
\mathbb{P}\left(Z_x\right||\Psi_{\Sigma}\rangle) =
\frac{\langle \Psi_{\Sigma}||\hat{L}_x(Z_x)|^2|\Psi_{\Sigma}\rangle}{\langle \Psi_{\Sigma}|\Psi_{\Sigma}\rangle}
 = \frac{\langle \Psi_{\Sigma^+} |\Psi_{\Sigma^+}\rangle }{\langle \Psi_{\Sigma}|\Psi_{\Sigma}\rangle}.
\label{PROB}
\end{align}
The  above  condition ensures that  the  total probability  (\ref{PROB}) is  suitably normalized. This formula corresponds to the 
standard formula for the quantum probability of a ``generalized measurement" corresponding  to  the  measurement  of the  observable      associated  with the 
operator $\hat{L}_x$. The collapse outcomes thus occur with standard quantum probability.
 The  central  feature of these theories is that there is no   measurement   device,  no  observer,  and nothing like  an   ordinary   measurement process involved  in this spontaneous  nondeterministic   change in the quantum state\footnote{ The theory however accounts  for  the standard  kind of {\it ``measurement"}    processes,  as a result of the  fact that    when a simple microscopic   degree of freedom   ( usually treated    with   quantum theory)  become   entangled  with a   very large    number of   degrees  of freedom as a result of  the interactions  involved in ant   standard measurement", the  overall   probability of   a    single collapse   affecting the whole system increases  dramatically, leading   to what  one  normally regards as an  instantaneous jump  in   the standard version of the theory.}  

In \cite{Bedingham:2016aus} it is shown that, provided  that   certain   microcausality conditions  involving the collapse  generating operators hold,  
then  (i) given the set of collapse locations 
$\{x_j\}$ occurring between hypersurfaces $\Sigma_i$ and $\Sigma_f$, and a complete set of collapse outcomes at these locations $\{Z_{x_j}\}$, 
the state dynamics leads to an unambiguous and foliation-independent change of state between $\Sigma_i$ and $\Sigma_f$; and  the probability rule  that  specifies the joint probability of complete sets of collapse outcomes 
is  independent of  the spacetime foliation, given  the state on the initial surface $\Sigma_i$.


 The resulting state histories with respect to different foliations are merely different descriptions of the same events 
 \cite{myrvold2002peaceful}. 
 
The covariant form of the collapse dynamics, together with the absence of any foliation dependence,  provides the foundation  for an adequate  relativistic collapse model. To fully  realize such a model,  both  the   rules determining the   density  of collapse events  $\mu$,  as well as the   specific  and general  form for a collapse operator $L_x$  satisfying  the above requirements  must  be specified. The proposals  that have been  considered so  far are   of the general form:  
\begin{align}
\hat{L}_x(Z_x) = \frac{1}{(2\pi\zeta^2)^{1/4}}\exp\left\{-\frac{(\hat{B}(x) - Z_x)^2}{4\zeta^2}\right\},
\label{L}
\end{align}
where $\hat{B}(x)$ is some  hermitian operator, and $\zeta$ is a new fundamental parameter. This collapse operator describes a quasi projection  of the state  of the system  onto  an approximate  eigenstate of $\hat{B}(x)$ about the point $Z_x$ meaning that, if the state  previous  to the  collapse   event  was represented in terms of eigenstates of $\hat{B}(x)$, the collapse effect is to diminish the relative   amplitude of eigenstates whose eigenvalues are far from $Z_x$ with respect to those that have eigenvalues close to  $Z_x$. The effect of many such collapses is to drive the state towards an $\hat{B}(x)$-eigenstate.

This collapse operator must satisfy a completeness condition, which  ensures  a  self  consistent   assignment of probabilities is possible. The microcausality conditions, analogous to those  imposed  on  any interacting quantum filed theory,  are satisfied if
$\left[\hat{B}(x), \hat{B}(y)\right] =0 $
for space-like $x$ and $y$. 
For simplicity, we  will  focus  on  the case of a  scalar field $\hat \phi$,   but the   general approach can be  generalized to  other  types of   fields.

One of the open questions in the  construction  of a viable   realistic   collapse theory  is what exactly plays the role  of a  collapse operator.
The fact that  the  ``mass density"  operator $\hat \rho_M$  has been used   rather  successfully in the  non-relativistic versions of the theory \cite{Ghirardi:1985mt, Pearle:1988uh} strongly suggests    that such  operator  ought to be constructed in   terms of the  energy momentum tensor. This, of course, requires dealing  with the  highly nontrivial  renormalization issues. Besides that  we must face te fact that we are  dealing with a tensorial object.  One  might consider  scalar objects   that  reduce in the non-relativistic limit to   $\hat \rho_M $,  such as  the  trace  of the energy momentum tensor  $ T_{a}^a $,    or $\sqrt{ T_{ab} T^{ab}}$ . One  might  even   consider    the  complete  energy momentum  tensor itself  $T^{ab} $  (in which case  we  will have  several  collapse operators, i.e.,   the various   components  controlling the  collapse  dynamics), something that  can be expected to  lead  to a rather complex behavior, due to the fact that    those components do not, in general, commute  with each other. This idea  was  first   described  briefly in \cite{Bengochea:2020efe}. 
It is   widely expected that  some   sort of  smearing of   quantum fields  and/ or their  derivatives  must be used,  because,  otherwise, the  average energy change in the  state  of  the  field  will, as a result of the collapse dynamics, be  divergent in  a continuous  spacetime \cite{myrvold2017relativistic}.
Such problem could  be addressed    by    several  strategies:    One  might   take  spacetime to have   a  fundamental discrete  structure (which  should not  enter in  conflict with  special  relativity \cite{Bombelli:1987aa}). One might  use  some  natural  region  for the smearing with  some    appropriate   weight function constructed,  for instance,  out of  the Alexandrov set  associated  with a  pair of randomly chosen events,  or the  use of the    underlying space-time geometry to   identify  such region and  weight function.
It is expected   that with appropriate choices for the parameters of the theory, as well as for  the specific operator   $\hat{B}(x)$ driving the spontaneous  collapse  dynamics   one would be able to  construct a specific model satisfying the various  requirements: One the one hand,    the collapse effects  on  macroscopic objects are sufficiently rapid to account  for the  corresponding  macro-objetivizaton, while the average energy increase is sufficiently small to  avoid conflict  with  the  ever  tighter experimental  bounds \cite{Feldmann:2012}, as well as dealing successfully with the  cosmological   problem to which these ideas have been  applied, and  which have  been mentioned  in the introduction.    
Alternatively, one might make  use of a new type  of quantum  field to mediate the collapse process with the effect of preventing infinite energy increase (see  a  specific  construction in  \cite{Bedingham:2010hz}) .

 Note, however,  that any  collapse event at point $x$ on the surface $\Sigma$ converts  the pure state into another pure state,   with a  smaller uncertainty in the  operator $\hat B(x)$.  On the other hand,   as the   specific  state  is  stochastically determined, it is  often  convenient to move   to a description in terms of  ensembles.
\begin{align}
\hat\rho_{\Sigma} = \frac{|\Psi_{\Sigma}\rangle\langle \Psi_{\Sigma}|}{\langle \Psi_{\Sigma}| \Psi_{\Sigma}\rangle}.
\end{align}
 That  is,  we consider  the   statistical mixture  representing the  ensemble of  a large number of identical  systems  characterized  by  the  same state  just before the collapse  event.  Making these aspects  clear and transparent is  in my view an   essential  requirement of any  proposal to address the  black hole information puzzle.  
 
 Putting together   everything  we have   described  so far, leads to the  evolution equation  for the  density matrix  describing  the   ensemble through  one  individual  collapse  event.
 This  is    described  by:   
\begin{align} 
\hat\rho_{\Sigma}\rightarrow  \hat\rho_{\Sigma^+} = \int dz \mathbb{P}\left(z\right||\Psi_{\Sigma}\rangle)
\frac{\hat{L}_x(z)\hat\rho_{\Sigma}\hat{L}_x(z)}{{\rm Tr}[\hat{L}_x(z)\hat\rho_{\Sigma}\hat{L}_x(z) ]}
=\int dz  \hat{L}_x(z)\hat\rho_{\Sigma}\hat{L}_x(z).
\end{align}
This equation, thus,  explains   both   the  meaning of the  statement and the specific   manner   in which   a pure state might be  taken to   evolve into a   mixed state.
We should not lose  sight  of the fact that collapse dynamics in  any  particular  individual case actually  transforms a pure state into other pure   state.   Thus,  the  change in the statistical   density matrix  operator, must be regarded as being    of purely epistemic  nature,    describing our  ignorance   about the individual  collapse   outcomes. 
If we choose a foliation of space-time  parametrized by $t$, with lapse function $N$ and spatial metric on the time-slices $h_{ij}$, and assume that there is a spacetime collapse density of $\mu$, then we can write\cite{Elias2}
\begin{align}
\frac{d}{dt}\hat\rho_t = -i\int d^{d-1} x N\sqrt{h}[\hat{\cal H}_{\rm int}(x),\hat\rho_t]-\int d^{d-1} x N\sqrt{h} \frac{\mu}{8\zeta^2} \left[\hat{B}(x),[\hat{B}(x),\hat\rho_{t}]\right],
\label{PHICOLL}
\end{align}
where   $h$  stands for the determinant of the  components  of the metric  $h_{ij}$  in the coordinates $\lbrace   t, x_i \rbrace $. The  first term corresponds to  the unitary dynamics of the interaction Hamiltonian,  which  would  vanish  in the case of a  free  field  theory. The  second term can be recognized  as giving the  evolution   equation in  the standard  Lindblad  form,
ensuring linearity and  the Markovian nature of the  spontaneous collapse process\cite{Lindblad:1975ef}.


  

The coupling parameter $\gamma = \mu/8\zeta^2$ is usually 
 taken as a constant,  but as  first  suggested in \cite{Okon:2013lsa}  we  will   assume 
 it is a  local function of   
 curvature scalars.  For  concreteness,  we take   
$ \gamma  = \gamma( K^2)$    where $\gamma (.)$ is an  
increasing  function of its  argument,   and $K^2$ is the Kretschman  scalar  ($K^2 = R_{abcd} R^{abcd}$,  ( $ R_{abcd} $ is
the Riemann  tensor of the   
spacetime  metric $g_{ab}$).   This  feature  ensures not only that the collapse  effects will be  much 
larger  in regions of  high  curvature  than in regions  where  the spacetime is   close  to flat,  but it 
might  also be used to ensure that the    
completely flat regions  where,  among other things,  the matter content  corresponds to  the  vacuum (which  in   such flat region is  well defined) , the effects of 
collapse  would disappear completely. 

  Note that the individual  collapse process will not,  however, lead to a precise  eigen-states of  $B(x)$,  or even  its smeared  version, simply because the collapse dynamics  is  only  designed to  narrow the  uncertainty in that quantity,  and  the free dynamics of the field can be expected  to  cause dispersion of the state, in competition with the collapse.
 

On the other   hand,  as  we  will be assuming  that the collapse rate increases  with curvature  in  an unbounded  fashion, we can expect that collapse effects will  accumulate   and dominate over  any dispersion,   in the high curvature region near  the black hole  singularity, --or  more  precisely  speaking,   as   the  quantum gravity region-- is approached, and, thus, we will assume (as a simplifying  idealization) that the collapse process leads   essentially  to  one of the   collective  eigenstates  of   the infinite  set of operators  $ B(x)$   for  all  $x$   on the    hypersurface   that  lies  just  to the past of   $S_{QG} $ ( small  deviations  from this  will have  essentially no impact on  overall conclusions).     We   will denote  a complete  basis   of states (restricted to the Hilbert space of the quantum field  $\phi$  associated with that hypersurface)   as  $\lbrace  | b_i \rangle  \rbrace,  i\in  I  $. 


 \subsubsection{ The  back-reaction    of spacetime}
 
 The  discussion  above,   despite being  rather  general,   has been made  assuming the collapse  affects the dynamics of  matter fields that  are  described  quantum mechanically  on a given  space-time,  but  that, of course can not be  the  whole  story.  Even  if we  are   using a semiclassical  language  in  which   spacetime  is described (at least in the regions  and to  the extent in which  that  it is possible) by a spacetime  metric, so that  we have a clear characterization  of the fact that we are   dealing with   something like a  black hole, the  evaporation  process requires that we  
 take into  account the fact  that,  if the  state of the matter  undergoes  a spontaneous  collapse   resulting  in   a change in the   expectation value of the energy momentum  tensor,  we must be  able,  in principle,  to analyze its   effect  on the  spacetime.
 Actually, several  arguments  have been put forward indicating that
 any such  semiclassical treatment is  simply  non-viable  such as \cite{Page:1981aj}. Such arguments have  led to   an   extensive  controversy  involving a  large   range of  positions \cite{Diosi:1999py,Mattingly:2006pu,Carlip:2008zf,Huggett:2001vyi,Struyve:2017mff}. We will not  offer here  anything close to  
 an   appropriate   description of the arguments and   refer the reader to the corresponding literature of which  we  just  offered   some  examples.    Our approach  has been to   work  towards making  dynamics collapse       
 coexists  with  semiclassical  gravity by  introducing   suitable   modifications in the latter.  
 This is  a highly nontrivial subject  on which    some    initial steps have been taken,  but  which is   far  from being   completely settled.
 
 The basic approach we have taken to  deal with the issue is to assume that,  in regions of spacetime in which  no  collapse  takes  place,   we have a self consistent   semiclassical  description in which    the quantum field theory is adapted to the spacetime metric,   while the   particular  state of the quantum field is  such that   semiclasical  Einstein equation  holds  with aa   suitably renormalized  energy momentum tensor   expectation value corresponding to   that    state as a source. The first  detailed formal  steps  and  application of this  idea  have  been presented in \cite{Diez-Tejedor:2011plw},  and  work is  underway  to ensure  the  general   viability of the procedure, in the  sense of ensuring that, given  a specific   collapse theory  appropriate  initial   data  and the  complete characterization of the   specific  realization of  the  stochastic processes, one  has  a  well posed initial value  problem\cite{Juarez-Aubry:2019jon}.

 A    rather  complex  issue that   arises   in this context  is the following:
  Given a space-time   event $x$,  and  a local  observable $ O(x)$ (or more appropriately  a  local  beable )   associated with it (or a neighborhood  of $x$), and the fact   that there  are infinite Cauchy  hypersurfaces   passing through $x$ and,  thus,  infinitely many  states of the quantum field (one  associated with each Cauchy  hypersurface). Thus,   we  need to find  a  recipe   for selecting,  in a  covariant manner,  one   such  state  from  which a value of   $ O(x)$   (or  some  other characterizations  like the  expectation value or  the uncertainty)  might be extracted.  This issue is of  great relevance if we are to  based our analysis on a  semiclassical treatment of gravity where  we  would need to compute  $ \langle T_{ab}  (x) \rangle$.   
   Here we should mention the proposal made in \cite{ghirardi1994outcome},  in which the state to be employed  for that purpose is the state  associated with the boundary of the causal past of $x$,  $ \partial J^{-}  (x) $. Such proposal, of course,  suffers  from  some  drawbacks.
   The first  is that,  in  general,  $ \partial J^{-}  (x) $ is not a Cauchy hypersurface. This  might be remedied  if we  assume we are given, for instance, the initial state of the universe on an initial  hypersurface  $ \Sigma_{initial}$, and,   thus,  joining   the  $ \partial J^{-}  (x) $  with the part of    $ \Sigma_{initial}$ that lies  outside $ J^{-}  (x) $. A  second problem is that  such hypersurface would in general,  fail to be smooth,  particularly at $x$,  a fact that would probably interfere  with the required    renormalization of   $ \langle T_{ab}  (x) \rangle$. This last aspect might  conceivably be resolved by  modifying the recipe    by one involving suitable limits of  the   quantity of interest  obtained    from states  associated  with   smooth  hypersurfaces,  which have  as a limit the surface  described above.  All these  are issues requiring further study. 
   We  note that  similar issues  occur   in the  approach  favored  by Alejandro, and that  they  have been  dealt  with effectively, by letting the   BH rest frame play  a   fundamental role in addressing them. This is for instance the case in the definition of black hole entropy as the $UV$ contribution to entanglement entropy where a choice of `rest frame' is necessary in the separation of low from long wavelength contributions as discussed in Section \ref{uv-ir}.

\subsection{Application to the Black Hole Evaporation Process}

 The process  we  want to consider  is that  corresponding to the formation of a black hole   by   the  gravitational collapse of
 a  situation involving an  initial 
 essentially flat  and  stationary  space-time,   and   an initial matter   distribution  characterized  by a pure quantum state $ |\Psi_{0}\rangle $  describing a  relatively   localized  excitation of  the  field  $\hat  \chi $
 and something very close to the  vacuum for the  field  $ \hat \phi$.  For simplicity,  we    will consider the  situation is     from the start  spherically  symmetric\footnote{However, it is clear, that as  a result of the randomness in the collapse   dynamics,  specific realizations  of the stochastic elements   will  generically break  that exact   symmetry.  We  will   disregard  that   in  the rest of  the analysis,  to  keep  things  simple enough, and because, as  the  theory itself   does not  break the symmetry  explicitly,  the situation  can be  expected to remain very  close to  spherically symmetric, in  particular,  when one   looks  at it    through  a coarse grained optics.}.
  
   The  resulting   space-time  is  supposed to be  described   by  a manifold  $M$   with a metric   $ g_{ab}$   defined    on $M$ except  for a  compact set $S_{QG}$ corresponding  to  the   region   where a   full quantum gravity  treatment   is required,  and that is  taken to  just  surround the  location of  the  classical  singularity.
   This   characterizes the  formation and evaporation of an essentially  Schwarzschild   black hole\footnote{ We  should  emphasize that, while  the   spherical  symmetry is  introduced for simplicity of the discussion,  there are powerful reasons to accept that,  in  any  event, this  is  nothing  more than  a  rough   sketch,  which we make   very   explicit for the sake of  clarity,  as  any  attempt at a  realistic  analysis  would  require  the    addressing  of the  complex  back reaction problem,   which,  of course, should  incorporate  the fact that  an evaporating  black hole  can not be stationary.},   supplemented by the  region $S_{QG}$, that  is not  susceptible   to a  metric  characterization, and  where  a  full quantum theory of  gravity  is  needed to provide a suitable description.  We  assume that  $\partial S_{QG}$ is  a  compact   boundary   surrounding the   
quantum gravity region,    which,   by  assumption,  corresponds to that  region  where otherwise (i.e.,  in the absence of   a  radical   modification of GR  due to QG  effects)  we would  
have encountered the black  hole singularity.  See for instance Figure \ref{PD-nous1}.
   
We take the $M-S_{QG}$  to be  foliated  by   some convenient  one parameter set of {\it almost} Cauchy hypersurfaces \footnote {Which we  might define as  those hypersurfaces  $\Sigma$  that  are   such  that any  inextendible causal  curve   intersects  $\Sigma$  either once,  or alternatively reaches $\partial S_{QG}$} .
 
  We  will  further make  some  relatively mild (and  rather common) assumptions   about  quantum   gravity.
  
   i) The first   is that the definite theory of  quantum gravity   will cure  the    singularities  of  general  relativity,   however,   in doing so,  we will   require  that there   would   be   regions   where the  standard  metric  characterization of spacetime simply   does not apply, and  which  can only  be   described in terms of    fully quantum     gravity     degrees  of freedom. That is, the   usual  emergence of  classical space-time  does not  happen  for  that region,  which,   in  our case, corresponds  to   what  is  referred  to  as   the set $S_{QG}$
   
   ii)  We  will assume that Quantum  Gravity does not   lead,    at the effective level, to   dramatic  and   arbitrarily large violation  of   conservation laws,   such as    energy or  momentum,  although as  discussed  in  \cite{Maudlin:2019bje},   some  level of  violation seems  unavoidable.
   
   iii) We  will assume that   the    space-time region   that   results ``at the other side" of the QG region  is  a  reasonable   and  rather   simple,   space-time (i.e., a spacetime that is  topologically  $ R^3\times R^{+} $   with  a metric  that  is  very close to  the   Minkowski one).  
  In this  sense, the  picture that emerges    from the  present approach,   is  quite similar, at the macroscopic level, to the one  advocated  by Alejandro. As  we  will see the  central  differences concern  the  microscopic  descriptions.
   
With  these assumptions,   we   can  already   make     some   simple predictions about the nature of the   full space-time.

Given that, by assumption, the  effects of  the collapse dynamics  will be   strong only in the region with  high   curvature, ( more explicitly in the regions   where the value of  the  Kretschman scalar
$ K^2 \equiv  R_{abcd} R^{abcd} $,  or   $ W^2 \equiv  W_{abcd} W^{abcd} $ where $W_{abcd}$ is Weyl the   curvature  tensor) \footnote{ The idea  of  using some  measure   of  the  Weyl tensor    in  determining the  strength of the collapse  dynamics    is  inspired   by  some   of   R.  Penrose's ideas. },   has  a rather  large value,   the   dynamics   
characterizing the   early   evolution  of our initial cloud of matter will be   essentially the same as that found in  the  standard accounts of  black hole formation and   evaporation: The  cloud of matter    will   contract  due to  its  own gravitational pull,  and, as follows from   Birkoff's  theorem, the 
 exterior  region   will be  described  by  the Schwarzschild  metric;  the original matter excitation  will  eventually  cross  the corresponding Schwarzschild  radius, and  
  generate  a Killing horizon 
  for the   exterior time-like Killing field $\xi^a$.   The   early  exterior  region,  and   even the  region to the interior of the  Killing  horizon  but close to it  at early times,  are regions of small curvature,  and thus  the    picture   based on standard  quantum field theory in curved  spacetime  that leads to  Hawking radiation  will remain essentially unchanged.   This, by itself, indicates that   almost  all the   initial   ADM mass of the spacetime  would  be radiated in the form of Hawking radiation   and will  reach  ${\cal  J}^+$ (asymptotic  null infinity) before $u_0$ of   Figure \ref{PD-nous1}).

  Next  let  us consider  the  space-time  that emerges  at the other  side of the singularity.  Given  that essentially  all the   initial energy has been radiated to ${\cal  J}^+$,   and in light of   assumption ii)  above, the resulting spacetime  should  correspond to  one   associated with an almost  vanishing mass (this  would be the Bondi mass  corresponding to   a  spacetime hypersurface  lying   to the future of region $S_{QG}$  and  intersecting ${\cal  J}^+$  in  a  segment  to the future of that containing  the Hawking flux) the point $u_0$ of Figure \ref{PD-nous1}. This  conclusion, together with assumption iii)  indicates that   this spacetime   region   should be close to  a rather simple  
  space-time, essentially  in vacuum, which  we take, for simplicity, to correspond to  a  flat Minkowski region (perhaps  with  some  small amounts of radiation).  At this  point our  picture   differs  in an important manner   from that  advocated by Alejandro, because  we take the space-time  in that region as not being    very different  from  Minkowski space-time  at the macro-description,  and  to whatever  corresponds to it at, the level of the  quantum gravity  description.  As  we  saw,  the point of view  advocated  by Alejandro requires  that region  of  space-time to differ  substantially  from what corresponds  to   ordinary Minokwski  spacetime  at the  quantum gravity level, even  though, at the  macroscopic  level  it is  very close to it (i.e.,  it must be filled   with  certain  spacial  quantum gravity excitations or  defects).
  
Let  us  now  focus on the state of the quantum fields\footnote{Again, this might be regarded  as a relatively explicit  sketch or  ``cartoon"  for the reasons    previously mentioned.}. For simplicity of the description, we  will consider   two  quantum fields, the field $ \hat \phi$ which  will  be the focus of the Hawking evaporation,   and  a  second   field $ \hat \chi$ whose role  will be to  characterize the initial matter  distribution  which   will undergo  gravitational collapse and  lead to the formation of  the black hole (we  might take it to be  very massive,  so that  its  contribution to the Hawking evaporation process  might  be  neglected  at  every but the  very last stages  when,  QG effects might  become visible, i.e., the   region  where the  would be  singularity becomes  ``naked). That is,  we  will  be    concentrating on  different aspects for each  field,  although, in principle, the   picture might be adapted to either  a single quantum  field  or a  very large multiplicity  thereof.
  We thus  take the initial situation,   as given on  a   very early  hypersurface, to correspond to
   an essentially flat   and  almost  stationary space-time, i.e.,    something very close to Minkowski space-time.   The matter fields are characterized   as follows   $\hat \phi$: will be  essentially in the  Minkowski   vacuum   state, while the  $ \hat \chi $  field  will  be in a state  that represents   a   low density  cloud  corresponding to  some of the  large scale- low energy-modes  being  in a relatively high state of excitation,  so that  the  system, left to its own dynamics,  will undergo  gravitational collapse   and the eventual   formation of a  large  black hole.

 
  The   state can  be  represented  in the   first QFT  construction   $ Q1$ (which is  based on a mode  characterization  at 
  $\Sigma_1$ as   as:
 \begin{align} \label{initial state}
|\Psi_0\rangle    =    |{\rm cloud}\rangle^{in}_{\chi}  \otimes  | 0\rangle^{in}_{\phi},
\end{align}
 where the    first  term    represents the  high  degree of excitation of  the few  modes  
 associated  with the   matter cloud,    while the  second represents the  state of the $\hat  \phi$  quantum field. 
   
After a large number of  collapse  events,  taken for simplicity  to have occurred   in the   black hole  interior, the state  of the  system  will 
be  in a pure  state, which,  in the outside,  will begin to resemble    one  of the  elements  of  a thermal  ensemble.

At late times,  corresponding to  the situation where the black hole has   completely  evaporated,  we  might  consider the quantization of the fields  either in   null infinity,   or, if say, the  field $  \chi$   is massive,    at a   late   hypersurfece 
$\Sigma_2$ covering the region  to the   chronological future of the singularity, as  well as  approaching  null infinity.   We denote  such construction   as $Q2$.

We  might, for simplicity,  describe   such construction in  terms  of two   quantum field   theory  constructions: Region I)   corresponding to the region of  {$\cal{J}^{+}$} that lies to the past of  the point 
of the   causal future of the   singularity  or QG   region (i.e., point $u_0$  of  Figure \ref{PD-nous1}) , and Region II)  of    {$\cal{J}^{+}$} that lies  to its future.

The point is that  any particle  or wave  packet mode   that  crossed the   singularity would    either go to  $i^+$ or  to  region  II) of   {$\cal{J}^{+}$}.

Now,  had the  QG  region replacing the singularity  be   a  normal space-time,  we  could have   evolved  modes  from  this  very   late time  QFT  
  construction to the  very early times thus establishing a correspondence  between the sets of  creation and anhilation operators. As it is,  and  under the assumptions we are working with,    no such  general  mode evolution is   defined,  simply because   the   QG region  we do not have a  metric characterization of the gravitation  DOF,   making it  rather difficult to  discus   some of the issues we  want to talk about.

\begin{figure}
\centering
\includegraphics[scale=0.5]{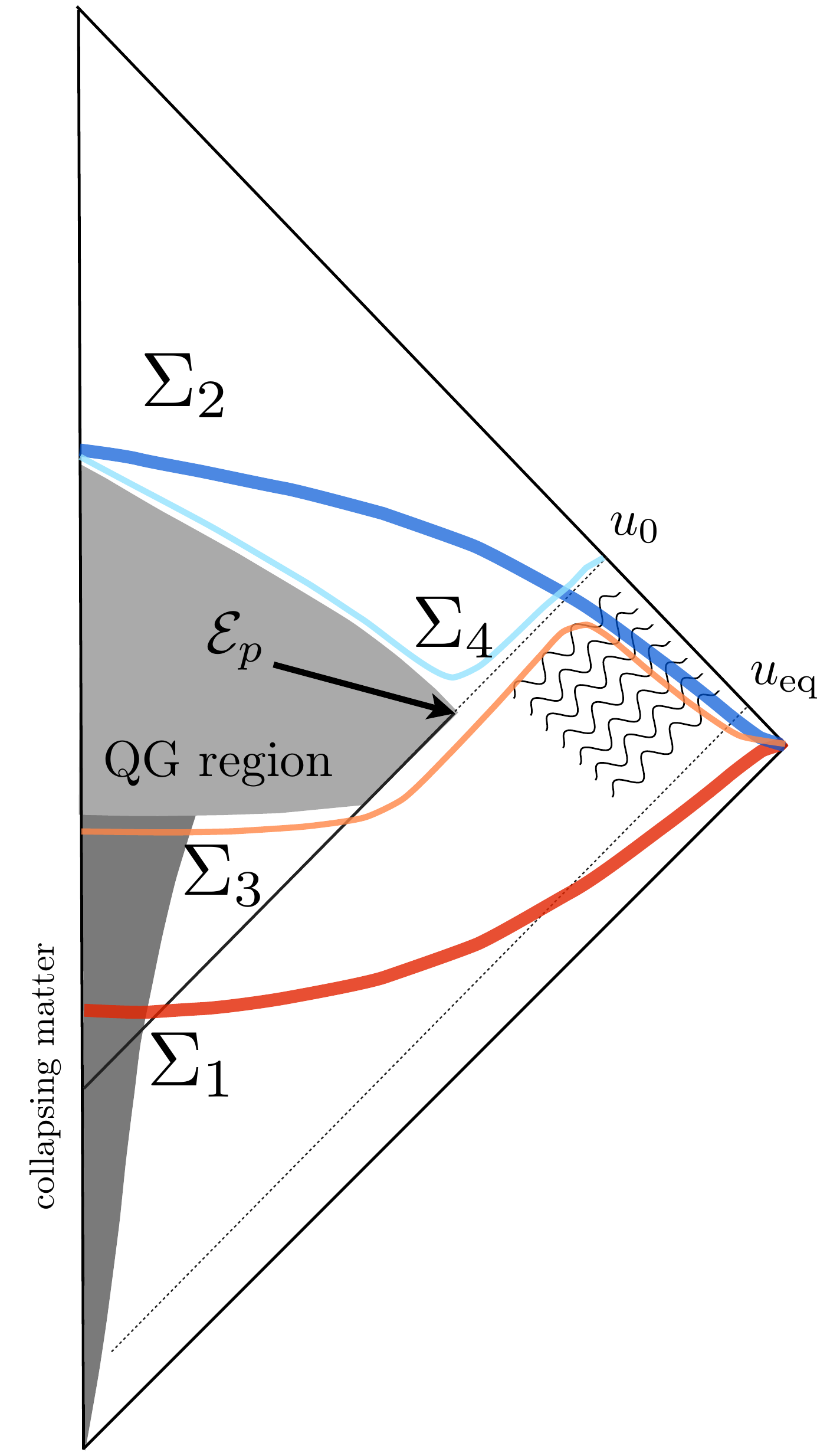}
\caption{ }
\label{daniel-pd}
\end{figure}

We can, however, consider, instead,  a 
   third    quantum field   construction $ Q3$  for the field  $\hat \phi$, that   will be   based on  a  mode characterization associated  with the  hypersurface $\Sigma _3$,   which is made of two parts, one lying just before the   QG region, and   one  joining the  ``Evaporation Point" $ E_P$  with  $i_0$ (see Figure \ref{daniel-pd}).Thus, the quantization  
    can be  done  in two parts: one    corresponding  to the   usual   one  associated with  the  positive  frequency modes   associated    with   timelike  Killing field  in the  neighborhood  of  Region  I  of  null infinity, slightly  deformed  to  connect with the  first segment.   
   We  label the  positive norm modes of this  quantization   by ${\lbrace \alpha \rbrace}$,  which we,  take to be highly peaked  on   energy $ \omega_\alpha$, at some   angular momentum  quantum  numbers  $ (l,m)$  as well as  being  relatively well localized  in the  
     radial direction.
     The  occupation number  of the mode  $\alpha $ in the  Fock space   basis  is   denoted  $n_\alpha $,  and the  general set of 
     occupation numbers  characterizing the   element of the Fock basis as $ F$.  We also  take  a rather  arbitrary, but  convenient  choice of modes  for the   black hole  interior,  such that the  interior and exterior pairs are maximally entangled.


 This  later   QFT  construction  will  involve  specific   creation an annihilation operators that  we use one can use a   define  a fiduciary notion  ``particle''  for the BH interior. The  vacuum  state of  the first  QFT  construction   for the field $\hat \phi $   can be  written as:

\begin{align}
  | 0\rangle_{\phi,   in} = {\cal N} \sum_{F} e^{-\beta_H E_F /2} |F\rangle^{\phi,  int}\otimes |F\rangle^{\phi,  ext},
\end{align}
   where  the  sum is  over the  sets   of   occupation number  for all modes $ F=\lbrace  F_{n_\alpha} \rbrace$ 
   (which  indicates that the  mode $ \alpha $ is  excited  by  $n_\alpha $ quanta),  $ E_F = \sum \omega_\alpha {n_\alpha} $ is the  total energy of the   state  according to the   notion  of    
   energy associated  with  the  asymptotic region, $ \beta_H $ is the Hawking   
   thermal  coefficient,  and ${\cal  N}$   is  a normalization   constant.
   
    On  the other hand,  we have  the   excitations of the  field $\chi$   associated  with  the original   cloud of   matter  $|{\rm cloud} \rangle^{\Xi } $ that  led to the formation of the black hole, and  which can be taken to
   have  propagated  completely  into  the interior  region  of the  Black Hole   Horizon,  so that   we can   write the state (\ref{initial state})  simply as: 
 	\begin{align}
 |\Psi_0\rangle  = {\cal N} \sum_F e^{-\beta_H E_F /2}   |{\rm cloud} \rangle^{\chi } \otimes  
 |F\rangle^{\phi ,  int}\otimes |F \rangle^{\phi,  ext}
\end{align}
   Note that,  at late times,  the first two  terms  correspond to   modes that lie   inside  the    black hole  region  while the   last   term   refers   to modes  that  would   propagate   all the  way to  part  I  of  {$\cal{J}^{+}$}.

   Of  course,   the  above  picture    still lacks the  effects  of  
      the  collapse  dynamics,  as well  as the  effects    associated  with  quantum gravity  that will take  place  when modes enter  the  QG   region.  Consideration of those  effects    will next allow   us   to provide  a    description of  the  rest of  the   evolution   and  a  characterization of the    state  of the  quantum fields at  very late   times,    including the  region that lies  to the future  of   the  QG  region that replaces what would  have been  the  singularity (were it not  for   the QG  effects).
      
\subsection{Collapse dynamics and the black hole evaporation}.

One of the  assumptions  that    underlies   the   present  approach to deal  with  the information   question  during  the Hawking evaporation of the black hole  is  that the  collapse  dynamics,  although    valid  everywhere,   will   depend   on   curvature, and that   its effects   will  become highly  significant   only  in  regions where that  is large. Thus, the  quantum  evolution of the   state of the fields   would  differ  quite  dramatically     from the     unitary   evolution of  standard  quantum theory deep in the  black hole  interior.
This  idea  will be implemented  by assuming    that  the     parameter   $\gamma$ controlling the  strength of the  collapse modifications  is a function of  the  Kretchman  scalar  $K =  R_{abcd} R^{abcd}$,  or,  alternatively,  the     square of the  Weyl curvature  $W^2  =  W_{abcd} W^{abcd}$.  For  simplicity, we   will, therefore, ignore the modification of  the quantum  state of the field  resulting from the   collapse dynamics  in the exterior region.

Now,  given that the  collapse dynamics contains  explicit stochastic  elements,    one  can not,   even in  principle,    predict  with  certainty   what will be the evolution  taking   place in the black hole  interior.  In fact,   given that    we  do not know   what is the  nature of the  quantum gravity theory   required to describe the region  $ S_{QG}$,  we can not  describe  what  goes on  inside that  region  either.  Actually  it is most natural to     consider that,  in reality, the boundary between  those regions in  the deep black hole interior is  not a  sharp one,  and   that
 part of what  we  will take to   be  occurring  in the  region preceding that corresponding to  the  QG  regime might  also take place  within  the later.  However, in order  to  simply the presentation of the  overall picture,  we  will make  an assumption    of mainly pedagogical  and aesthetic  nature. We will take the   collapse  dynamics  to  perform    the  full  modification of the quantum state of which it is  capable,  before   the  QG region is  reached,   while  we  reserve  for the QG theory the  tasks  which, at this  point,  do not seem to possible to  associate  with   the type of collapse theories  that have  been considered so far.  This  will  become  clear  in the discussion   around    equation \ref{Post-singularity-vacuum}.

       As noted, the collapse  dynamics  involves  rather    strong   stochastic elements,  and,  thus,   it is  convenient  to move to a  characterization of the  situation in terms of  an  ensemble of    identical  copies of  the initial setting, and to then   follow the  evolution of  that ensemble.  We take the initial state  characterizing the state of the matter  at the initial  hypersurface  
      expressed in terms of the pure,    density 
   matrix  (indicating   all  elements  in the   ensemble  are in the same initial state)  as
represented  in the  language of the    quantization $ Q2$: 
\begin{align}
\rho  (\Sigma_0) = {\cal N}^2 \sum_{F, F'} e^{-\beta_H (E_F+E_{F'})/2 } |F\rangle^{\rm int}\langle F'|^{\rm int} \otimes |F\rangle^{\rm ext} \langle F|^{\rm ext}.
\end{align}
 We  can now    simplify  things   using the   basis  of  eigen-states of  collapse  operators,    which we   will refer to as   {\it the  collapse basis}.  We  thus rewrite the  above   density matrix  in the form  
\begin{align}
\rho  (\Sigma_0) = {\cal N}^2 \sum_{ij}\sum_{F,F'} e^{-\beta_H (E_F+E_{F'})/2 } \langle F'|b_j\rangle^{\rm int} \langle b_i |F\rangle^{\rm int}  |b_i \rangle^{\rm int} \langle b_j|^{\rm int} \otimes |F\rangle^{\rm ext} \langle F|^{\rm ext}.
\end{align}

 Next,  we  use the fact that,   in the collapse  region,    especially in the   late  part thereof,  the    collapse  dynamics    becomes  extremely strong and  effective, and thus  drives the  state of the system to  one of  the  eigen-states of the collapse operators. This  allows  us to  write   the  state  representing  an ensemble of systems  initially prepared in the  same   state \ref{initial state},    at  any  hypersurface  $\Sigma_1$  lying  just  before the  would-be  classical singularity --or more precisely the  quantum  gravity region,   after the  complete collapse process   has  taken place  as,  
\begin{align}
\rho (\Sigma_1) &= {\cal N}^2 \sum_{i}\sum_{F, F'} e^{-\beta_H (E_F+E_{F'})/2 } \langle F'|b_i\rangle^{\rm int} \langle b_i |F\rangle^{\rm int}  |b_i \rangle^{\rm int} \langle b_i|^{\rm int} \otimes |F\rangle^{\rm ext} \langle F'|^{\rm ext}.
\end{align}

Finally,  we   need to consider  the    system as  it emerges on the  other side of the quantum gravity region, i.e.,  the  state   describing the     ensemble   after  the  would-be  classical singularity. 
As  we have  discussed in the introduction   we are   assuming  that    quantum gravity  
would resolve the singularity and lead  on the other side,  to some  reasonable 
spacetime  and state of  the quantum fields. We  now    consider    
the characterization of the system  on  an hypersurface  lying just to the future of  the   
would-be  classical singularity (which we  will often  refer  simply as the   {\it "singularity"}).  Such  a hypersurface,  dented  by  $  \Sigma_4$  in  figure \ref{daniel-pd},   would  not  be a Cauchy  
hupersurface,  as it would  intersect  ${\cal J}^+$ rather than $i^0$.   As  such,   one   
can  partially characterize   the  state of fields  on  it  by  the value of the Bondi mass.   
It  is clear,   as we have argued  in the   the introduction,  that if we assume that Quantum 
Gravity   does not lead  to large violations of   energy and  momentum conservation 
laws,   the only possible  value for this  Bondi mass  would have to be  the mass of the 
initial   matter shell  minus the energy emitted   as Hawking   radiation,  which is  
present to the   past of the {\it ``singularity"} on ${\cal J}^+$. This  remaining mass  will  thus have to be   
very small.  

The task for quantum gravity,  according to the  present approach,  is to turn the {state characterizing the physical  situation on,  say, the  interior part of $ \Sigma _3$ (which involves   both complex quantum  states  characterizing the fields   $ \chi$  and  $\phi$ and  the  rather  extreme spacetime    geometry  characterized by    very large  curvatures)  into an ordinary  low energy state in the post-singularity region}. For simplicity, we  assume that this state  is,  for   every element of ten  ensemble,  the vacuum in a  flat  spacetime region\footnote{The  overall picture would hardly change if  we  took  the  post-singularity   state  emerging  to the future of the    QG  region  were  slightly different in  each realization,  as when consider tracing over the corresponding  DOF  in order to describe  the  conditions  to the of the {\it ``singularity"} on ${\cal J}^+$  we  would sill end  up  with  a thermal   state describing the Hawking radiation.}
\begin{align}\label{Post-singularity-vacuum}
|b_i\rangle^{\rm int}|{\rm cloud}\rangle_L \rightarrow |0^{\rm post-sing}\rangle,
\end{align}
the particular $|b_i\rangle^{\rm int}$ being selected  by the  whole  series of  collapse processes that result from the collapse dynamics deep in the black hole  interior. 

This means,  (taking into account the joint probability densities  
) that the final state characterizing  the  ensemble of systems (on $\Sigma_f$)    should be of the form: 
\begin{align}
\rho_{\rm final} &= {\cal N}^2\sum_{i}\sum_{F, F'} e^{-\beta_H (E_F+E_{F'})/2 } \langle F'|b_i\rangle^{\rm int} \langle b_i |F\rangle^{\rm int}  |0^{\rm post-sing} \rangle \langle 0^{\rm post-sing}| \otimes |F\rangle^{\rm ext} \langle F'|^{\rm ext}
\nonumber\\
&={\cal N}^2\sum_{F,F'} e^{-\beta_H (E_F+E_{F'})/2 } \langle F'|F\rangle^{\rm int}  |0^{\rm post-sing} \rangle \langle 0^{\rm post-sing}| \otimes |F\rangle^{\rm ext} \langle F'|^{\rm ext}
\nonumber\\
&={\cal N}^2\sum_{F,F'} e^{-\beta_H (E_F+E_{F'})/2 }\delta( F'-F) |0^{\rm post-sing} \rangle \langle 0^{\rm post-sing}| \otimes |F\rangle^{\rm ext} \langle F' |^{\rm ext}
\nonumber\\
&={\cal N}^2\sum_{F} e^{-\beta_H E_F }  |0^{\rm post-sing} \rangle \langle 0^{\rm post-sing}| \otimes |F\rangle^{\rm ext} \langle F|^{\rm ext}. \nonumber\\
&= |0^{\rm post-sing} \rangle \langle 0^{\rm post-sing}| \otimes  \rho_{\rm thermal}^{\rm ext}
\end{align}
That  is,   the    system has   evolved   from  an initially pure state  into a  state representing   the proper thermal state of radiation  on  the
 early part of ${\cal J}^+$,
 and the vacuum state  afterwards.
 
  The  general picture is then  that loss of unitarity  and fully  irreversible physics   is  just part of fundamental physics  when  described   in terms of spatio-temporal notions.  The underlying   nature of the  fundamental  theory   from which    this   emerges is  the same  physics underlying the emergence of    spacetime.   Loss of unitarity  occurs   very seldom when a  few degrees of freedom are  involved,  as in delicate   few particle laboratory experiments, but becomes   an  essential  aspect  if one    wants to  account for  macro-objectification (i.e., part of the  measurement problem  underlying quantum theory),  and becomes   even more  dramatic when black holes   are concerned , not just   due to the  high number of degrees of freedom involved,  but also as a result of the high  curvatures   occurring,   which   according  to our proposal,  enhance the  collapse  dynamics   basic  strength.

\subsection{Some open issues}

\subsubsection{Energetics}
We will expand now on a   point we briefly touched before. 
 One  of the most  serious challenges  one   faces  when  attempting     to construct   relativistic models of   spontaneous   dynamical   reduction of the wave function, either of  the  discrete or  continuous  kind,   
is  their  intrinsic tendency   to  predict    the violation  of energy   conservation by  infinite   
amounts\cite{myrvold2017relativistic}: 
The problem  is resolved   in the non-relativistic  setting where  one can easily control the magnitude of that kind of effect,  by  relying on a suitable  spatial  smearing of the collapse operators,   usually taken to be the   position operators  for the individual particles that make  up the system.


     This  issue  becomes relevant in the present  context  at two places. First and foremost,   at the point where one  wants  to  consider, in some detail,  the back reaction    of the spacetime  metric  to the   changes   in the quantum state of the field  $\hat \phi$  induced  by the collapse dynamics. Here, one must  ensure that  the collapse dynamics takes Hadamard  states   into Hadamard  states\cite{Juarez-Aubry:2017ery}.   The second place  where the issue  appears is the point  where   one     considers   the role of   the quantum gravity region.  In a previous treatment  it was  argued  that,   provided   that  quantum gravity did not result in  large violations of energy conservation,  one can  expect the  state  after the   quantum gravity region  to   correspond energetically   to  the   content of the  region just before the   would be singularity, and that  this region    would have almost  vanishing  energy content   being   made  up of  the  positive   energy contribution of the    collapsing matter shell and the   negative energy contribution of the   in-falling   counterpart  to the Hawking flux. 
We  would face a serious  problem  with this  argument if  the  region just before the    would-be singularity   could  contain  an   arbitrarily large  amount of  energy   as  a result of the   
unboundedness of the  violation of   energy  conservation brought about by the collapse dynamics.

There are various  approaches offering paths to  deal   with   this issue.
i)  We  might  consider a fundamental discreteness of spacetime (which, however,  as  discussed in\cite{PhysRevLett.93.191301}, should  not  be tied to violations of  special relativity).

ii)  We might  adjust  the choice  of   collapse operators and    provide  a   sensible   space-time   smearing scheme   dependent  on  the  energy momentum of the matter fields,  or  on the  geometric  structure of   the    curved  spacetime.  In this    context  it is  worth   noting that,   when   one   considers  that  the   parameter  controlling the   strength  or  intrinsic   rate   of the   collapse dynamics   depends on   spatial curvature   through,  say, the   Weyl scalar  $W^2  =  W_{abcd}W^{abcd}$,   one  might assume that   in flat space-times the   collapse rate   actually vanishes,    thus removing  most   concerns  about    the stability of the vacuum  in these theories.   In that case, one  would adopt the position that the collapse  associated with  individual particles  in  the non-relativistic  quantum mechanical   context  is  actually derived  from the  small   deformation of flat space-time associated  with that  same  particle. That  is,  one  would consider that   the particles  energy momentum   curves the spacetime  and this in turn,   turns-on  the  quantum collapse    dynamics.  How  exactly      would this  work  deep in the black  hole  interior region 
requires  further exploration. 

iii)  We  might rely on  the  effective  smearing  provided  by  the  use of  the    auxiliary pointer field   as   a   way to      introduce  the smearing    procedure without   seriously affecting the    simplicity of the   treatment,  as    discussed in the Appendix  of   reference \cite{Bedingham:2010hz}.  

iv) Finally, we  should mention a   recently proposed version of collapse theories that involves  both  energy increases 
 (associated with the  increased  localization of  ordinary  wave packets), as  well as an intrinsic fundamental  dispersion. This reduces   the overall level  of   energy  creation in   ordinary  situations  and   generates some  overall energy dissipation in others\cite{Smirne:2014paa}.

\subsubsection{States  Hypersurfaces and Local Beables}

 As is  well  known  in  quantum field theory   one usually  assigns  a quantum  state  to  every   Cauchy  hypersurface,  and the  relation among the various  states  is encoded   in  a Functional   Schr\"odinger equation. This  is   clear   when  the theory is presented in   the  Schr\"odinger picture. When  working  in  the Heisenberg picture  one  could  say the  same,   with the understanding that the   same state is assigned to all hypersurfaces  because all evolution has  been passed   on to   the operators. In most   practical applications,  one   uses the  so called   {\it Interaction picture}  \footnote{Despite the fact  that  Haag's   theorem  shows this to be,  strictly  speaking,  incorrect.}.   
 There are, in  general, no   doubts  regarding  what   state to use in computing  
any   quantity of  physical  interest  at  some point,  except,  of course, for the fact that  quantum fields are distributional objects, and  some  level of  smearing is required in order  to obtain   well defined operators  acting  on the  theory's Hilbert space.
This  seems  to  raise no problem  when dealing with  the simplest operators,   but, as it  is well known,  one  must rely on renormalization   
when dealing with more complex ones such as the energy momentum tensor.   However,  all this obscures the fact that  the theory is famously  rather unclear about  its ontology:  What are the  theory's  
 local beables,  representing what,  according to the theory, exists in the  four dimensional  world  out there ?
  We  will not enter this complex  discussion  here  and point the reader to some relevant  literature  
  (see for instance \cite{sep-quantum-field-theory} and references therein),  
   but   note that,   when  choosing to work with  semi-classical gravity,   one must include the expectation value of the energy momentum tensor as one of the local beables. After all, that is the  quantity that determines the   Ricci  tensor of the spacetime.  
 Here, one must confront  the  rather  delicate  problem that  emerges  when incorporating  spontaneous collapses of the quantum  states  into the  mix\footnote{A  similar problem   would   arise if one  were to  consider  measurement induced  collapses}:  
 Given a local  beable, say, $A(x)$ 
 associated with a  spacetime  point  $x$ (ignoring, for simplicity of the  discussion,  the need for smearing),
  there are  infinite   number of Cauchy hypersufaces  
 $\lbrace \sigma_{\alpha} \rbrace$  
 that pass  through  it, and  associated  to each one  there is, in principle,  different state     
$\lbrace \Psi_{\alpha} \rbrace$. Which of those  states    should  one use in extracting  the relevant   information    about    the  local beable?  The point is that, as these  states are not connected by a unitary  evolution,  different choices  could  lead to different results.  This problem   has  been noted 
 in  \cite{aharonov1981can, aharonov1984usual}, and  further discussed in  \cite{myrvold2002peaceful}.  
 In our case,  the issue  applies  directly to the   quantity $ \langle T_{ab}^{Ren} (x)  \rangle$, the  renormalized  value of the  expectation value of the  energy momentum tensor  corresponding to one of the  states   $\Psi_{\alpha_0}  $, but  which one? One  would expect a reasonable theory to provide a  covariantly defined  prescription, and it is not clear what    it should  be.   As noted above, an interesting proposal  \cite{ghirardi1994outcome} is to take  the state  associated with
  hypersurface defined by the boundary of the causal  past  of  $x$, namely  $\partial J^{-} (x) $. That  proposal  is   attractive  for its covariance but seems  to require some tampering  to  deal with two   undesirable  features.  In general,  
  $\partial J^{-} (x) $ is not a Cauchy hypersurface, and   it is not smooth  at $x$,  which might be  a problem    when   attempting to  carry out the renormaliztion  procedure.   There  are   some ideas being played with at the time, but  the issue   certainly   requires   further in depth   investigation.

 \subsection{DISCUSION}
 

 
 \subsubsection{On The Nature of space-time to the future of the    Quantum  Gravity Region.}

   DANIEL:  According to your view,  the  space-time  that  emerges  after the  BH has  evaporated    looks  very much  like the   Minkowski that   existed before the BH formed,   but is  somehow  different, right?  
  
  ALEJANDRO: Yes,   the  space-time   to the  future  of the QG region  is a superposition of states  which   macroscopically  seem   indistinguishable  from   Minkowski's space-time,   and   a   vacuum  state for the quantum fields  (and, thus,  compatible  with the energy available  according to the estimate   obtained  at  $ u^0 $   in    $ { \cal J}^{+} $). However, each  element of the  superposition   is  entangled  with the  Hawking radiated  particles  via  hidden quantum   numbers characterizing  aspects  of  quantum gravity that have neither geometric  nor effective  field theory description. One  might  want to call them  quantum gravity  defects  when  described in    macroscopic  language.

   DANIEL:  Those  quantum gravity  defects  might  be   regarded  as analogous   to    what,  in my picture,  are the  collapse events.  However,
  they seem also to have substantial  differences. On one   hand, in  your picture they play   a  most prominent role in the  post   QG  region,  i.e.,   they  must  be  present there in order to   be entangled   with the   matter   DOF  so  as to  purify   the  very late  quantum  state,  although, I presume, they might start  forming in the  black hole  interior    region  before the  full quantum  gravity regime  is reached.  
  Another difference is that, while  the   collapses   are essentially   space-time  events,  that is, they have  no   persistence or permanence in time,  your  defects  must, once  formed,   persist in time  forever,  in order to ensure that, in  any late  Cauchy hypersurface, the state is pure,   right? 
  
  ALEJANDRO I am not sure if my defects need to `stay in time' in the way you put it, even when this is probably the simplest scheme that our intuition can represent. What I have in mind, is that at the fundamental level quantum gravity, is timeless. This is particularly clear in the canonical approach to quantization where one sees that (due to general covariance) there is no Schr\"odinger-like evolution equation, but rather only constraints. Thus, the quantum physical states of the theory are states
annihilated by the quantum constraints representing by themselves a whole history of spacetime. Spacetime causality and localization is information to be extracted via Dirac observables encoding this in a diffeomorphism invariant fashion (the explicit construction of which represents an obvious technical and conceptual problem in quantum gravity, referred to as the problem of time) Therefore, from the quantum perspective - that we have decided to drop in telling our versions of the schemes in favour of a semiclassical account - my defects are written in each and every of the physical states whose linear superposition represents the physical situation at hand, namely, black hole formation and evaporation. From this perspective, one can envisage defects as spacetime events like your collapse events. There spacetime location will have to be constructed in terms of a subsidiary ``mean field semiclassical spacetime geometry, like the one we are using to tell our individual proposals. This is also the reason why I like the representation of Figure \ref{funo}.

  DANIEL: Do you envision  those  defects  becoming  diluted  with  the further  passing  of   time,    so that  at very late times   your spacetime  has  such a low  density of  defects that is essentially  (and  at the microscopic  level)  like the Minkowski  spacetime?
  
   ALEJANDRO:This is a possibility if the defects are persistent in time. However, the previous perspective where the quantum correlations   are  encoded in defects in the physical states might offer more subtle possibilities. The very least I can say is that there does not  seem  to be a single  spacial  state that we can safely call {\bf the}  Minkowski spacetime in quantum gravity. I think this is the central point in my view: one cannot define what a state that looks like Minkowski all the way to the microscopic level is,  in quantum gravity.

  \subsubsection{Firewalls  and  such.}
  
  ALE:  In order to  avoid the firewall  problem,    you need   the  collapses  to   occur in  a very particular way  and place?
  
  DANIEL:   The collapses  dictate  the  changes in the quantum  state  on  any hypersurface,  however, in    order to  complete the theory,    one  must  prescribe  how  to  compute the    {\it local beables} of the theory  (see for instance\cite{Maudlin_2014}) .  In the view we  adopt  when  using  semiclassical gravity,  the main   local beable is   the  expectation value of the energy momentum tensor at  a point  $x$  (if one is  working, say, in  some adapted version of semi-classical gravity). Thus,   one must be careful   regarding the   choice of the     hypersurface  and  corresponding state one  uses,    because  as    discussed above, through   any  $x$  there passes   an infinite  number of hypersurfaces. I   think  an attractive   recipe is  the one   considered  in \cite{ghirardi1994outcome},
    which  uses the state associated with the hypersurface  $ \partial  J^{-} (x)$ (a  related  version    is the proposal    by Penrose  where the  state  is updated  on  the past  light cone). Note that  in this approach, and  ignoring the  collapses that take place  outside the   horizon,     for no  point on the  Event Horizon   there is   anything   strange  concerning the   expectation value of the energy momentum tensor,  and, thus,  no `` fire wall".  Of course,  one  still needs to make  sure that  collapses, in general,  do  not lead to  violations  of energy momentum conservation   that are too large (and  are  excluded  by experiments \cite{Feldmann:2012} ).

\subsubsection{Time reversal invariance}

 ALE: Why,  if the  fundamental  theory  is not   time reversal invariant,  do   classical theories,  which  are very   often  an excellent approximation,   seem to  be  time  reversal invariant?

  DANIEL: As I see it,  this is  because life is  possible  in regions of the universe with a high degree  of  predictability (at least as  far as  gross features are  concerned). Humans focus on  situations of high predictability  because they are  useful  and  susceptible  to manipulation,  etc.   We  generally  refer to  the  part of nature that can be  described in terms of highly  predictable  theories,    classical   (with exceptions,   such as   statistical mechanics, where   we work  with  approximations which are required   due to    the sheer complexity of the situation under study,  or  due  to our   own epistemic  limitations), and  the  characterization of   quantum mechanical  we reserve for the phenomena involving  features  that   seem fundamentally unpredictable (and  include  aspects that  are not predictable even as approximations (note, however, that in  the Bohmian Mechanics approach to  quantum theory, the
lack of predictability  is  all ascribed to  fundamental epistemic  limitations)\footnote{One  might  also point  at the   T  violation in the electro-weak sector,  responsible for the  famous violation    of CP  in the neutral  Kaon,  and  other neutral  heavy meson   systems,  as an exception,  but, in  all honesty, I do not see how that  could   modify in a significant way  the  broad   time  reversal invariance at the level of physical laws  that you are alluding to.}.

Now, it seems  quite clear that  the  spontaneous  collapse theories   involve   a  serious  departure from   time reversal invariance:   Given a  state  at  one time,   one has  a certain level of predictability   encoded in the probabilities assigned    to  the collapse  events, just to the future of it.   However, retrodictability  seems to be   much weaker:    that is,  given a   state  at one time, the  set of  states  from which  it might have    evolved   through some   series  of collapse  events is extremely large,  and very   little   can be said  about their assigned  probabilities. On the  other hand,  and   I  am not trying to      use   {\it  what-aboutery}  as a  defense,    your approach also   seems   to involve some departure   of  time  reversibility to be inherent to the  theory:   your  space-time defects  are created  but not destroyed,  is that  right?

 ALEJANDRO:  No, in the  approach I am presenting  the fundamental laws are assumed to be time reversal invariant. The apparent irreversibility is only emergent as a feature of the macroscopic world. The evolution is unitary at
the fundamental level, and, so, defects can be created but also destroyed: if you take the final state of BH evaporation and you evolve back with $U^{\dagger}$you get back to the initial state that led to the BH formation and subsequent evaporation. The apparent arrow of time emerges only for us (coarse grained) macroscopic observers just like in usual situations involving many degrees of freedom that escape precise tracking.
 My view  is  essentially that  depicted in  Penrose's picture  in Figure \ref{second} . 
 As far as the (apparent) violation of time reversal invariance is concerned the tale is the same as for more familiar systems, like the burning of a newspaper or the breaking of a perfume bottle.

ALEJANDRO:   What is your view regarding  the nature of BH  Entropy.

 DANIEL:  That is  a  question on which I  confess to  be extremely  confused.  To  start  with,  I think  one  must  clarify   the  precise notion of   the   entropy   one  associates  to  a black hole.  Should  we   be   referring to a  Boltzmann  Entropy?,  a   Gibbs Entropy?,   a  Von Newman Entropy?,   and  we should also  indicate what  precisely is  this   entropy   being associated with?   with  the black hole   interior?   with    the  event  horizon?,  or  are  we referring to  the entanglement between interior and  exterior  DOF? Should  one take the view that  the  entropy is  to  be    associated  with a state,  which is  itself  associated  with    a Cauchy hyperurface? 
 That  would  be,  after  all, the  analogous thing to the  assignment of  entropy  with  a physical system at a given time.   Perhaps,  we  should  extend that   to  include   a general   spatial section (any  closed  3-d  space-like   hypersurface,  whether  it  does or  does not have a border), and the  generically  mixed state of quantum fields  associated to it? , or should  we  only associate entropy  to  such regions  when viewed  as fully immersed  in a   complete space-time ?  (here,  we have in mind the  problematic   fact that  the  location   of  the  horizon might be   undetermined,  even if  complete   data  are  given on a  generic  hypersurface, as  in the   example  of the  Schr\"odiger black hole  considered  in \cite{Sorkin:1999yj, Corichi:2000xf}, where  a  situation  is created involving a quantum   superposition of   having  a BH horizon intersecting a  Cauchy hypersurface  and  having  no  horizon  there ).
 I  should  here  emphasize that,  in my view,     entropy  should  be  assigned   to  black holes,  in    general,    and  not just to  stationary  or quasi-stationary  black holes, because,  in its  standard  usage,   entropy  is a characterization  of a general  macrostate  of a  system,  and it  is   not just assigned   a value  to  equilibrium  or  quasi-equilibrium  situations.  In   fact,  equilibrium  situations  are supposed to be those that maximize  the   entropy  (given  certain   constraints)  which makes  sense  only  if   entropy is  defined  for all states. 
 
 Thus, in dealing with  black holes,  
 we should not limit  our considerations to stationary or static situations, otherwise 
 Hawking's Area Law would lose much of its relevance, as far as entropy is concerned. 
  Needless is to say, that most  works  that,  starting  with a proposal  for a  quantum gravity theory,  claim to have  obtained  the result  $ S_{BH} = A/4$,  have  focused  on just  those  equilibrium or quasi-equilibrium situations, and  there is  no  clear  path  for the kind of  generalization    that I  think  would  be  required, in order to  justify  the  claims that the issue has  been  resolved. 
 
The  next issue that causes me great confusion are  the  reasons  that could  be  behind  the  validity of the  
 generalized   second law, namely the  claim that the  quantity  that  is  always  non decreasing is  $ S_{G}   = S_{matter outside}  + (1/4)  \Sigma_{i}   A_{ BH_i}$.   However,  as discussed  in  detail  in \cite{sorkin1997statistical},  one would expect that if  $S = A/4$   is the entropy of a black hole, it somehow  represents  the number of  possible internal 
 states $N_{in} $ (so 
$S =  ln(N_{in}) $ (setting  Boltzmann's  constant $ k_B =1$)   compatible with its macroscopic characterization (as seen from the outside) (i.e.,  in terms of  say  $(M, A, J, Q ...) $.
Then, the generalized second law (GSL)  would follow from consideration of the number of totally available states of exterior and interior  $N$,
  with $ N \approx  N_{in} N _{Out} $  it would follow that $ S_{total }=    ln (N)  = ln N_{in} +ln N_{out}= A/4 + S_{out} $  

But there are reasons to doubt all this:
   
i) In principle,  $N_{in}$  (unless  somehow  restricted  beyond the requirement that it matches the outside macro-state)  would  be unbounded, even a  complete universe of arbitrary size might be attached to the  interior !  
       
ii)  The number  $ N_{in} $  might be expected to  be a simple  expression of the macroscopic parameters    $M,  A,  J ,  Q $  only   in  or close to equilibrium,  but the interior is  not close to equilibrium in any sense.  So,  why should  the entropy  be   simply $ S=  A/4$?  Strikingly, even  for BH's in which  the exterior is out of equilibrium!!.
   
iii)   There are good  reasons to doubt that,   in general,  $ N_{in} $   could  be  determined   by  the  parameters  that characterize the BH  exterior    $M,  A,  J ,  Q $,   simply because there are   infinite ways to reach the same   external situation,  from  a   direct collapse of  various  kinds of matter configuration,   a slow  build  up,  or  even   a  scheme where one  lets  it evaporate and continuously  feeds   it  back  with matter at the appropriate rate for as long   as  wanted. 
     
  And,  to me,  any claim that    one    understands   what  is  the entropy   of a   black hole (and why  does  it have the  value   it seems to have) must be accompanied  with an understanding  of the reasons  for the  validity of the generalized  second  law.  So, to me,  the whole  thing   remains  very  mysterious, indeed.

  DANIEL :  Besides information, there are other  things usually taken to be conserved,  but which,   in the case of the  formation and  evaporation  of a black hole,  their conservation is not  longer  assured. For instance,  Barion number  B)\footnote {More precisely,  Barion minus Lepton number  B-L,  which is  an  exactly conserved  number  in the   standard model of particle physics. Barion number, by itself,  is  known to be  violated via Sphaleron mediated processes}. I guess  that   in  your approach also  B  (or B-L)   conservation   is    violated in BH  formation and  evaporation, right?  So,  why  is  it that  you are  so attached to unitarity  and  information conservation,  while  you seem not to be  bothered  by these other  novel violations of conservation laws?

 ALEJANDRO: I do not have a solid argument to say that unitarity must hold and I believe that your proposal is a possibility. I am only trying to argue that it is also possible to have an account where unitarity holds in a very natural way with extremely conservative assumptions. My view, I believe, shows that the question could have a simple answer that is the same that we give in a vast variety of situations of more familiar nature. Concerning other conservation laws, I do not see in what way the ones discovered empirically in the construction of the standard model of particle physics should continue to hold at the fundamental level of quantum gravity. However, time reversal invariance and predictability is a universal feature of everything we have understood so far and produces a picture of the world that fits my philosophical stand. I admit it is a belief, one of the many that we all have when trying to explore how to push the boundaries of well understood physics.
 
 DANIEL:  In both our approaches,    we require something rather dramatic to  take place  in the  QG   region, namely the  negative  energy excitation  of the field $\hat \psi $  must combine      with  the  positive energy of the matter degrees of freedom    that lead to the BH formation in the first place,   and  in a sense  annihilate   each other.  In  my approach, the only thing  remaining  from this process    ought to be  something like the vacuum.  In your approach,    you need this to    lead to the generation of the defects  that must be entangled  with the  Hawking particles, so as  to lead,  at late times, to a pure state  unitarily related to the initial state.
 
  However,  it seems to me that,    in  your case,   the required process  needs to be  more  precisely fine  tuned  and  extremely  robust.   The defects   must  form in the precise amount, and become  precisely entangled   with the Hawking particles,  otherwise   the  picture will not work. It seems to me, that in  the  picture I   advocate,   one needs  much less precision  and robustness  for things  to work.  For instance, if  some type of  matter   remnants  emerge    into  the  post QG  region,  we would face no problems,  and the essential picture  would  still work, as long as the   late time  state is Hadamard (which  is a   much weaker requirement). Of  course,  for all we know,   QG  might be a theory  naturally  incorporating   such  fine tuning and   robustness,  but one  approach  seems   to be    asking  more from the  theory than the other.  Do you think I am missing something? 
 
ALEJANDRO: Yes, I think we are all missing  crucial  aspects  of what the exact form the theory of   quantum gravity takes, and specially  its  effects in  near the would-be-singularity. What you call fine tuning is, in my view, just the details of the microscopic dynamics. The situation is like the one of the ship and the bulbous bow reducing its drag (mentioned in the introduction). From far away, we understand what is happening when comparing the situation with and without the bulbous bow in terms of the destructive interference of waves. But drag or friction is happening locally at the surface of the ship. Somehow, the presence of the bulbous bow changes the dynamics of the fluid, and this must have a microscopic description where friction is reduced by a change in the local interaction between the fluid molecules and the surface atoms. From this perspective, it seems that there must be an extremely precise fine tuning of events allowing for what (from the microscopic perspective) seems as a miracle. But there is no miracle nor fine tuning, we believe that a sufficiently powerful computer would be able to compute and show that the reduction of the drag follows from simple dynamical rules of the individual constituents once the macroscopic features of the problem are suitably incorporated.
  
 DANIEL:   Your  construction  is  clearly motivated  by a  deep commitment to unitary  evolution  and  the corresponding   preservation --at the fundamental level-- of  all information.  So, do you think  that   the kind of   special   quantum  gravitational  DOF that   appears  in your  scenario   in the post QG  region,  also  appears  in connection with  ordinary  measurements,   such  as, say, a   Stern Gerlach  experiment--- where information  about the initial state of the particle seems  to be lost, and  the evolution  seems to depart  from   strict unitarity?. Would  you also  consider that  when  an observer    in   the asymptotic  region detects  a Hawking particle,   similar  DOF  would  be involved? 
 
ALEJANDRO: I agree that the measurement problem remains an open issue in standard quantum mechanics. I also believe that this issue is particularly serious in the context of quantum gravity where one is describing the universe as a whole with a theory that is meant to be the fundamental theory, and, thus, reserves no place to classical auxiliary structures or mysterious observers for which the dynamical rules should be different from the fundamental ones. However, in my understanding of the issue that concerns us here, these questions do not play a central role. In fact, my perspective is that we should be  able to come up with a description where these important questions are left for a future development. Basically, I think that quantum theory is not the last and definite fundamental picture, and, in that, the two of us agree, I believe, Quantum theory, as it stands, is (as Einstein would put it) incomplete and quantum gravity pushes us, in my view, to face that incompleteness and solve it. I do not know how this difficult issue will be resolved, and your perspective, where an active collapse feature is included, might be a possibility. At present, and lacking of a better view, I am inclined to believe in something perhaps approximately realized in (non-local) hidden variable theories, like Bohmian mechanics, where there is no collapse at the fundamental level, but only an effective one. I confess that it is tempting to imagine that the microscopic Planckian degrees of freedom that I evoke in my scenario for black hole evaporation could actually be related to the  hidden variables in a possibly deterministic fundamental theory. However, this is something that should be regarded as a rather wild speculation at the moment.

\subsubsection{ On the nature of the collapse operator}

ALEJANDRO: Could you comment more on the nature of your operator B where things collapse to in your picture?

 DANIEL:  That is,   of course, one of the key  open  issues in the spontaneous collapse  theory   program. What  seems  to  be most solidly  established  is the indication that,   at  the non-relativistic level,  what  seems  to  work, is for  this operator to  closely resemble  a kind of  {\it smeared}  mass density  operator $\hat \rho(x)$,  and, thus, it is expected that the  general   version must  reduce to something   close to that in the appropriate  limit. However, when facing the search  for a suitable  relativistic  generalization,  in  particular  a general  relativistic  case,   expressed in terms of  the language  of quantum  field theory,    one faces     rather serious challenges.   For starters, if attempting to   respect the  spirit of general covariance,   one  would  need  to   device  a generally covariant recipe  for  determining the   smearing region. That   is  highly nontrivial  as,   when limiting  consideration to the  simple   case of special  relativity,  one faces the problem that   there are no  compact sets   that   are   invariant under Lorentz  transformations.   One might attempt to follow  a  strategy analogous to   that  developed  by the  Causal Set program  and  chose the smearing  regions  through  some   stochastic  procedure that   will not  produce  individually invariant regions,    but that     would  do so   {\it in the  average}.  A concrete  example  would  involve     choosing  for each  individual  collapse   event (in a  GRW like theory  involving  discrete   collapses)  a  pair of events in spacetime $ p \& q$  and   fix the   smearing region to be $J^{+} (p) \cap  J^{-} (q)$  (when such set is  nonempty).
 Unfortunately, this path  seems  to be  blocked, at least when relaying only on  ordinary type of degrees of  freedom,   by the arguments  put forward in \cite{myrvold2017relativistic}   concerning the stability of the  vacuum. Another option  would be to  restrict the collapse  dynamics to act  on curved  space-times ( recall the whole approach is  designed to  be     applied in   a  semi-classical context)   and make  use of the  features  encoded  in the curvature tensor  to  determine  the  smearing region.  The  point being that   it seems  perfectly reasonable to   assume that   in exactly  flat  space-times (which would of course be devoid of  matter  in the  semiclassical sense  i.e., $ \langle T_{\mu\nu} \rangle =0$)  the  collapse  dynamics  would   shut  off.   Once  the    issue of the  smearing region is  determined,  the next    part of the question is  what  is  the operator to be smeared   in     the  prescribed neighborhood of  $x$ in  order to generate $B(x)$. Here,  we  seem to  have a  very     large set of covariant    possibilities, starting   with the  trace of the energy momentum tensor $T=T^{\mu}_{\mu}$, 
or suitable   contractions of that  tensor taken  with an  appropriate  power  (so  as to reduce to  $\rho(x)$ in the non-relativistic  limit )  such as $ (T^{\mu\nu} T _{\mu\nu} )^{( 1/2)}$, etc.   An  even   broader set of possibilities opens up if one considers combining the  energy momentum tensor of matter with some  curvature  scalars  (or even  direct  contractions of  the energy momentum tensor with  curvature tensor, or including   things  like torsion into the theory, after all  fermions and  torsion seem to    go   together  rather naturally in  various formulations of variants  of GR).  But, of course,  the   concise answer  is that, at this point,  we do not  really know  what  works and  what does not,  and the study  of this important   question is  underway  on    several fronts.

\subsubsection{Timelessness of quantum gravity}

 DANIEL:  Your last couple of replies reflect something that is  very hard for me  to absorb, not just about  your specific approach,  but also 
  about the  general  program adopted by  most   workers in,  say,   Loop Quantum Gravity, which,  as I understand,  you take essentially  as  an inspirational source. The fact that the  resulting theory is timeless, and thus,  the  question of recovering  spacetime,   what it means,   
  and how is one to  envision it,  irrespective of the technical details, seems to  creep up  in almost all discussions (often only after some   careful examination). 
    For example,   you emphasize that the essentially flat spacetime, that one  finds  to  the ``future" of the quantum gravity region, is  a  rather different kind of  Minkowski spacetime,   as compared with the one   that existed before the black hole formed, with the former  filled  with those  non-geometric type of   defects (which serve  as  {\it ancilla} to purify the  late time Hawing radiation), while the   latter is a  cleaner (i.e. defect free)  type of Minkowski  spacetime.  That   is a story I can  understand,  but one   which I  simply can no longer follow  when  you remove the temporal component of  spacetime  from it.  For instance,  when you  state  that, as  in  the Loop Quantum Gravity program,  there is no spacetime, and  so that  your  defects  could   be taken as spacetime  events  (just like the collapses  in the approach I favor),  I cannot  avoid  thinking that a more accurate description would be  to say that those  defects  are   points in a timeless- space, rather  than  spacetime  events,  simply because  when  one  reverts to  the   spacetime  description which presumably would  emerge in the treatment  of our problem in  LQG program, those  defects  would indeed  have to  persist in time,  don't you  agree?

ALEJANDRO:   
I  agree that,  when  expressing things in a  spacetime  language,   something like  what you  describe must take place. However, while  we have,    in this discussion employed  a  classical language to talk about spacetime geometry, (at least outside of the so-called quantum region)  I do not see  that  as a fundamental issue. 
 The  point is  that the  general covariance of the theory implies that, on any given Cauchy surface, the data for matter and geometry satisfying the constraints of general relativity contain all the information about what will or has happened  in a way that is, in principle, available without any spacetime representation (or evolution). Of course in practice, it is convenient to fix lapse and shift fields and evolve such initial data to reconstruct a spacetime geometry with matter on it. There is complete arbitrariness in your choice where different choices will lead to spacetime configurations of the basic fields related by a diffeomorphism. The reason why it is convenient to reconstruct a spacetime out of the data is that, with a spacetime at hand (written in some coordinates), we can extract physical dynamical information (coordinate independent information) using the notion of test fields and test observers. We can, for instance, answer  the question of how many years (turns around the sun) will pass between two total solar eclipses in France on Earth. This is coordinate independent information which  encodes  things    we  usually  talk  about in terms  of ``where "and  `when" without making reference to any external time variable. In practice, this information can be   extracted most readily  from the spacetime representation; but the structure of general relativity is such that this is also written in terms of initial data satisfying the constraint on any  Cauchy surface. Now,  the way this information is written when employing  different Cauchy surfaces of  a  single  spacetime depends on the  chosen Cauchy surface even when the information  encoded  is  really the same.

Thus, when I say that the flat vacuum-geometry state in the past is different from the flat vacuum-geometry state in the future in Section \eqref{vgs}, I am  implicitly making reference to such  distinction. In particular, as we have agreed to  use semiclassical language here, and taking advantage of the additional background  structure  of special relativity and quantum field theory on flat spacetime  recovered in the asymptotic regions near $\sI^-$ and $\sI^+$, it is possible to say in what sense the time-less information is encoded differently in the comparison of, say, the initial and final Cauchy slices, which we can idealize here by $\sI^\pm$, respectively. Concretely, those degrees of freedom that are UV quantum geometry defects, which end up being entangled with the Hawking radiation in the future near $\sI^+$, are simply QFT modes entangled among themselves in $\sI^-$, so that the state is seen as the vacuum according to the positive frequency notion of observers in $\sI^-$. The persistence of the degrees of freedom that you evoked is there, but the character of these changes when you change the Cauchy surface with respect to which you read the physical data, which in my underling,    fundamental  picture,  is just time-less. 

In fact, the most basic phrasing of my perspective is intended to be taken as  fundamentally  timeless. The timeless character of the degrees of freedom in general relativity reveals the deep deterministic nature of the classical theory. The quantum side of such feature is timeless as well (in my picture of course), and this is the most fundamental statement of unitarity (which is only a wording imported from the time-full description of quantum fields on a background spacetime). The semiclassical description that we have agreed to give in Section \ref{language} together with the asymptotically flat assumption allowed to describe my picture in terms of an `evolution' from $\sI^-$ to $\sI^+$ that is unitary. However, all this must   be  considered as  resulting  from  a timeless account in the full quantum gravity description.  

DANIEL:  Actually, 
   if I understand  correctly,  the  program (at least as advocated by  followers of the  Loop  Quantum Gravity  program)   envisions  the recovering  of time  through, a  so, called  ``relational approach".  Such  approach,  as I understand,  is  based on the  idea of considering a   wave function of  several variables $\Psi( g_\alpha, m_i )$  (including geometric ones which  I have   labeled  generically  
   by $  g_\alpha$ and matter  related ones, which I have labeled   generically by $m_i$)  and  then select one  of those, say $ m_0$,    and  make    it play the role of  a ``subsidiary time",  a  step that  is followed  by  focusing  on  ``the relative  wave function"   characterizing the reminder of the  variables  ``when" the variable  $m_0$ takes the value  $t$.   This  sounds  too  close to  relying (implicitly)  on  something like the projection postulate, and, thus,   bringing in the   measurement problem  into the story, something that,   as I understand,  you  do not want to  do. Moreover,  the  step seems  fraught with  dangers:  Consider, for instance, a   pair of  fermions $(1,2)$  in a   singlet state $ |\Psi \rangle = \frac{1}{\sqrt{2}} [ | +z \rangle^{(1)} | -z \rangle^{(2)} - | -z \rangle^{(1)} | +z \rangle^{(2)} ] $, and    using on that   simple  system   a similar logic:  Consider the   $z$ component  of the  spin   of particle  $2$ as  subsidiary  time, and  then consider the relative  wave  functions at the two times  ``when" the variable  $z^{(2)}$ takes the value  $ +1/2$,  namely  $\Psi^{\rm rel}  = | -z \rangle^{(1)}$,   and  that  " when" the variable  $z^{(2)}$ takes the value  $-1/2$ namely  $\Psi^{\rm rel}  = | +z \rangle^{(1)} $ .     
    Both  results suggest  the system  has  spin 1/2 (pointing in some definite  direction), 
     while  the complete  system, of course,  has no  such  feature\footnote{ At this point one might  reply  with  the  counterargument  that the previous  characterization had  such problems only because  one has failed to  include the spin of particle  $2$  in the picture,  and that  a  more accurate  description  will make  use of  the 2  particle  wave functions, namely
   $  | +z \rangle^{(1)} | -z \rangle^{(2)}$ in one case,  and  $ | -z \rangle^{(1)} | +z \rangle^{(2)}$  for the other, both having    $J_z=0$. However,  then  one  would have to confront the  fact that   both such  2 particle  states have a component with total  angular momentum  $ J=+1$, while the original state  had  $ J=0$. I  can  imagine the list of `dangers" increasing as  one considers  more complex situations,  where things such as the Kochen Specker theorem \cite{kochen1975problem} ---exhibiting the fundamental  contextuallity of  quantum theory---might create  further difficulties.   Needless is to say that the general  idea   of recovering time  from   a relative   wave function,  and  its connection  with  all   the  above mentioned    issues  deserves  a  detailed  and  careful consideration, something  clearly outside the scope of this  work.   My aim here  is   just  to  note the fact  that  the strategy of  avoiding the measurement problem  is not so easily implemented, as the issue  has this  tendency to reenter through the back  door  while  one is  carefully closing  the  front one.}. Moreover,    you mentioned that   one  should   focus  on the  Dirac Observables, i.e., those that   commute  with   all constraints,  which,  as  I  understand, are  hard to come by  in any  canonical   approach to quantum  gravity.    What  I  am getting at   is, that it  seems  that,  as  much as  your approach  is  based on attempting to  take   distance   from confronting the  conceptual  problems of quantum theory, and  leave   those  to   be dealt with  in a future  and  separate study (a very understandable aim,  by the way, and one  supported  by the  old adage ``divide  and  conquer"),  they seem to  come   back at you with a vengeance. So, could  this  not be  a  clue  indicating  that   the two  issues    ought to be  confronted  simultaneously?  By the  way,  and  in all fairness,  I must acknowledge  that your approach  aims to   deal  with the BH information issue,  while  confronting the  problem of quantizing gravity,   a  subject  that, in the approach  I follow is left  for  {\it  a  future separated  study}. In this  sense,  we  are   both using the ``divide  and  conquer" strategy, and it is  mainly at  where  do we chose to  divide, and  where do we chose  to  work at conquering,  that  our approaches  diverge.   The  considerations  above aim to   explain my preferences.
   
ALEJANDRO: We are entering into an area where many things have been proposed and consensus is, in my view, still lacking. The setup of relational time you use in your example is  one of  the  possible paths to recover a notion of time (and  if following it,   one  would  have to  address the  issues you  noted),   but I am not sure it is the one that best fits my view. 
 I will thus  comment on your point using my personal perspective,  which does not necessarily reflect the point of view of other   colleagues.

I think the key aspect of your question concerns the measurement problem in quantum mechanics which is particularly severe, I agree, in the context of quantum gravity.  As this problem is hard and complex, in my approach  I try to rely  as much as possible on  the standard quantum framework  without entering into the difficult issue of the measurement problem (``divide and conquer'' as you say). Your question  is  one  that aims   to {\it  `drive   me  out of such calm shores into the turbulent open sea"}.   So let me,   at least in this first paragraph,  answer  while  staying  { \it ``safe  and close to  the  shore"}.   My personal view is that in quantum gravity we have to accept to renounce to the availability of a notion of time (in the sense of time coordinates in general relativity, and, of course, of space in the sense of coordinate location as well)  and that dynamical evolution will have to take a much more basic form in the more fundamental framework. This would not be surprising, after all, the timeless nature of general relativity is already responsible for the fact that there is no local notion of energy and momentum of the gravitational field, nothing like an energy momentum tensor of gravity makes sense.  Therefore, this  should  not  be taken as  indicating a problem but rather simply a feature  tied to  general covariance. Physical understanding is not impeded by the absence of a local notion of energy and momentum of gravitational degrees of freedom;  one just has to learn to do without that. Similarly, I believe that the more basic description will be essentially timeless.  Thus, I do not believe that questions of the type ``when'' something takes place  " then" something else happens will be possible  at the fundamental level, and the  kind of strategy based on a relative  wave function, at least  as you  described it,  will  need to  be replaced  by something else. {\it What   exactly?}, I can not  say at this  point.

 If such  strategy  is to work,  quantum states would  have to be interpreted as quantum states of the timeless universe,    encoding  features concerning   timeless Dirac operators (i.e. gauge invariant observables), even when such observables might not necessarily be measurable in any  practical sense. It is,  thus, likely that  such  description   will  not admit, in most situations,  a clear dynamical interpretation (something that we can phrase in terms of a spacetime with matter on it). The fact   that  such  states  encode the true fundamental degrees of freedom of quantum gravity would imply their eigenvalues are just the (timeless) quantum numbers labeling the possible quantum physical states of the theory. More precisely,  one   should envision that a complete set of commuting Dirac observables, characterizing such true degrees of freedom of quantum gravity,  ought to  be   available to be  used to construct (in terms of superposition of the corresponding eigenstates) suitable states with suitable semiclassical features: for instance, states that admit the interpretation of asymptotically flat semiclassical geometries with matter that admits a semiclassical interpretation in terms of QFT. 
 These states will not be semiclassical in any complete sense as they will generically develop strong curvature regions, like in gravitational collapse, or hide UV features that escape the scrutiny of low energy probes (like my defects here) and that are the basis of the often  called UV-completion of quantum gravity. Things we can measure will be a subclass of Dirac observables which act (in these subclass of semiclassical states) as quantities that we can identify with operations with a meaning that allows the separation of the system and the observer (outside the system). For instance, perhaps such observables will be available for asymptotic    
observers which (as such) might be considered outside the system. Maybe this is what would eventually allow for the application of something like a standard Copenhagen interpretation, yet where the separation between system and observer is made only under certain circumstances (like near the $\sI^\pm$ that such special states might be able to accommodate).  It seems to me that one might be able to push the standard quantum mechanical paradigm to such extreme case and it could be that this is where our theoretical modeling of nature might take us to. If so, the measurement problem might remain open and the hopes you hold, namely that 
somehow quantum gravity considerations might help resolving it, might just not be realized. I believe this is (unfortunately) a possibility.

But, of course, what I described  above   is not something  we can, at   this point, be  sure will work out, and  I am not completely convinced that  the  approach  I am putting forward here is the one that will result in a  satisfactory resolution of all related  issues, yet I believe it is one  worth pushing as far as possible to see where it leads, with the implicit hope of finding some nontrivial insight that could  eliminate the present incompleteness of the standard interpretation of the quantum theory. I  also believe that your perspective is an interesting one that ought to be explored for the very same reasons. In fact, we  should  acknowledge  the existence of   other possibilities which are quite appealing in  various  respects. For instance, the measurement problem could be resolved in a way perfectly compatible with the standard quantum paradigm as (one might argue is the case) in a scheme  similar to that   provided  by  Bohmian mechanics (BM) for non-relativistic quantum mechanics. In other words, by adding  new ingredients (analogous to the adding of the {``hidden variables"}  which  are  guided by the wave function in BM) that is yet completely compatible with the previous picture (Hilbert spaces, Dirac observables, etc.). 
In my opinion, the resolution of the issues that concern the interpretation of quantum theory as a theory of the universe (forced upon us  quite clearly  when considering  quantum gravity) and the related issues of interpretation often  discussed in  terms of the collapse  or  reduction of the wave function,  might be resolved in a way where the perspective I am taking in addressing the issue of information (i.e. maintaining unitarity)  remains unchanged in its basic structure, with the  new ingredients  complementing the   current picture in a  manner that  eliminates  the present   tensions  connected   with the measurement problem.

It is, of course, possible that nature is best modelled by a completely different description. It is possible that something like what you propose is the correct thing. All perspectives must be explored. I have taken one that, in my view, deserves attention,  but I remain open to novelty.

ALEJANDRO: The previous discussion  focuses on an issue that is    quite   challenging  for non perturbative (yet standard from the point of view of quantum theory) approaches to quantum gravity. 
I would like to know what form does it take in the path you follow?

DANIEL: As I  have  noted, the approach I follow  takes the question of  quantization of gravity as  something left for the future, when  hopefully we   would  have acquired  some   useful  clues from  the studies at the   {\it ``effective level"} we have  set  out to explore, so  it  is not  surprising we     cannot offer anything resembling a  definite answer to this difficult   question. Having said  that,  I think    it is  noteworthy that  we can say something that,   in my view, seems  quite appealing. The point  was first made in \cite{Okon:2013lsa}, and   is based on the  simple observation that, when including spontaneous collapses into the theory, the information encoded in different   Cauchy   hypersurfaces  of a given spacetime is  not longer the same. Information is  both  erased and  created by the collapse events  taking place in between those   hypersurfaces,   so that  the  requirement that   the   quantum state  for one is physically equivalent to that on the other  is  simply gone. As it is precisely   that  physical equivalence what  lies  in   ultimate instance behind the    timelessness of the    standard    theory, it seems very natural to expect   that  once,  a  spontaneous collapse   has been    incorporated into the theoretical framework, the   issue might  simply  disappear, and the  theory will    naturally incorporate space-time  notions.  Needless  is to say  that   this is not something    we can,  at this  point,   assert  with full confidence, but  nonetheless,  it seems clear  that  elements with  the potential  of removing the  difficulty are there   from the start.

In  any  event, I  agree wholeheartedly   with   you that,  in this  exploration, we must keep an open mind,  search  and explore the various   approaches  in looking  for clues or  problems,   and  attempt to extract  useful lessons from the exercise.  Scientists are not supposed to  become  irredeemably  attached to their own  preferred  theories. It is not as  if we are  ship captains  expected to  go  down  with their ships.   
\section{Conclusions}
\label{Con}

We have  reviewed the nature of the puzzle  known as the black hole information  loss paradox, offering a  rather   precise  characterization  of the   assumptions   one is   explicitly or implicitly accepting in any argument   indicating that there  is in fact a paradox. 

 We have  clarified  several  issues that  often  accompany the  discussions on the    subject,   and that,  more often than not,  are at the course  of  some   of the  strong    disagreements  among  colleagues interested in the    question.  
 
 We have,   then, presented    our  respective perspectives   regarding what we  think is the most likely path  towards  addressing the puzzle, 
 while   indicating,   either at the point of presentation, or in our discussion,  what  we see  as the  weakness,  or the aspects  that  
  require most   clarification  and  further work  on  each one  of  the two  proposals. It  it   noteworthy   how  despite our different perspectives the diagrams   depicting  the whole black hole   formation and  evaporation process,   have so  much in common\footnote{Including  an aspect  that  R. Penrose  noted,  seems a bit  unsettling, namely, the fact  that  so called  quantum gravity   region    has  a finite ( possibly long lasting)  shadow  on  $\sI^+$,  indicating that asymptotic  observers  would   perceive   this region,    which for them  will be   something akin to   naked singularity, not  as an instantaneous  burst  but as  an enduring  feature.}.  
  
   We have  strong  disagreements  that   have  been made  evident in the presentation  and in the   final discussion, starting of course,  with  the    opposite  answers  we  give to the question of whether or not  information is   lost, and    whether the initial  and final states  are  unitarily related,  but it is  perhaps  worthwhile  to point  out also  some  convergences.  We  both  call upon novel   and   discrete features  of nature, be them  the    defects in the geometry of Alejandro's  scheme, or the   collapse  events in the picture advocated by  Daniel.  It   is   worthwhile noting that  these entities are   very   different,   as far as  their spatio-temporal nature is  concerned.  Alejandro's  defects  are created  in the  QG  region  that replaces  the classical singularity, and then  they remain as entities that  follow  some  kind of  world-lines (most likely causal  or time-like), so each one of them might  be  considered  as  effectively  one-dimensional   entities  in space-time.   The   collapse  events   are,  as  their  name    suggest, essentially point like      entities  in four dimensional space-time,   so  they  do not  have  any  sort of permanence, a  feature intimately associated  with their capacity to erase information and   prevent  unitary  evolution.  
   
Daniel's  proposal connects the  black information  question directly 
 with the measurement problem of quantum theory,  although  gravity does play an important role in  enhancing the 
  collapse  dynamics,   so   as to ensure that the    very small level  of departure from unitary evolution affecting, say, individual particles in ordinary situations, becomes intense enough to   ensure complete  information  destruction  before the  quantum gravity region is reached, leaving to   quantum gravity   only the task of 
   covering the  result of those  collapses  together with  the  matter that  originally led to the information of  the  black hole  into a suitable ``vacuum", so that  space-time    after the QG  region would be  essentially  Minkowski  both  at the macro,  and  fundamental level.   In Alejandro's  proposal,  quantum  gravity plays  a much more important part of the story, generating the
     ``defects" that  remain  past the QG  region, and    do  so  in a  state of  complete entanglement  with the  Hawking radiation, so that 
the   state of the full system (quantum fields  and  quantum gravity DOF)  remains pure.  The measurement problem is 
viewed as  something completely separated  from the problem at hand.

Since the publication of Hawking's analysis, more than forty years ago, the issue of 
black hole information loss has remained a central topic of analysis and debate in theoretical physics.  
The fact is that, after all this time, the discussion continues, often hindered and exacerbated by confusion and misunderstanding among participants. 
We hope the present article helps clarify some aspect of the discussion, and highlights some of the attractive features of our own takes on the subject, illustrating at the same time,  the  serious  issues that lie   at the core of our  mutual  disagreements.

Finally,  needless is to say it,   in our  adoption of the  style used  by S. Hawking  and R. Penrose  to  air their  divergent  points of view on  similar subjects,  we   do  not intent to  put ourselves  in the  shoes  of  either of those  giants,  but merely to honor them both,  follow their example, and  continue to constructively  engage  and  explore  our   differences,  while   seeking to deepen our understanding of the workings of nature.

\section*{Acknowledgments}
We thank  Prof.  Roger  Penrose for the  permission to use  Figures 4 and  10 of his authorship.   We  thank  Guillermina Cabral  and   Bruno  Costa  for carefully reading the  manuscript, pointing out mistakes in the first drafts, and helping make it more  clear  and readable. AP  acknowledges support from I$\varphi$U of Aix-Marseille University.
DS  acknowledges partial financial support from  PAPIIT-DGAPA-UNAM project IG100120 and  CONACyT project 140630.   He is grateful for the support provided by the grant FQXI-MGA-1920 from the Foundational Questions Institute and the Fetzer Franklin Fund, a donor advised by the Silicon Valley Community Foundation.

\providecommand{\href}[2]{#2}\begingroup\raggedright\endgroup


\begin{thebibliography}{100}

\bibitem{Hawking:1996ny}
S.~W. Hawking and R.~Penrose, ``{The Nature of space and time},'' Sci. Am. {\bf
  275} (1996) 44--49.

\bibitem{Okon:2017pvc}
E.~Okon and D.~Sudarsky, ``{Losing stuff down a black hole},'' Found. Phys.
  {\bf 48} (2018), no.~4, 411--428,
\href{http://arXiv.org/abs/1710.01451}{{\tt arXiv:1710.01451}}.

\bibitem{Strominger:1996sh}
A.~Strominger and C.~Vafa, ``{Microscopic origin of the Bekenstein-Hawking
  entropy},'' Phys. Lett. B {\bf 379} (1996) 99--104,
  \href{http://arXiv.org/abs/hep-th/9601029}{{\tt arXiv:hep-th/9601029}}.

\bibitem{G.:2015sda}
J.~F. Barbero~G. and A.~Perez, ``{Quantum Geometry and Black Holes},''
  arXiv:1501.02963; to appear in the World Scientific series (2015).

\bibitem{Bombelli:1986rw}
L.~Bombelli, R.~K. Koul, J.~Lee, and R.~D. Sorkin, ``{A Quantum Source of
  Entropy for Black Holes},'' Phys.Rev. {\bf D34} (1986)
373--383.

\bibitem{Sudarsky:2002yi}
D.~Sudarsky, ``{A Schroedinger black hole and its entropy},'' Mod. Phys. Lett.
  A {\bf 17} (2002) 1047--1057.

\bibitem{Bonder:2017jcu}
Y.~Bonder, C.~Chryssomalakos, and D.~Sudarsky, ``{Extracting Geometry from
  Quantum Spacetime: Obstacles down the road},'' Found. Phys. {\bf 48} (2018),
  no.~9, 1038--1060, \href{http://arXiv.org/abs/1706.08221}{{\tt
  arXiv:1706.08221}}.

\bibitem{d2018conceptual}
B. d'Espagnat,  {\em Conceptual foundations of quantum mechanics}.
\newblock CRC Press, 2018.

\bibitem{Wald:1995yp}
R.~M. Wald, {\em {Quantum Field Theory in Curved Space-Time and Black Hole
  Thermodynamics}}.
\newblock Chicago Lectures in Physics. University of Chicago Press, Chicago,
  IL,
1995.
\newblock

\bibitem{Maudlin:2017lye}
T.~Maudlin, ``{(Information) Paradox Lost},''
  \href{http://arXiv.org/abs/1705.03541}{{\tt arXiv:1705.03541}}.

\bibitem{Pearle:1988uh}
P.~M. Pearle, ``{Combining Stochastic Dynamical State Vector Reduction With
  Spontaneous Localization},'' Phys. Rev. A {\bf 39} (1989) 2277--2289.

\bibitem{Ghirardi:1989cn}
G.~C. Ghirardi, P.~M. Pearle, and A.~Rimini, ``{Markov Processes in Hilbert
  Space and Continuous Spontaneous Localization of Systems of Identical
  Particles},'' Phys. Rev. A {\bf 42} (1990) 78--79.

\bibitem{Diosi:1998px}
L.~Diosi, N.~Gisin, and W.~T. Strunz, ``{NonMarkovian quantum state
  diffusion},'' Phys. Rev. A {\bf 58} (1998) 1699,
  \href{http://arXiv.org/abs/quant-ph/9803062}{{\tt arXiv:quant-ph/9803062}}.

\bibitem{Gisin:1989sx}
N.~Gisin, ``{Stochastic quantum dynamics and relativity},'' Helv. Phys. Acta
  {\bf 62} (1989) 363--371.

\bibitem{Wald:1984rg}
R.~Wald, {\em General Relativity}.
\newblock University of Chicago Press, Chicago, 1984.

\bibitem{Ashtekar:2004cn}
A.~Ashtekar and B.~Krishnan, ``{Isolated and dynamical horizons and their
  applications},'' Living Rev. Rel. {\bf 7} (2004) 10,
\href{http://arXiv.org/abs/gr-qc/0407042}{{\tt arXiv:gr-qc/0407042}}.

\bibitem{Bousso:2015qqa}
R.~Bousso and N.~Engelhardt, ``{Proof of a New Area Law in General
  Relativity},'' Phys. Rev. D {\bf 92} (2015), no.~4, 044031,
  \href{http://arXiv.org/abs/1504.07660}{{\tt arXiv:1504.07660}}.

\bibitem{Ashtekar:2020ifw}
A.~Ashtekar, ``{Black Hole evaporation: A Perspective from Loop Quantum
  Gravity},'' Universe {\bf 6} (2020), no.~2, 21,
  \href{http://arXiv.org/abs/2001.08833}{{\tt arXiv:2001.08833}}.

\bibitem{Blau}
M.~Blau, ``{Lecture Notes on General Relativity},''
http://www.blau.itp.unibe.ch/GRLecturenotes.html.

\bibitem{Hawking:1973uf}
S.~W. Hawking and G.~F.~R. Ellis, {\em {The Large Scale Structure of
  Space-Time}}.
\newblock Cambridge Monographs on Mathematical Physics. Cambridge University
  Press, 2, 2011.

\bibitem{Hawking:1975vcx}
S.~W. Hawking, ``{Particle Creation by Black Holes},'' Commun. Math. Phys. {\bf
  43} (1975) 199--220. [Erratum: Commun.Math.Phys. 46, 206 (1976)].

\bibitem{Josset:2016vrq}
T.~Josset, A.~Perez, and D.~Sudarsky, ``{Dark energy as the weight of violating
  energy conservation},'' Phys. Rev. Lett. {\bf 118} (2017), no.~2, 021102,
\href{http://arXiv.org/abs/1604.04183}{{\tt arXiv:1604.04183}}.

\bibitem{Perez:2017krv}
A.~Perez and D.~Sudarsky, ``{Dark energy from quantum gravity discreteness},''
  Phys. Rev. Lett. {\bf 122} (2019), no.~22, 221302,
\href{http://arXiv.org/abs/1711.05183}{{\tt arXiv:1711.05183}}.

\bibitem{Perez:2018wlo}
A.~Perez, D.~Sudarsky, and J.~D. Bjorken, ``{A microscopic model for an
  emergent cosmological constant},'' Int. J. Mod. Phys. {\bf D27} (2018),
  no.~14, 1846002,
\href{http://arXiv.org/abs/1804.07162}{{\tt arXiv:1804.07162}}.

\bibitem{Perez:2019gyd}
A.~Perez and D.~Sudarsky, ``{Black holes, Planckian granularity, and the
  changing cosmological 'constant'},''
\href{http://arXiv.org/abs/1911.06059}{{\tt arXiv:1911.06059}}.

\bibitem{Perez:2020cwa}
A.~Perez, D.~Sudarsky, and E.~Wilson-Ewing, ``{Resolving the $H_0$ tension with
  diffusion},''
\href{http://arXiv.org/abs/2001.07536}{{\tt arXiv:2001.07536}}.

\bibitem{Amadei:2021aqd}
L.~Amadei and A.~Perez, ``{Inflation from the relaxation of the cosmological
  constant},'' \href{http://arXiv.org/abs/2104.08881}{{\tt arXiv:2104.08881}}.

\bibitem{BarberoG:2015xcq}
J.~F. Barbero~G. and A.~Perez, {\em {Quantum Geometry and Black Holes}},
  pp.~241--279.
\newblock WSP, 2017.
\newblock \href{http://arXiv.org/abs/1501.02963}{{\tt arXiv:1501.02963}}.

\bibitem{Perez:2017cmj}
A.~Perez, ``{Black Holes in Loop Quantum Gravity},'' Rept. Prog. Phys. {\bf 80}
  (2017), no.~12, 126901,
\href{http://arXiv.org/abs/1703.09149}{{\tt arXiv:1703.09149}}.

\bibitem{Wald:1975kc}
R.~M. Wald, ``{On Particle Creation by Black Holes},'' Commun. Math. Phys. {\bf
  45} (1975) 9--34.

\bibitem{Perez:2014ura}
A.~Perez, ``{Statistical and entanglement entropy for black holes in quantum
  geometry},'' Phys.Rev. {\bf D90} (2014), no.~8, 084015,
\href{http://arXiv.org/abs/1405.7287}{{\tt arXiv:1405.7287}}.

\bibitem{Bianchi:2012br}
E.~Bianchi, ``{Horizon entanglement entropy and universality of the graviton
  coupling},''
\href{http://arXiv.org/abs/1211.0522}{{\tt arXiv:1211.0522}}.

\bibitem{rovelli_2004}
C.~Rovelli, {\em Quantum Gravity}.
\newblock Cambridge Monographs on Mathematical Physics. Cambridge University
  Press, 2004.

\bibitem{thiemann_2007}
T.~Thiemann, {\em Modern Canonical Quantum General Relativity}.
\newblock Cambridge Monographs on Mathematical Physics. Cambridge University
  Press, 2007.

\bibitem{Bombelli:1987aa}
L.~Bombelli, J.~Lee, D.~Meyer, and R.~Sorkin, ``{Space-Time as a Causal Set},''
  Phys. Rev. Lett. {\bf 59} (1987)
521--524.

\bibitem{Ashtekar:1997yu}
A.~Ashtekar, J.~Baez, A.~Corichi, and K.~Krasnov, ``{Quantum geometry and black
  hole entropy},'' Phys.Rev.Lett. {\bf 80} (1998) 904--907,
\href{http://arXiv.org/abs/gr-qc/9710007}{{\tt arXiv:gr-qc/9710007}}.

\bibitem{Unruh:1988in}
W.~G. Unruh, ``{A Unimodular Theory of Canonical Quantum Gravity},'' Phys. Rev.
  {\bf D40} (1989)
1048.

\bibitem{Smolin:2010iq}
L.~Smolin, ``{Unimodular loop quantum gravity and the problems of time},''
  Phys. Rev. {\bf D84} (2011) 044047,
\href{http://arXiv.org/abs/1008.1759}{{\tt arXiv:1008.1759}}.

\bibitem{Smolin:2009ti}
L.~Smolin, ``{The Quantization of unimodular gravity and the cosmological
  constant problems},'' Phys. Rev. {\bf D80} (2009) 084003,
\href{http://arXiv.org/abs/0904.4841}{{\tt arXiv:0904.4841}}.

\bibitem{Chiou:2010ne}
D.-W. Chiou and M.~Geiller, ``{Unimodular Loop Quantum Cosmology},'' Phys. Rev.
  {\bf D82} (2010) 064012,
\href{http://arXiv.org/abs/1007.0735}{{\tt arXiv:1007.0735}}.

\bibitem{Amadei:2019ssp}
L.~Amadei and A.~Perez, ``{Hawking's information puzzle: a solution realized in
  loop quantum cosmology},''
\href{http://arXiv.org/abs/1911.00306}{{\tt arXiv:1911.00306}}.

\bibitem{Amadei:2019wjp}
L.~Amadei, H.~Liu, and A.~Perez, ``{Unitarity and information in quantum
  gravity: a simple example},'' Front. Astron. Space Sci. {\bf 8} (2021) 46,
  \href{http://arXiv.org/abs/1912.09750}{{\tt arXiv:1912.09750}}.

\bibitem{Perez:2014xca}
A.~Perez, ``{No firewalls in quantum gravity: the role of discreteness of
  quantum geometry in resolving the information loss paradox},'' Class. Quant.
  Grav. {\bf 32} (2015), no.~8, 084001,
\href{http://arXiv.org/abs/1410.7062}{{\tt arXiv:1410.7062}}.

\bibitem{Page:1976df}
D.~N. Page, ``{Particle Emission Rates from a Black Hole: Massless Particles
  from an Uncharged, Nonrotating Hole},'' Phys. Rev. D {\bf 13} (1976)
  198--206.

\bibitem{Fabbri:2005mw}
A.~Fabbri and J.~Navarro-Salas, ``{Modeling black hole evaporation},''
  Published in London, UK: Imp. Coll. Pr. (2005) 334 p.

\bibitem{Bojowald:2001xe}
M.~Bojowald, ``{Absence of singularity in loop quantum cosmology},''
  Phys.Rev.Lett. {\bf 86} (2001) 5227--5230,
\href{http://arXiv.org/abs/gr-qc/0102069}{{\tt arXiv:gr-qc/0102069}}.

\bibitem{Ashtekar:2006wn}
A.~Ashtekar, T.~Pawlowski, and P.~Singh, ``{Quantum Nature of the Big Bang:
  Improved dynamics},'' Phys. Rev. {\bf D74} (2006) 084003,
\href{http://arXiv.org/abs/gr-qc/0607039}{{\tt arXiv:gr-qc/0607039}}.

\bibitem{Ashtekar:2005cj}
A.~Ashtekar and M.~Bojowald, ``{Black hole evaporation: A Paradigm},''
  Class.Quant.Grav. {\bf 22} (2005) 3349--3362,
\href{http://arXiv.org/abs/gr-qc/0504029}{{\tt arXiv:gr-qc/0504029}}.

\bibitem{Liberati:2019fse}
S.~Liberati, G.~Tricella, and A.~Trombettoni, ``{The information loss problem:
  an analogue gravity perspective},'' Entropy {\bf 21} (2019), no.~10, 940,
\href{http://arXiv.org/abs/1908.01036}{{\tt arXiv:1908.01036}}.

\bibitem{Olmedo:2016ddn}
J.~Olmedo, ``{Brief review on black hole loop quantization},'' Universe {\bf 2}
  (2016), no.~2, 12, \href{http://arXiv.org/abs/1606.01429}{{\tt
  arXiv:1606.01429}}.

\bibitem{Gambini:2013hna}
R.~Gambini, J.~Olmedo, and J.~Pullin, ``{Quantum black holes in Loop Quantum
  Gravity},'' Class. Quant. Grav. {\bf 31} (2014) 095009,
  \href{http://arXiv.org/abs/1310.5996}{{\tt arXiv:1310.5996}}.

\bibitem{Chen:2014jwq}
P.~Chen, Y.~C. Ong, and D.-h. Yeom, ``{Black Hole Remnants and the Information
  Loss Paradox},'' Phys. Rept. {\bf 603} (2015) 1--45,
  \href{http://arXiv.org/abs/1412.8366}{{\tt arXiv:1412.8366}}.

\bibitem{Unruh:2012vd}
W.~Unruh, ``{Decoherence without Dissipation},'' Trans.Roy.Soc.Lond. {\bf 370}
  (2012) 4454,
\href{http://arXiv.org/abs/1205.6750}{{\tt arXiv:1205.6750}}.

\bibitem{Christodoulou:2016tuu}
M.~Christodoulou and T.~De~Lorenzo, ``{Volume inside old black holes},'' Phys.
  Rev. {\bf D94} (2016), no.~10, 104002,
\href{http://arXiv.org/abs/1604.07222}{{\tt arXiv:1604.07222}}.

\bibitem{Banks:1983by}
T.~Banks, L.~Susskind, and M.~E. Peskin, ``{Difficulties for the Evolution of
  Pure States Into Mixed States},'' Nucl.Phys. {\bf B244} (1984)
125.

\bibitem{Unruh:1995gn}
W.~G. Unruh and R.~M. Wald, ``{On evolution laws taking pure states to mixed
  states in quantum field theory},'' Phys.Rev. {\bf D52} (1995) 2176--2182,
\href{http://arXiv.org/abs/hep-th/9503024}{{\tt arXiv:hep-th/9503024}}.

\bibitem{Modak:2014vya}
S.~K. Modak, L.~Ortiz, I.~Pe\~na, and D.~Sudarsky, ``{Non-Paradoxical Loss of
  Information in Black Hole Evaporation in a Quantum Collapse Model},'' Phys.
  Rev. {\bf D91} (2015), no.~12, 124009,
\href{http://arXiv.org/abs/1408.3062}{{\tt arXiv:1408.3062}}.

\bibitem{Modak:2014qja}
S.~K. Modak, L.~Ort\'\i{}z, I.~Pe\~na, and D.~Sudarsky, ``{Black hole
  evaporation: information loss but no paradox},'' Gen. Rel. Grav. {\bf 47}
  (2015), no.~10, 120, \href{http://arXiv.org/abs/1406.4898}{{\tt
  arXiv:1406.4898}}.

\bibitem{Bedingham:2016aus}
D.~Bedingham, S.~K. Modak, and D.~Sudarsky, ``{Relativistic collapse dynamics
  and black hole information loss},'' Phys. Rev. D {\bf 94} (2016), no.~4,
  045009, \href{http://arXiv.org/abs/1604.06537}{{\tt arXiv:1604.06537}}.

\bibitem{Okon:2016qlh}
E.~Okon and D.~Sudarsky, ``{Black Holes, Information Loss and the Measurement
  Problem},'' Found. Phys. {\bf 47} (2017), no.~1, 120--131,
\href{http://arXiv.org/abs/1607.01255}{{\tt arXiv:1607.01255}}.

\bibitem{Modak:2017yth}
S.~K. Modak and D.~Sudarsky, ``{Collapse of the wavefunction, the information
  paradox and backreaction},'' Eur. Phys. J. C {\bf 78} (2018), no.~7, 556,
  \href{http://arXiv.org/abs/1711.01509}{{\tt arXiv:1711.01509}}.

\bibitem{Perez:2005gh}
A.~Perez, H.~Sahlmann, and D.~Sudarsky, ``{On the quantum origin of the seeds
  of cosmic structure},'' Class. Quant. Grav. {\bf 23} (2006) 2317--2354,
  \href{http://arXiv.org/abs/gr-qc/0508100}{{\tt arXiv:gr-qc/0508100}}.

\bibitem{DeUnanue:2008fw}
A.~De~Unanue and D.~Sudarsky, ``{Phenomenological analysis of quantum collapse
  as source of the seeds of cosmic structure},'' Phys. Rev. D {\bf 78} (2008)
  043510, \href{http://arXiv.org/abs/0801.4702}{{\tt arXiv:0801.4702}}.

\bibitem{Leon:2010fi}
G.~Leon and D.~Sudarsky, ``{The Slow roll condition and the amplitude of the
  primordial spectrum of cosmic fluctuations: Contrasts and similarities of
  standard account and the 'collapse scheme'},'' Class. Quant. Grav. {\bf 27}
  (2010) 225017, \href{http://arXiv.org/abs/1003.5950}{{\tt arXiv:1003.5950}}.

\bibitem{Leon:2010wv}
G.~Leon, A.~De~Unanue, and D.~Sudarsky, ``{Multiple quantum collapse of the
  inflaton field and its implications on the birth of cosmic structure},''
  Class. Quant. Grav. {\bf 28} (2011) 155010,
  \href{http://arXiv.org/abs/1012.2419}{{\tt arXiv:1012.2419}}.

\bibitem{Leon:2011hs}
G.~Leon and D.~Sudarsky, ``{Novel possibility of nonstandard statistics in the
  inflationary spectrum of primordial inhomogeneities},'' SIGMA {\bf 8} (2012)
  024, \href{http://arXiv.org/abs/1109.0052}{{\tt arXiv:1109.0052}}.

\bibitem{Landau:2011aa}
S.~J. Landau, C.~G. Scoccola, and D.~Sudarsky, ``{Cosmological constraints on
  non-standard inflationary quantum collapse models},'' Phys. Rev. D {\bf 85}
  (2012) 123001, \href{http://arXiv.org/abs/1112.1830}{{\tt arXiv:1112.1830}}.

\bibitem{Diez-Tejedor:2011plw}
A.~Diez-Tejedor and D.~Sudarsky, ``{Towards a formal description of the
  collapse approach to the inflationary origin of the seeds of cosmic
  structure},'' JCAP {\bf 07} (2012) 045,
  \href{http://arXiv.org/abs/1108.4928}{{\tt arXiv:1108.4928}}.

\bibitem{Landau:2011ljv}
S.~Landau, G.~Le\'on, and D.~Sudarsky, ``{Quantum Origin of the Primordial
  Fluctuation Spectrum and its Statistics},'' Phys. Rev. D {\bf 88} (2013),
  no.~2, 023526, \href{http://arXiv.org/abs/1107.3054}{{\tt arXiv:1107.3054}}.

\bibitem{Canate:2012nwv}
P.~Ca\~nate, P.~Pearle, and D.~Sudarsky, ``{Continuous spontaneous localization
  wave function collapse model as a mechanism for the emergence of cosmological
  asymmetries in inflation},'' Phys. Rev. D {\bf 87} (2013), no.~10, 104024,
  \href{http://arXiv.org/abs/1211.3463}{{\tt arXiv:1211.3463}}.

\bibitem{Leon:2017yna}
G.~Le\'on, A.~Majhi, E.~Okon, and D.~Sudarsky, ``{Expectation of primordial
  gravity waves generated during inflation},'' Phys. Rev. D {\bf 98} (2018),
  no.~2, 023512, \href{http://arXiv.org/abs/1712.02435}{{\tt
  arXiv:1712.02435}}.

\bibitem{Leon:2016ysi}
G.~Le\'on, A.~Majhi, E.~Okon, and D.~Sudarsky, ``{Reassessing the link between
  B-modes and inflation},'' Phys. Rev. D {\bf 96} (2017), no.~10, 101301,
  \href{http://arXiv.org/abs/1607.03523}{{\tt arXiv:1607.03523}}.

\bibitem{Okon:2013lsa}
E.~Okon and D.~Sudarsky, ``{Benefits of Objective Collapse Models for Cosmology
  and Quantum Gravity},'' Found. Phys. {\bf 44} (2014) 114--143,
  \href{http://arXiv.org/abs/1309.1730}{{\tt arXiv:1309.1730}}.

\bibitem{Okon:2016pty}
E.~Okon and D.~Sudarsky, ``{A (not so?) novel explanation for the very special
  initial state of the universe},'' Class. Quant. Grav. {\bf 33} (2016),
  no.~22, 225015, \href{http://arXiv.org/abs/1602.07006}{{\tt
  arXiv:1602.07006}}.

\bibitem{Rodriguez:2017rjh}
S.~Rodriguez and D.~Sudarsky, ``{Revisiting Higgs inflation in the context of
  collapse theories},'' JCAP {\bf 03} (2018) 006,
  \href{http://arXiv.org/abs/1711.04912}{{\tt arXiv:1711.04912}}.

\bibitem{Sudarsky:2020wts}
D.~Sudarsky, ``{Spontaneous Collapse Theories and Cosmology},'' Fundam. Theor.
  Phys. {\bf 198} (2020) 291--320.

\bibitem{Bassi:2012bg}
A.~Bassi, K.~Lochan, S.~Satin, T.~P. Singh, and H.~Ulbricht, ``{Models of
  Wave-function Collapse, Underlying Theories, and Experimental Tests},'' Rev.
  Mod. Phys. {\bf 85} (2013) 471--527,
  \href{http://arXiv.org/abs/1204.4325}{{\tt arXiv:1204.4325}}.

\bibitem{Pearle:1979vm}
P.~M. Pearle, ``{TOWARD EXPLAINING WHY EVENTS OCCUR},'' Int. J. Theor. Phys.
  {\bf 18} (1979) 489--518.

\bibitem{Ghirardi:1985mt}
G.~C. Ghirardi, A.~Rimini, and T.~Weber, ``{A Unified Dynamics for Micro and
  MACRO Systems},'' Phys. Rev. D {\bf 34} (1986) 470.

\bibitem{Penrose:1996cv}
R.~Penrose, ``{On gravity's role in quantum state reduction},'' Gen. Rel. Grav.
  {\bf 28} (1996) 581--600.

\bibitem{Hawking:1976ra}
S.~Hawking, ``{Breakdown of Predictability in Gravitational Collapse},''
  Phys.Rev. {\bf D14} (1976)
2460--2473.

\bibitem{maudlin1995three}
T.~Maudlin, ``Three measurement problems,'' topoi {\bf 14} (1995), no.~1,
  7--15.

\bibitem{Schlosshauer:2003zy}
M.~Schlosshauer, ``{Decoherence, the Measurement Problem, and Interpretations
  of Quantum Mechanics},'' Rev. Mod. Phys. {\bf 76} (2004) 1267--1305,
  \href{http://arXiv.org/abs/quant-ph/0312059}{{\tt arXiv:quant-ph/0312059}}.

\bibitem{Adler:2001us}
S.~L. Adler, ``{Why decoherence has not solved the measurement problem: A
  Response to P. W. Anderson},'' Stud. Hist. Phil. Sci. B {\bf 34} (2003)
  135--142, \href{http://arXiv.org/abs/quant-ph/0112095}{{\tt
  arXiv:quant-ph/0112095}}.

\bibitem{Okon:2015fgr}
E.~Okon and D.~Sudarsky, ``{Less Decoherence and More Coherence in Quantum
  Gravity, Inflationary Cosmology and Elsewhere},'' Found. Phys. {\bf 46}
  (2016), no.~7, 852--879, \href{http://arXiv.org/abs/1512.05298}{{\tt
  arXiv:1512.05298}}.

\bibitem{Bedingham:2010hz}
D.~J. Bedingham, ``{Relativistic state reduction dynamics},'' Found. Phys. {\bf
  41} (2011) 686--704, \href{http://arXiv.org/abs/1003.2774}{{\tt
  arXiv:1003.2774}}.

\bibitem{Tumulka:2006}
R.~Tumulka, ``A Relativistic Version of the Ghirardi Rimini Weber Model,''
  Journal of Statistical Physics {\bf 125} (Dec, 2006) 821840.

\bibitem{Pearle:2014tda}
P.~Pearle, ``{Relativistic dynamical collapse model},'' Phys. Rev. D {\bf 91}
  (2015), no.~10, 105012, \href{http://arXiv.org/abs/1412.6723}{{\tt
  arXiv:1412.6723}}.

\bibitem{Durr:2013asa}
D.~D\"urr, S.~Goldstein, T.~Norsen, W.~Struyve, and N.~Zangh\`\i{}, ``{Can
  Bohmian mechanics be made relativistic?},'' Proc. Roy. Soc. Lond. A {\bf 470}
  (2013) 20130699, \href{http://arXiv.org/abs/1307.1714}{{\tt
  arXiv:1307.1714}}.

\bibitem{myrvold2002peaceful}
W.~C. Myrvold, ``On peaceful coexistence: is the collapse postulate
  incompatible with relativity?,'' Studies in History and Philosophy of Science
  Part B: Studies in History and Philosophy of Modern Physics {\bf 33} (2002),
  no.~3, 435--466.

\bibitem{Bengochea:2020efe}
G.~R. Bengochea, G.~Le\'on, P.~Pearle, and D.~Sudarsky, ``{Discussions about
  the landscape of possibilities for treatments of cosmic inflation involving
  continuous spontaneous localization models},'' Eur. Phys. J. C {\bf 80}
  (2020), no.~11, 1021, \href{http://arXiv.org/abs/2008.05285}{{\tt
  arXiv:2008.05285}}.

\bibitem{myrvold2017relativistic}
W.~C. Myrvold, ``Relativistic Markovian dynamical collapse theories must employ
  nonstandard degrees of freedom,'' Physical Review A {\bf 96} (2017), no.~6,
  062116.

\bibitem{Feldmann:2012}
W.~Feldmann and R.~Tumulka, ``Parameter diagrams of the GRW and CSL theories of
  wavefunction collapse,'' Journal of Physics A: Mathematical and Theoretical
  {\bf 45} (Jan, 2012) 065304.

\bibitem{Lindblad:1975ef}
G.~Lindblad, ``{On the Generators of Quantum Dynamical Semigroups},'' Commun.
  Math. Phys. {\bf 48} (1976) 119.

\bibitem{Page:1981aj}
D.~N. Page and C.~D. Geilker, ``{Indirect Evidence for Quantum Gravity},''
  Phys. Rev. Lett. {\bf 47} (1981) 979--982.

\bibitem{Diosi:1999py}
L.~Diosi, N.~Gisin, and W.~T. Strunz, ``{Royal road to coupling classical and
  quantum dynamics},'' Phys. Rev. A {\bf 61} (2000) 22108,
  \href{http://arXiv.org/abs/quant-ph/9902069}{{\tt arXiv:quant-ph/9902069}}.

\bibitem{Mattingly:2006pu}
J.~Mattingly, ``{Why Eppley and Hannah's thought experiment fails},'' Phys.
  Rev. D {\bf 73} (2006) 064025, \href{http://arXiv.org/abs/gr-qc/0601127}{{\tt
  arXiv:gr-qc/0601127}}.

\bibitem{Carlip:2008zf}
S.~Carlip, ``{Is Quantum Gravity Necessary?},'' Class. Quant. Grav. {\bf 25}
  (2008) 154010, \href{http://arXiv.org/abs/0803.3456}{{\tt arXiv:0803.3456}}.

\bibitem{Huggett:2001vyi}
N.~Huggett and C.~Callender, ``{Why Quantize Gravity (Or Any Other Field for
  That Matter)?},'' Phil. Sci. {\bf 68} (2001), no.~S3, S382--S394.

\bibitem{Struyve:2017mff}
W.~Struyve, {\em {Towards a Novel Approach to Semi-Classical Gravity}},
  pp.~356--374.
\newblock 2017.
\newblock \href{http://arXiv.org/abs/1902.02188}{{\tt arXiv:1902.02188}}.

\bibitem{Juarez-Aubry:2019jon}
B.~A. Ju\'arez-Aubry, T.~Miramontes, and D.~Sudarsky, ``{Semiclassical theories
  as initial value problems},'' J. Math. Phys. {\bf 61} (2020), no.~3, 032301,
  \href{http://arXiv.org/abs/1907.09960}{{\tt arXiv:1907.09960}}.

\bibitem{ghirardi1994outcome}
G.~Ghirardi and R.~Grassi, ``Outcome predictions and property attribution: the
  EPR argument reconsidered,'' Studies in History and Philosophy of Science
  Part A {\bf 25} (1994), no.~3, 397--423.

\bibitem{Maudlin:2019bje}
T.~Maudlin, E.~Okon, and D.~Sudarsky, ``{On the Status of Conservation Laws in
  Physics: Implications for Semiclassical Gravity},'' Stud. Hist. Phil. Sci. B
  {\bf 69} (2020) 67--81, \href{http://arXiv.org/abs/1910.06473}{{\tt
  arXiv:1910.06473}}.

\bibitem{Juarez-Aubry:2017ery}
B.~A. Ju\'arez-Aubry, B.~S. Kay, and D.~Sudarsky, ``{Generally covariant
  dynamical reduction models and the Hadamard condition},'' Phys. Rev. D {\bf
  97} (2018), no.~2, 025010, \href{http://arXiv.org/abs/1708.09371}{{\tt
  arXiv:1708.09371}}.

\bibitem{PhysRevLett.93.191301}
J.~Collins, A.~Perez, D.~Sudarsky, L.~Urrutia, and H.~Vucetich, ``Lorentz
  Invariance and Quantum Gravity: An Additional Fine-Tuning Problem?,'' Phys.
  Rev. Lett. {\bf 93} (Nov, 2004) 191301.

\bibitem{Smirne:2014paa}
A.~Smirne and A.~Bassi, ``{Toward an energy-conserving model of spontaneous
  wavefunction collapse},'' \href{http://arXiv.org/abs/1408.6446}{{\tt
  arXiv:1408.6446}}.

\bibitem{aharonov1981can}
Y.~Aharonov and D.~Z. Albert, ``Can we make sense out of the measurement
  process in relativistic quantum mechanics?,'' Physical Review D {\bf 24}
  (1981), no.~2, 359.

\bibitem{aharonov1984usual}
Y.~Aharonov and D.~Z. Albert, ``Is the usual notion of time evolution adequate
  for quantum-mechanical systems? II. Relativistic considerations,'' Physical
  Review D {\bf 29} (1984), no.~2, 228.

\bibitem{Maudlin_2014}
T.~Maudlin, ``What Bell did,'' Journal of Physics A: Mathematical and
  Theoretical {\bf 47} (oct, 2014) 424010.

\bibitem{Sorkin:1999yj}
R.~D. Sorkin and D.~Sudarsky, ``{Large fluctuations in the horizon area and
  what they can tell us about entropy and quantum gravity},'' Class. Quant.
  Grav. {\bf 16} (1999) 3835--3857,
  \href{http://arXiv.org/abs/gr-qc/9902051}{{\tt arXiv:gr-qc/9902051}}.

\bibitem{Corichi:2000xf}
A.~Corichi and D.~Sudarsky, ``{When is S = A / 4?},'' Mod. Phys. Lett. A {\bf
  17} (2002) 1431--1444, \href{http://arXiv.org/abs/gr-qc/0010086}{{\tt
  arXiv:gr-qc/0010086}}.

\bibitem{sorkin1997statistical}
R.~D. Sorkin, ``The Statistical Mechanics of Black Hole Thermodynamics,'' 1997.

\end{thebibliography}
\end{document}